\documentclass[letterpaper,12pt]{article}
\pdfoutput=1

\usepackage{jheppub}
\usepackage{amsmath}
\usepackage{graphicx}
\usepackage{subfig}
\usepackage{cancel}

\def\({\left(}
\def\){\right)}
\def\[{\left[}
\def\]{\right]}
\def\ph1{\phantom{1}}

\newcommand{\labell}[1]{\label{#1}}

\newcommand{\reef}[1]{(\ref{#1})}

\newcommand{\mt}[1]{\textrm{\tiny #1}}

\newcommand{\be}{\begin{equation}}
\newcommand{\ee}{\end{equation}}
\newcommand{\beq}{\begin{equation}}
\newcommand{\eeq}{\end{equation}}
\newcommand{\ba}{\begin{aligned}}
\newcommand{\ea}{\end{aligned}}
\newcommand{\beqa}{\begin{eqnarray}}
\newcommand{\eeqa}{\end{eqnarray}}
\newcommand{\beqar}{\begin{eqnarray*}}
\newcommand{\eeqar}{\end{eqnarray*}}
\newcommand{\eg}{{\it e.g.,}\ }
\newcommand{\ie}{{\it i.e.,}\ }

\newcommand{\cL}{\mathcal{L}}

\newcommand{\la}{\lambda}

\newcommand{\Gn}{G_\mt{N}}

\newcommand{\sbh}{S_\mt{BH}}

\newcommand{\ga}{\gamma}

\newcommand{\G}{\Gamma}

\newcommand{\pd}[2]{\frac{\partial#1}{\partial#2}}

\newcommand{\ban}[1]{\begin{align}#1\end{align}}
\newcommand{\coords}[1]{\left\{#1 \right\}}

\DeclareMathOperator{\sgn}{sgn}

\newcommand{\bulkc}{\gamma} %{x}

\arxivnumber{1408.4770}

\title{Holographic Holes and Differential Entropy}

\author[a]{Matthew Headrick,}
\author[b]{Robert C. Myers,}
\author[b,c]{and Jason Wien}

\affiliation[a]{Martin Fisher School of Physics, Brandeis University, Waltham, Massachusetts 02453, USA}
\affiliation[b]{Perimeter Institute for Theoretical Physics, Waterloo, Ontario N2L 2Y5, Canada}
\affiliation[c]{Department of Physics \& Astronomy and Guelph-Waterloo Physics Institute,\\
University of Waterloo, Waterloo, Ontario N2L 3G1, Canada}

\emailAdd{mph@brandeis.edu}
\emailAdd{rmyers@perimeterinstitute.ca}
\emailAdd{jswien@gmail.com}

\abstract{Recently it has been shown that the Bekenstein-Hawking entropy formula evaluated on certain closed surfaces in the bulk of a holographic spacetime has an interpretation as the differential entropy of a particular family of intervals (or strips) in the boundary theory \cite{hole,jun}. We first extend this construction to bulk surfaces which vary in time. We then give a general proof of the equality between the gravitational entropy and the differential entropy. This proof applies to a broad class of holographic backgrounds possessing a generalized planar symmetry and to certain classes of higher-curvature theories of gravity. To apply this theorem, one can begin with a bulk surface and determine the appropriate family of boundary intervals by considering extremal surfaces tangent to the given surface in the bulk. Alternatively, one can begin with a family of boundary intervals; as we show, the differential entropy then equals the gravitational entropy of a bulk surface that emerges from the intersection of the neighboring entanglement wedges, in a continuum limit.}

\preprint{BRX-TH-6283}

\begin{document}
\maketitle

%%%%%%%%%%%%%%%%%%%%%%%%%%%%%%%%%%%%%%%%%%%%%%%%%%%%%%%%%%%%%%%%%%%%%%%%%%%%%%%%
%%%%%%%%%%%%%%%%%%%%%%%%%%%%%%%%%%%%%%%%%%%%%%%%%%%%%%%%%%%%%%%%%%%%%%%%%%%%%%%%
%%%%%%%%%%%%%%%%%%%%%%%%%%%%%%%%%%%%%%%%%%%%%%%%%%%%%%%%%%%%%%%%%%%%%%%%%%%%%%%%
\section{Introduction}

The Bekenstein-Hawking formula \cite{beks,hawk0} describes how the geometry of spacetime
encodes the entropy of an event horizon,
 \be
S_\mt{BH} = \frac{{\cal A}}{4\,\Gn}\ .
 \labell{prop0}
 \ee
One perspective on this formula (see \eg \cite{areaent}) is that $S_\mt{BH}$ characterizes the entanglement of the 
underlying degrees of freedom associated to the interior and exterior of the horizon. 
Recently, it was suggested that this idea actually applies much more broadly than just 
to event horizons. More precisely, the spacetime entanglement conjecture of \cite{new1}
states that, in a theory of quantum gravity, any state describing a smooth spacetime geometry
manifests the following property: for any sufficiently large region, there is a (finite) gravitational entropy
which is characteristic of the entanglement between the degrees of freedom describing
the given region and those describing its complement; furthermore, the leading contribution to this entropy is given by the Bekenstein-Hawking
formula \reef{prop0} evaluated on the boundary of the region. Of course, an implicit assumption is that the usual Einstein-Hilbert action
(including, possibly, a cosmological constant term) emerges as the leading
contribution in the low-energy effective gravitational action. As demonstrated in \cite{misha9},
higher curvature corrections to the gravitational action will also control the subleading contributions
to this entanglement entropy, which take a form similar to those in the Wald entropy \cite{WaldEnt}.

The AdS/CFT correspondence \cite{revue} provides a natural framework where one might examine this proposal.
In particular, in a strong coupling limit of the boundary theory, the bulk theory reduces to
Einstein gravity with a negative cosmological constant (and matter fields), and for suitable boundary states,
the bulk geometry is just a classical solution of the corresponding equations of motion; \eg the CFT
vacuum is described by anti-de Sitter space. Hence an obvious question is: Are there boundary observables
corresponding to $S_\mt{BH}$ evaluated on general surfaces in the bulk spacetime?

Of course, one such observable is the entanglement entropy of boundary regions
as calculated by the Ryu-Takayanagi (RT) prescription \cite{rt1,rt2,rt3}. In particular, the 
entanglement entropy for a specified spatial region $A$ in the boundary is evaluated as
 \be
S(A) = \mathrel{\mathop {\rm
ext}_{\scriptscriptstyle{a\sim A}} {}\!\!} \[\frac{{\cal A}(a)}{4\Gn}\]
 \labell{define}
 \ee
where $a\sim A$ indicates that the bulk surface $a$ is homologous
to the boundary region $A$ \cite{head,fur06a}. The symbol `ext'
indicates that one should extremize the area over all such surfaces
$a$. The RT prescription was tested in a variety of interesting ways, \eg \cite{rt3,head,highc}
and a general argument verifying this prescription was recently provided 
in \cite{aitor}.\footnote{We should add that the RT prescription 
was originally discussed in the context of static states of the boundary theory or for 
static background geometries in the dual gravity theory. Further, such a static situation is implicit
in the general argument of \cite{aitor}. Holographic entanglement entropy was first considered in
dynamical situations by Hubeny, Rangamani, and Takayanagi (HRT) \cite{hrt}. Their proposal
was essentially to extend eq.~\reef{define} to dynamical backgrounds
but it is fair to say that this HRT proposal has been subjected to fewer consistency tests, \eg
\cite{matcov,aron2}. We add this note here because in much of our analysis, \eg sections \ref{general} 
and \ref{new}, we are allowing for time-dependent
backgrounds and so implicitly we are applying the HRT prescription.}
Hence in this context, we are evaluating the Bekenstein-Hawking formula
\reef{prop0} on surfaces which generally do not correspond to a horizon in the
bulk.\footnote{An exception to this general rule arises for a spherical entangling surface on the boundary
of anti-de Sitter space in any dimension \cite{chm,eom1}.} 

Further, in considering AdS black holes, the usual AdS/CFT
dictionary equates an entropy on the boundary CFT to an entropy in the bulk gravity theory.
Hence it seems reasonable to infer that the Bekenstein-Hawking formula
in eq.~\reef{define} literally yields an entropy for the extremal surface in the bulk. 
In fact, a natural interpretation of recent work \cite{also} on one-loop corrections
to holographic entanglement entropy would be that the entanglement entropy in the boundary 
theory is associated with entanglement entropy in the bulk. Let us also add that there have been
previous speculations that evaluating eq.~\reef{prop0} on more general, \ie non-extremal, surfaces in the bulk
geometry may yield additional entropic measures of entanglement in the boundary
theory \cite{mark1,causal0}. 

Recently, observables in the boundary theory were constructed which yield the Bekenstein-Hawking
entropy \reef{prop0} of certain closed surfaces in the bulk spacetime \cite{hole,jun}.
This `hole-ographic' construction originated by considering closed curves in the bulk of AdS${}_3$ \cite{hole}, 
and was then extended to higher dimensions, to more general holographic
backgrounds (even beyond asymptotically AdS spacetimes), and to certain higher curvature bulk theories, 
including Lovelock gravity \cite{jun}. The key boundary quantity is the `differential entropy,'
 \be
E=\sum_{k=1}^n \[\, S(I_k)-S(I_k\cap I_{k+1})\,\]
\,, \labell{residue}
 \ee
where $S(I_k)$ is the entanglement entropy of a member $I_k$ of a family of intervals
that cover a time slice in the boundary. Then, it was shown
that applying the holographic prescription \reef{define} in a particular continuum limit
yields $E=\sbh$ for a corresponding surface in the bulk --- the details of the hole-ographic construction are reviewed in section \ref{time}. 
We should note that the higher dimensional extensions of \cite{jun} assume a simple planar boundary geometry
that is covered by strips $I_k$, each of a uniform width. As a result, the bulk surfaces which can be described in these
constructions have a profile that varies with only a single boundary coordinate. 

In this paper, we will further extend the hole-ographic construction to more general contexts. To do so, it will be useful to define a continuum version of the differential entropy, as follows:
\begin{equation}
E:=-\oint d\lambda\,\left.\pd{S(\gamma_L(\lambda'),\gamma_R(\lambda))}{\lambda'}\right|_{\lambda'=\lambda}\,.
\labell{left}
\end{equation}
Here $\gamma^a_{L,R}(\lambda)$ denotes the left and right endpoints of a family of intervals (or strips, in the higher-dimensional case) that depend periodically on the parameter $\lambda$. As we will discuss in section \ref{time}, in simple situations this expression \reef{right} can be derived as the continuum limit of eq.~\eqref{residue}; however, it can be applied to much more general families of intervals. To avoid giving the impression that the left endpoint is playing a distinguished role here, we note
that, after an integration by parts, eq.~\reef{left} becomes
\begin{equation}
E=\oint d\lambda\,\left.\pd{S(\gamma_L(\lambda),\gamma_R(\lambda'))}{\lambda'}\right|_{\lambda'=\lambda}\,.\labell{right}
\end{equation}
Note also that $E$ is invariant under (orientation-preserving) reparametrizations of $\lambda$.

Given the continuum definition of the differential entropy \eqref{right}, which does not refer to intersections of intervals, there is no particular reason to restrict the intervals to lie on a constant-time slice, or even a common Cauchy slice. (Of course, each interval must lie on a Cauchy slice in order to have a well-defined entanglement entropy.) The question thus naturally arises of whether there exists a covariant `holographic hole' correspondence, in which the bulk curve and corresponding boundary intervals are not restricted to lie on a constant-time slice of a static spacetime. We will investigate this question in section \ref{time} by studying time-varying holes in planar AdS${}_3$. Starting from a generic spacelike curve $\gamma_B(\lambda)$ in the bulk, we construct a family of boundary intervals by finding the geodesic tangent to it for each $\lambda$; the geodesic's endpoints on the boundary define $\gamma_L(\lambda),\gamma_R(\lambda)$. By explicit calculation, we find that the differential entropy of this family of intervals agrees with the gravitational entropy of $\gamma_B$. This shows that the hole-ographic correspondence is not restricted to constant-time slices of static spacetimes.

The agreement between the gravitational and differential entropies found in so many different contexts calls for a unifying explanation. We provide one in section \ref{general}. Specifically, using basic tools from classical mechanics, we show that, if for all $\lambda$ the extremal curve giving the entropy of the interval $[\gamma_L(\lambda),\gamma_R(\lambda)]$ is tangent to the bulk curve $\gamma_B$, then the differential entropy of the family of intervals equals the gravitational entropy of $\gamma_B$. The theorem naturally encompasses the higher-dimensional and higher-curvature cases studied in \cite{jun}, as well as time-varying bulk curves. It makes only limited assumptions about the geometry of the holographic background, \eg the latter need not be asymptotically AdS, and we make precise the generalized planar symmetry which is required in section \ref{general}. 

The hole-ographic constructions considered in section \ref{time}, as well as in \cite{hole,jun}, begin with a bulk curve, and then find the appropriate family of boundary intervals by constructing extremal surfaces tangent to it. However, in more complicated spacetimes than AdS${}_3$, such a `bulk-to-boundary' construction can potentially fail in two ways: first, the tangent extremal curve may not reach the boundary (\eg it may hit a singularity instead); second, even if it does reach the boundary, it may not be the \emph{minimal} extremal surface for the resulting boundary interval, and so may not correctly calculate the entanglement entropy. We are therefore motivated, in section \ref{new}, to establish a converse construction that starts with a family of boundary intervals and produces a bulk curve with gravitational entropy equal to their differential entropy. 

To successfully establish such a `boundary-to-bulk' construction, it turns out that we must address two issues. First, the extremal curves for the intervals are not in general tangent to a common bulk curve. Second, in some cases the differential entropy is negative, so it can't equal the area of any bulk surface. We resolve the first issue by showing that, in the proof of the above theorem, the tangency condition can be relaxed: at each intersection point, the vectors tangent to the extremal curve and to $\gamma_B$ need not be parallel, but can instead span a null plane. We then show how to construct $\gamma_B$ given a family of intervals (obeying a certain simple condition), such that this weaker condition is obeyed. However, there is a subtlety: it can happen that the two tangent vectors are oriented oppositely (\ie have negative dot product); whenever this happens, one finds that the area element of $\gamma_B$ must contribute \emph{negatively} to the gravitational entropy in order to give agreement with the differential entropy. Thus we are forced to generalize the notion of gravitational entropy, and define it as a \emph{signed} area, where certain parts of $\gamma_B$ contribute positively and others negatively. Indeed, by this definition the total gravitational entropy can be negative, thereby resolving the second issue above.

We conclude the paper with a brief discussion of our results and future directions in section \ref{discuss}. Appendix \ref{geom} provides explicit proofs of certain intuitive arguments which we used in section \ref{new} to establish our new geometric interpretation. Finally, we discuss the extension of our analysis to Lovelock gravity in appendix \ref{lovelock}.

%%%%%%%%%%%%%%%%%%%%%%%%%%%%%%%%%%%%%%%%%%%%%%%%%%%%%%%%%%%%%%%%%%%%%%%%%%%%%%%%
%%%%%%%%%%%%%%%%%%%%%%%%%%%%%%%%%%%%%%%%%%%%%%%%%%%%%%%%%%%%%%%%%%%%%%%%%%%%%%%%
%%%%%%%%%%%%%%%%%%%%%%%%%%%%%%%%%%%%%%%%%%%%%%%%%%%%%%%%%%%%%%%%%%%%%%%%%%%%%%%%
\section{Time-varying holes}
\labell{time}

In this section, we review the discussion of \cite{hole,jun} and generalize it to arbitrary spacelike bulk surfaces which can vary in time. This construction motivates the holographic lemma proved in section \ref{general}. 
To simplify the discussion, we only outline the construction explicitly for AdS$_3$; however, as we will see by the general argument of section \ref{general}, as long as the backgrounds possess a generalized planar symmetry, this procedure readily extends to higher dimensions, to other holographic backgrounds (\eg backgrounds that are not asymptotically AdS), and to
certain classes of higher-curvature gravity theories. The example of applying the hole-ographic construction to time-varying holes in higher dimensions can be found in \cite{thesis}.

\subsection{Setup}

Before proceeding with the above generalization, we begin with a general consideration of the continuum limit of
the differential entropy \reef{residue}, which in fact reveals the origin of this name. Recall that this limit was an essential step in establishing the equality $E=\sbh$ for a corresponding bulk surface. 
To begin, in the continuum limit, we replace the discrete label for boundary intervals $I_k$ by a continuous parameter $\lambda\in[0,1]$. That is, with $n$
intervals, we set $\lambda_k=k/n$ and then $\lambda$ becomes continuous in the limit $n\to\infty$.
As alluded to above, we specify the family of intervals by defining by two curves in the boundary, whose coordinates we denote $\gamma^a_L(\la)$ and $\gamma^a_R(\la)$, representing the left and right endpoints, respectively. Further, we assume the boundary conditions on $\la$ are periodic, \ie  $\gamma^a_{L,R}(0)=\gamma^a_{L,R}(1)$,\footnote{Implicitly, in the
cases considered in \cite{hole,jun}, the family of intervals covers an entire time slice in the boundary geometry.  As discussed in \cite{jun} for
Poincar\'e coordinates, we can then assume the spatial direction orthogonal to the boundary intervals is periodic.} and we denote the entanglement entropy of the interval at $\lambda$ by $S(\gamma_L(\la), \gamma_R(\la))$. Now consider eq.~\reef{residue}. As illustrated in figure \ref{LRintersect0}, the intersection $I_k\cap I_{k+1}$ corresponds to
the interval extending from $\gamma^a_L(\lambda_{k+1})$ of $I_{k+1}$ to $\gamma^a_R(\lambda_k)$ of $I_k$. 
Therefore we can write $S(I_k\cap I_{k+1})=S(\gamma_L(\lambda_{k+1}), \gamma_R(\lambda_k))$. 
Hence, written in terms of the endpoints of the intervals, eq.~\reef{residue} becomes\footnote{As discussed in \cite{jun}, 
if the bulk curve varies too rapidly in the
radial direction, the center of the intervals, \ie $(\gamma^a_L+\gamma^a_R)/2$, is not a monotonically increasing function of $\la$ and eq.~\reef{residue} must
be modified to accommodate this situation. However, written in terms of the endpoints of the intervals, the modified expression
still takes the form given in eq.~\reef{residue9}. \labell{wacky}} 
\beq
E=\sum_k\left[S(\gamma_L(\lambda_{k}), \gamma_R(\lambda_k))-
S(\gamma_L(\lambda_{k+1}), \gamma_R(\lambda_k))\right]\,.
\labell{residue9}
\eeq
Now,  in the continuum limit, we have
\beqa
\lim_{n\to\infty}\left[S(\gamma_L(\lambda_{k}), \gamma_R(\lambda_k))-
S(\gamma_L(\lambda_{k+1}), \gamma_R(\lambda_k))\right]&=&
S(\gamma_L(\la), \gamma_R(\la))-S( \gamma_L(\la+d\la), \gamma_R(\la))
\nonumber\\
=&&
-\frac{dS(\gamma_L(\la),\gamma_R(\la))}{d\gamma^a_L(\la)}\ \frac{d\gamma^a_L(\la)}{d\la}\ d\la
\labell{differ}
\eeqa 
and therefore, the differential entropy \reef{residue9} becomes
 \beqa
E&=&-\oint_0^1d\la\ \frac{dS(\gamma_L(\la),\gamma_R(\la))}{d\gamma^a_L(\la)}\,
\frac{d\gamma^a_L(\la)}{d\la}
\labell{left1}\\
&=&-\oint d\lambda\,\left.\pd{S(\gamma_L(\lambda'),\gamma_R(\lambda))}{\lambda'}\right|_{\lambda'=\lambda}\,.
\nonumber
\eeqa
That is, we have recovered eq.~\reef{left} as describing the continuum limit of eq.~\reef{residue}.  Alternatively,
we can shift the index in the second sum in eq.~\reef{residue} and consider
$E=\lim_{n\to\infty}\[\, S(I_k)-S(I_{k-1}\cap I_{k})\,\]$. It is straightforward to see that this approach yields
\beqa
E&=&\int_0^1d\la\ \frac{dS(\gamma_L(\la),\gamma_R(\la))}{d\gamma^a_R(\la)}\,
\frac{d\gamma^a_R(\la)}{d\la}
\labell{right1}\\
&=&\oint d\lambda\,\left.\pd{S(\gamma_L(\lambda),\gamma_R(\lambda'))}{\lambda'}\right|_{\lambda'=\lambda}\,.
\nonumber
 \eeqa
Hence this simple shift of the index in eq.~\reef{residue} has allowed us to recover eq.~\reef{right}.
Again the equality
of the two continuum expressions readily follows using integration by parts. In any event, either of these
expressions brings to light the `differential' character of differential entropy.

\begin{figure}[h!]
\begin{center}
\includegraphics[width=0.7\textwidth]{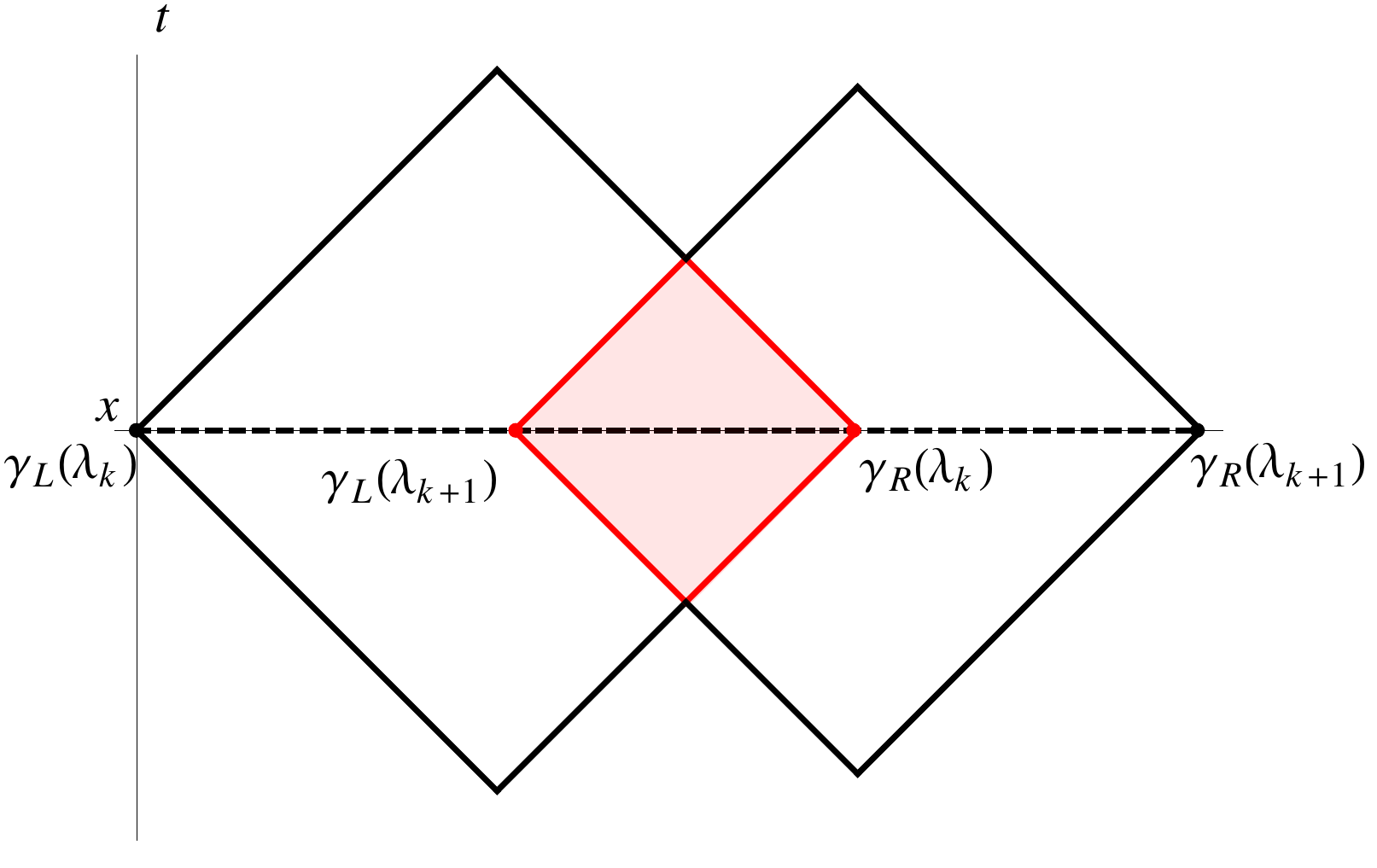}
\caption{(Colour online) The causal diamonds for two neighbouring intervals are drawn above:
$I_k$ with endpoints $\gamma_L(\la_k)$ and $\gamma_R(\la_k)$, and $I_{k+1}$ with $\gamma_L(\la_{k+1})$ and $\gamma_R(\la_{k+1})$. Red shading highlights the intersection region, which, of course, is the causal diamond for the interval $I_k\cap I_{k+1}$ with endpoints $\gamma_L(\la_{k+1})$ and $\gamma_R(\la_k)$.}\labell{LRintersect0}
\end{center}
\end{figure}

Now let us review the hole-ographic construction of \cite{hole,jun} using this new formalism. 
Given a spacelike curve in AdS$_3$, we must construct a family of boundary intervals whose differential entropy is equal to the gravitational entropy of the original curve. We will work in Poincar\'e coordinates with metric
\ban{ds^2= \frac {L^2}{z^2}\(dz^2 +dx^2-dt^2 \)\labell{3metric}}
where $L$ is the AdS radius. Let the initial curve in the bulk be specified by the parameterization $\bulkc_B(\la)=\lbrace Z(\la), X(\la), T(\la) \rbrace$ where $0\le\la\le1$. In addition, we impose periodic boundary conditions and rescale our parameterization so that $\bulkc_B(\la=0)=\bulkc_B(\la=1)$. As described in the introduction, we specify the corresponding
family of intervals on the asymptotic boundary at $z=0$ by the two endpoint curves: $\ga_L(\la)=\lbrace x_L(\la), t_L(\la) \rbrace$ and $\ga_R(\la)=\lbrace  x_R(\la), t_R(\la) \rbrace$.  Implicitly, here and throughout the paper, we are imposing that the $x$ direction is periodic with period $\Delta x=\ell$. One should think of the latter as some
infrared regulator scale, \eg it ensures that the proper length of the bulk curves considered here are finite. We assume that $\ell$ is always much
larger than the proper length of any of the intervals defined by $\ga_L$ and $\ga_R$.

The quantities we wish to compute are defined via volume functionals in Einstein gravity, and so this setup enjoys reparameterization invariance both for the Bekenstein-Hawking formula evaluated on the bulk surface and for the extremal-surface areas determining entanglement entropies in the boundary theory. Under reparameterization of $\bulkc_B(\la)$ via $\lambda \to \tilde \lambda$, the entropy of the hole given by the Bekenstein-Hawking formula \reef{prop0} is unchanged, as the volume functional keeps the same form, \ie
\ban{S_\mt{BH}=\frac 1 {4 G_N} \int_0^1  \sqrt{g_{\mu\nu}\pd{x^\mu}\lambda\pd{x^\nu}\lambda}\, d\lambda \ 
=\  \frac 1 {4 G_N} \int_0^1  \sqrt{g_{\mu\nu}\pd{x^\mu}{\tilde \lambda}\pd{x^\nu}{\tilde \lambda}}\, d\tilde \lambda\,.
\labell{reparam1}}
Similarly, we have reparameterization invariance for an extremal curve in the bulk, which determines the 
holographic entanglement entropy for an interval at fixed $\la$. Let $s$ be the `time' parameter
on these extremal curves, \ie  $\Gamma(s;\la)=\lbrace z(s;\la), x(s;\la), t(s;\la) 
\rbrace$ with the boundary conditions:  $\Gamma(s=0;\la)=\lbrace 0, 
\gamma^a_L(\la)\rbrace$ and $\Gamma(s=1;\la)=\lbrace 0, \gamma^a_R(\la)\rbrace$.
Then, since the volume functional is analogous to that above, reparameterizations $s\to\tilde s$ do not
 change the entropy of the interval at any given $\lambda$.

\subsection{Constant-$t$, constant-$z$ hole}
\label{VarTConstZ}
Next we show explicitly how to construct an appropriate family of intervals $[\gamma_L(\la)$, $\gamma_R(\la)]$ from the initial curve $\bulkc_B(\la)$ in the bulk, beginning with a re-derivation of the results of \cite{hole}. For each $\lambda$, we follow the extremal curve, \ie the geodesic, tangent to $\bulkc_B(\lambda)$ to the boundary, and the intersection of each geodesic with the boundary defines the endpoints $ \gamma_L(\lambda)$ and $ \gamma_R(\lambda)$. Stated in this way, this  prescription straightforwardly extends to more general cases.

For simplicity, let us first consider a bulk curve $\bulkc_B(\la)$ at constant $z=Z_0$ and $t=T_0$, \ie $\bulkc_B(\la) = \lbrace Z_0, \ell\la, T_0 \rbrace$   ---   recall that $\la\in[0,1]$ and $\ell$ is the period in the $x$ direction.
In this case the tangent curve is given by a semicircle parameterized by 
\ban{ \Gamma(s; \la) = \lbrace  Z_0 \sin s,\ell\lambda- Z_0 \cos s,T_0\rbrace  \labell{geod}}
where $s\in [0, \pi]$. Therefore we have 
\ban{\gamma_L (\lambda) = \coords{ \ell\lambda - Z_0,T_0} \hspace{0.5cm}  \gamma_R(\lambda)=\coords{\ell\lambda+Z_0,T_0} \labell{bound0} }
The general setup is illustrated in figure \ref{intervals}.

\begin{figure}[h!]
\begin{center}
\includegraphics[width=0.8\textwidth]{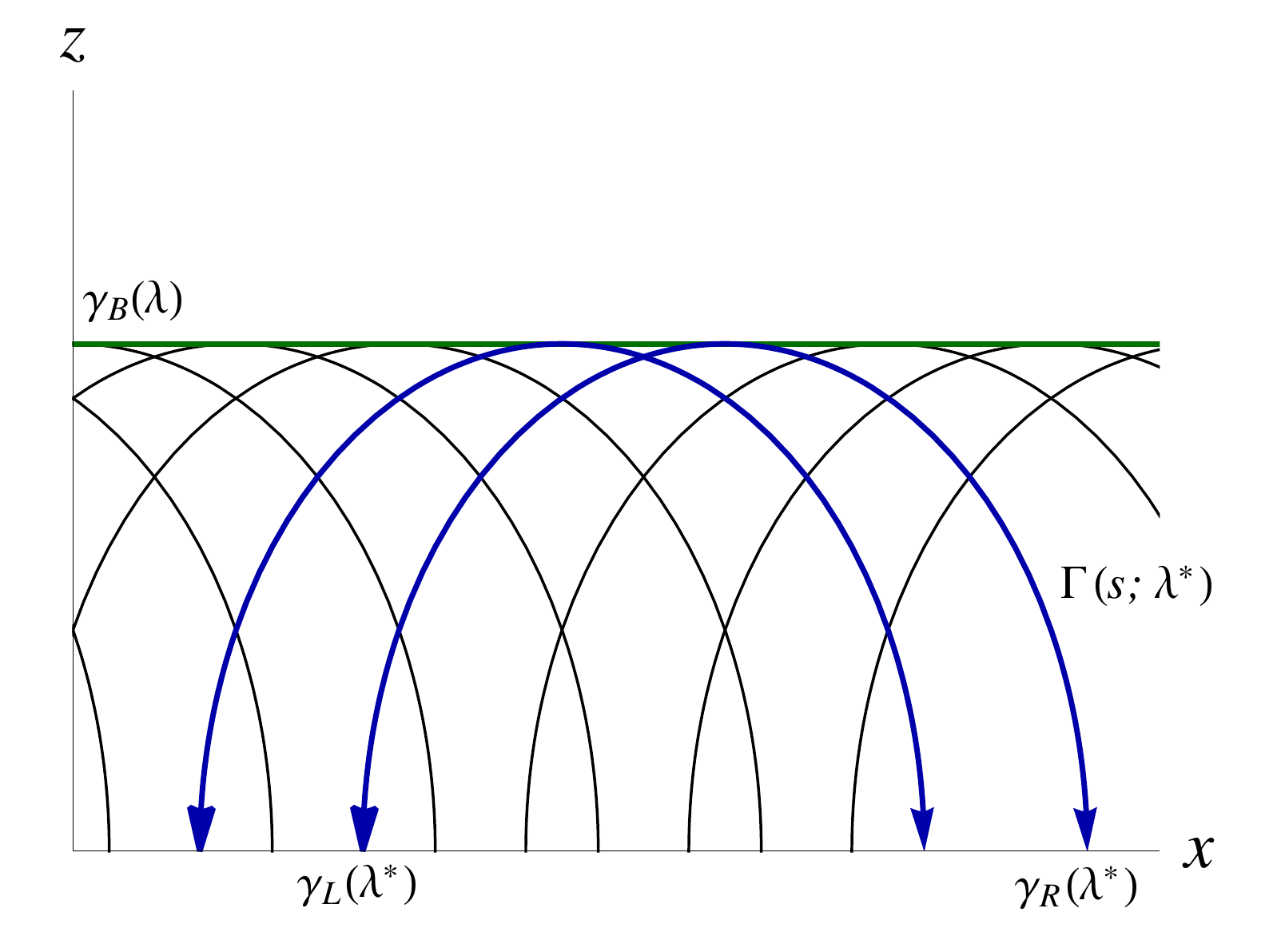}
\caption{(Colour online) The bulk curve $\bulkc_B(\la)$ is shown above in green, along with the tangent geodesics at each point. One such geodesic $\Gamma(s;\lambda^*)$ is highlighted in blue, along with a neighbouring geodesic at $\lambda^*-d\lambda$. The points $\gamma_L(\lambda^*)$ and $\gamma_R(\lambda^*)$ are explicitly drawn on the boundary at $z=0$. }\labell{intervals}
\end{center}
\end{figure}

The entanglement entropy of a single interval is given holographically by \cite{rt1}
\ban{S(\gamma_L(\lambda), \gamma_R(\lambda))=\frac L {2G_N} \log \[\frac {x_R(\lambda)-x_L(\lambda)}\delta
 \]=\frac L {2G_N} \log \[\frac {2Z_0}\delta \]\labell{entropy0}}
where $z=\delta$ is the position of the regulator surface in the AdS$_3$ geometry.\footnote{Of course, $\delta$ is also the short-distance cut-off for the boundary CFT.}
We can compute the differential entropy \reef{differ} to get
\ban{E= \frac L{4G_N}\, \int _0^1 \frac{\ell}{Z_0}\,d\lambda  \labell{diff0}}
Comparing this to gravitational entropy \reef{prop0} applied to $\bulkc_B$ we have
\ban{S_{BH}=\frac L{4G_N}\, \int_0^1 \frac{\ell}{Z_0}\,d\lambda \labell{area0}}
and hence $E=S_{BH}$.

\subsection{Time-varying, constant-$z$ hole}

We now let the bulk curve $\bulkc_B(\la)$ vary in time and be parameterized by 
\ban{
\bulkc_B (\lambda) = \lbrace Z_0, \ell\lambda, T(\lambda) \rbrace
}
For each point on the curve, we can construct the tangent extremal surface by following a geodesic in the direction of the tangent vector until it reaches the boundary. At a given $\lambda$, the tangent vector is proportional to 
\ban{
u(\lambda)= \lbrace 0,\ell ,T'(\lambda)\rbrace
}
To find the geodesic along this tangent vector, we take advantage of the Lorentz symmetry in the $(t,x)$-coordinates of the AdS$_3$ space \reef{3metric}. First we boost by angle $\beta(\la) = \log {\sqrt{\frac{\ell+ T'(\la)}{\ell- T'(\la)}}}$ so that the tangent vector has vanishing timelike component. In this boosted frame, the correct geodesic is simply given by $\Gamma^*(s;\la)=\lbrace Z_0 \sin s, \ell\lambda- Z_0 \cos s, T(\la)\rbrace$. 
Then we apply the inverse boost to construct the geodesic tangent to the curve in the original coordinate system.
\ban{
\G(s; \la)=\coords{ Z_0 \sin s,\ell\la-\frac{\ell\ Z_0 \cos s}{\sqrt{\ell^2-T'(\la)^2}},T(\la)-\frac{ T'(\la)Z_0 \cos s}{\sqrt{\ell^2-T'(\la)^2}}} \labell{geod1}
}
This extremal curve intersects the AdS boundary at $s=0$ and $s=\pi$, and so the family of intervals is given by 
\ban{\gamma_{R,L} (\lambda) = \coords{ \ell\lambda\pm \frac{\ell\, Z_0}{\sqrt{\ell^2-T'(\la)^2}}, T(\la)\pm \frac{ T'(\la)Z_0}{\sqrt{\ell^2-T'(\la)^2}}}\labell{family1}}
where the + and -- signs are chosen for $\gamma_R$ and $\gamma_L$, respectively.
\begin{figure}[h!]
\begin{center}
\includegraphics[width=0.7\textwidth]{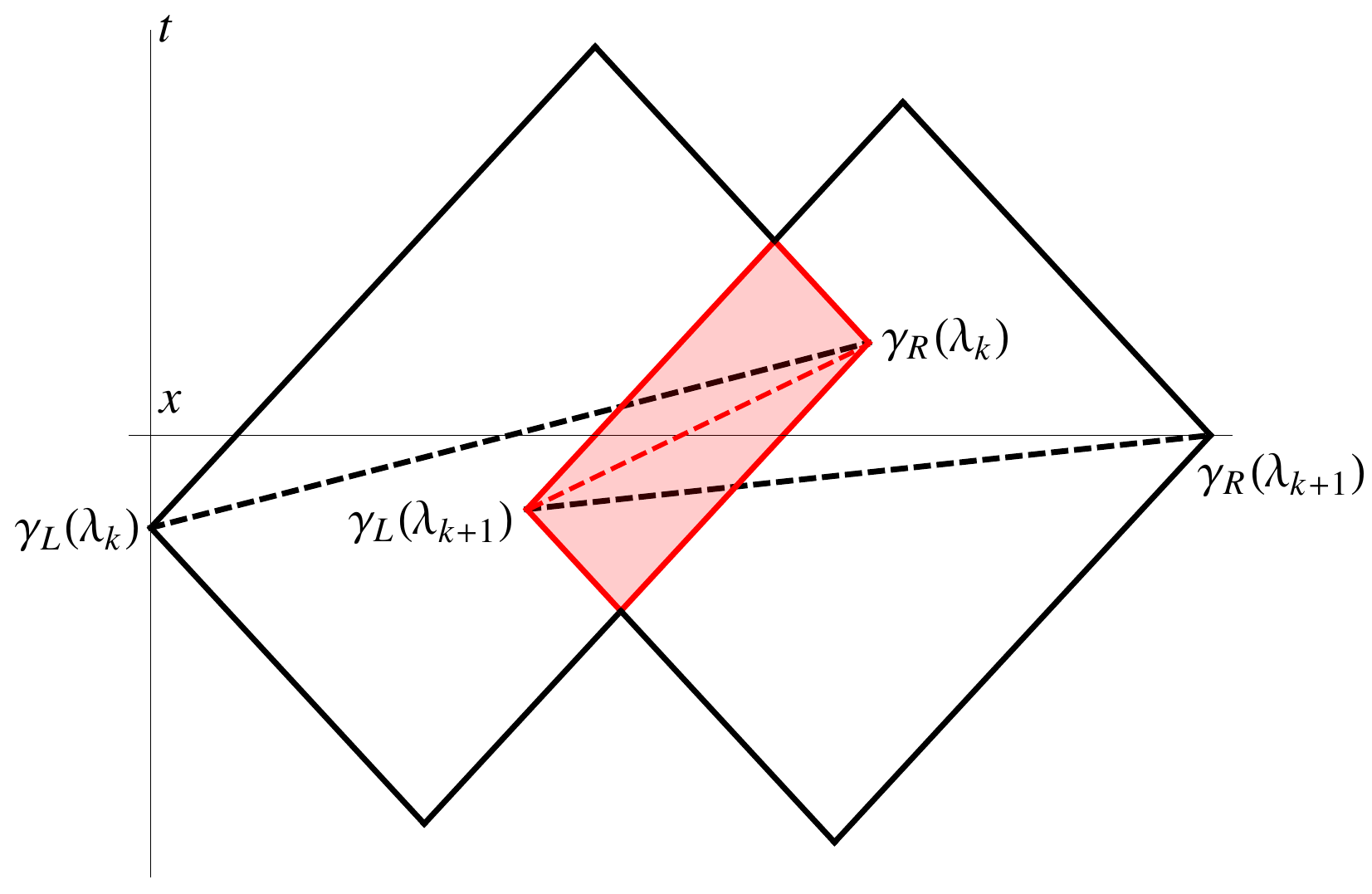}
\caption{(Colour online) The causal diamonds for two neighbouring intervals, $I_k$ and $I_{k+1}$, in the case of a time varying bulk curve. Their respective endpoints $\gamma_L(\la_k)$ and $\gamma_R(\la_k)$,	 and $\gamma_L(\la_{k+1})$ and $\gamma_R(\la_{k+1})$. The intersection of these two causal diamonds (highlighted with red shading) is the causal diamond for the interval with endpoints $\gamma_L(\la_{k+1})$ and $\gamma_R(\la_k)$.}\labell{LRintersect}
\end{center}
\end{figure}

To compute the entanglement entropy of each interval, we compute it in the boosted frame, where the result is known \reef{entropy0}, and carry it over to the original coordinates by Lorentz symmetry. Hence
\ban{S(\gamma_L(\lambda), \gamma_R(\lambda))=\frac L{2G_N}\, \log \[\frac {|\gamma_R-\gamma_L|}{\delta} \]=\frac L{4G_N}\, \log \[\frac {(x_R-x_L)^2-(t_R-t_L)^2}{\delta^2} \] \labell{Sboost}}
Substituting in \reef{family1} to our formulae, we can compute
\ban{E = \frac {L}{4G_N} \int _0^1 \frac 1{Z_0} \sqrt{\ell^2-T'(\la)^2}\ d\la\,. \label{ent1}}
Note that there is no total derivative contribution here since $Z'(\la)=0$  ---  compare with eq.~\reef{entropy2} in the following section.
The gravitational entropy of $\bulkc_B(\la)$ given by eq.~\reef{reparam1} is 
\ban{S_{BH}=\frac {L}{4G_N} \int _0^1 \frac 1{Z_0} \sqrt{\ell^2-T'(\la)^2}\ d\la \label{vol1}}
Comparing eqs.~\reef{ent1} and \reef{vol1}, we see that in this case $E=S_{BH}$. Note that $\bulkc_B(\la)$ is assumed to be spacelike everywhere, so $|T'(\la)^2|<\ell$. 

In closing, let us reconsider the original definition of differential entropy \reef{residue} for a moment. The time-varying holes
above and in the next section highlight the definition: $S(I_k\cap I_{k+1})=S(\gamma_L(\lambda_{k+1}), \gamma_R(\lambda_k))$, introduced above eq.~\reef{residue9}. In particular,
with the boundary intervals defined by following the geodesics tangent to each point along the bulk curve $\gamma_B(\lambda)$, we will typically find that neighbouring intervals $I(\lambda_k)$ and $I(\lambda_{k+1})$ are not on the same time slice, as illustrated in figure \ref{LRintersect}. In this case, the meaning of $I_k\cap I_{k+1}$ in eq.~\reef{residue}
becomes unclear. However, we can still naturally replace $I_k\cap I_{k+1}$ by the interval extending from $\gamma^a_L(\la_{k+1})$ of $I_{k+1}$ to $\gamma^a_R(\la_k)$ of $I_k$, as discussed above. This definition becomes intuitively clear if we picture the intersection of the corresponding causal diamonds on the boundary, shown in figure \ref{LRintersect}. Hence the continuum version \reef{left} of the differential entropy, which follows with this choice, consistently incorporates the case of time-varying bulk curves, as described above.

\subsection{Arbitrary hole}\labell{arbhole}

\begin{figure}[h!]
\begin{center}
\includegraphics[width=0.8\textwidth]{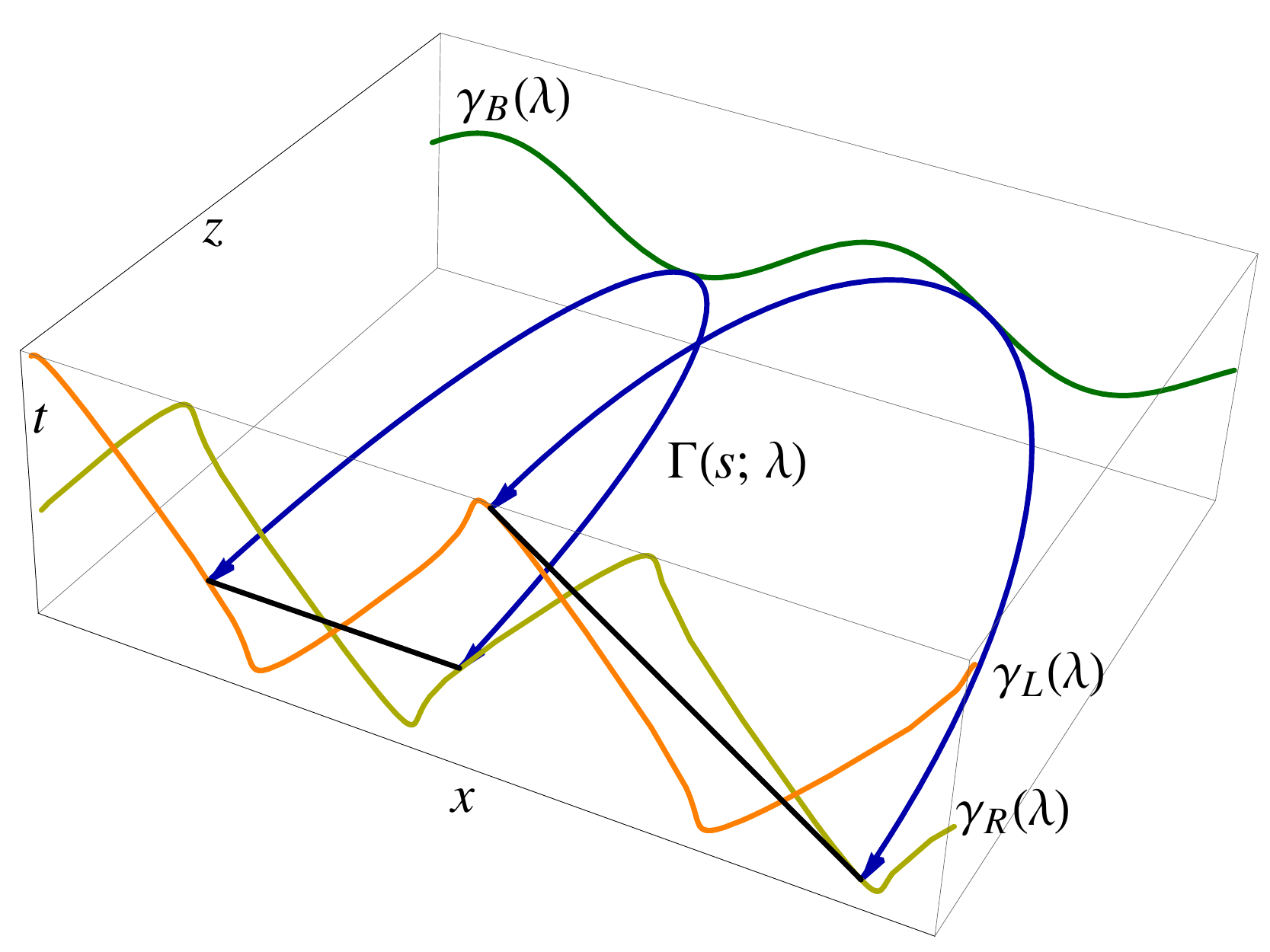}
\caption{(Colour online) For each point on the bulk curve $\bulkc_B(\la)$, the intersection of the extremal curve $\Gamma(s;\lambda)$ with the boundary defines an interval between $\gamma_L(\la)$ and $\gamma_R(\la)$. We take the family of intervals as described by the curves $\gamma_L(\la),\gamma_R(\la)$ shown in yellow and orange respectively. The differential entropy of this family of intervals equals the gravitational entropy of the bulk curve.}
\label{3Dintervals}
\end{center}
\end{figure}

We now consider an arbitrary bulk curve $\bulkc_B(\la)=\coords{Z(\lambda),X(\lambda), T(\lambda)}$ with the condition that its tangent vector is spacelike everywhere. To find the tangent extremal curve at a point, we again begin by boosting the tangent vector by $\beta(\la) = \log \sqrt{\frac {X'(\la)+T'(\la)}{X'(\la)-T'(\la)}}$ so it is completely spacelike. In the boosted coordinates, the tangent vector is proportional to 
\ban{u^*(\lambda)=\coords{Z'(\la), \sqrt{X'(\la)^2-T'(\la)^2}, 0} \labell{tan2}}
As constant time geodesics in AdS$_3$ are given by semicircles, we can just use Euclidean geometry in the ($z,x$)-plane to characterize the extremal curve. The tangent vector $u^*(\la)$ lies on a semi-circle, so following its normal vector $n^*(\la)$ to the boundary gives its center. We choose the (coordinate) length of $n^*(\la)$ such that $\bulkc_B(\lambda)+n^*(\la)$ lies on the boundary, so the coordinate radius of the semi-circle containing the geodesic is equal to $|n^*(\la)|$. We have
\ban{
n^*(\lambda)&= \frac{Z(\la)}{\sqrt{X'(\la)^2- T'(\la)^2}} \coords{  -\sqrt{X'(\la)^2- T'(\la)^2},  Z'(\la) ,0}
}
So $c^*(\lambda) \equiv \coords{0, X(\lambda) +\frac {Z(\la)Z'(\la)}{\sqrt{X'(\la)^2- T'(\la)^2}},T(\la)}$ is the center of the semi-circle in the boosted coordinates and $r^*(\lambda)\equiv Z(\la)\sqrt{1+ \frac { Z'(\la)^2}{X'(\la)^2-T'(\la)^2}}$ is the radius. Therefore we can parameterize this semicircle and boost back to the original coordinate system to get the tangent extremal curve as 
\ban{
\notag  \G(s; \la)= &\left\{r^*(\la)  \sin s,X(\la)+ \frac{Z(\la) Z'(\la)X'(\la) }{{X'(\la)^2- T'(\la)^2}}-\frac{X'(\la)\, r^*(\la)}{\sqrt{X'(\la)^2- T'(\la)^2}}\cos s,\right.\\
&\hspace{0.6cm}\left.T(\la)+ \frac{Z(\la) Z'(\la)T'(\la) }{{X'(\la)^2- T'(\la)^2}}-\frac{T'(\la)\, r^*(\la)}{\sqrt{X'(\la)^2- T'(\la)^2}}\cos s\right\}
\labell{housefire}}
An example of such a bulk curve and some tangent extremal curves are illustrated in figure \ref{3Dintervals}. 

Given this parameterization, it is straightforward to compute the differential entropy via \reef{differ} as
\ban{
\notag E= \frac L{4G_N}\int_0^1 d\la & \left( \frac 1 {Z(\la)}{ \sqrt{X'(\la)^2-T'(\la)^2+Z'(\la)^2}}\right. %\notag\\&\qquad
+\frac{Z''(\la)}{\sqrt{X'(\la)^2-T'(\la)^2+Z'(\la)^2}} \notag \\
&\qquad\qquad\left.+  \frac{Z'(\la)}{X'(\la)^2-T'(\la)^2}\frac{T'(\la)T''(\la)-X'(\la)X''(\la)}{\sqrt{X'(\la)^2-T'(\la)^2+Z'(\la)^2}}\right) \notag\\
\notag =\frac L {4G_N}\int_0^1d\la&  \frac 1{ Z(\la)}{ \sqrt{X'(\la)^2-T'(\la)^2+Z'(\la)^2}} 
% \\\notag&\qquad
 +\left.\frac L{4G_N} \sinh^{-1}\left({\frac{ Z'(\la)}{\sqrt{X'(\la)^2- T'(\la)^2}}}\right)\right|_{0}^1\\
=\frac L {4G_N}\int_0^1d\la & \frac 1 {Z(\la)}{ \sqrt{X'(\la)^2-T'(\la)^2+Z'(\la)^2}} \labell{entropy2} 
}
where the boundary term vanishes by the periodic boundary conditions for $\bulkc_B(\la)$. Computing the gravitational entropy for $\bulkc_B(\la)$, we have
\ban{
S_{BH}=\frac L{4G_N}\int_0^1 d\lambda  \frac 1 {Z(\la)} \sqrt{X'(\la)^2-T'(\la)^2+Z'(\la)^2} \labell{vol2}
}
Therefore we see that for any spacelike curve in AdS$_3$, $E=S_{BH}$. Note that in the case where $T'(\la)=0$, eqs. \reef{entropy2} and \reef{vol2} reduce to formulas found in \cite{jun} for constant-time bulk curves. In addition, this result extends straightforwardly to time-varying surfaces with planar symmetry in higher dimensions. Details for the latter can be found in \cite{thesis}.

From these constructions, we can see that an essential step involves choosing boundary intervals so that their extremal surfaces are tangent to the bulk surface. These examples also highlight the utility of describing the family of boundary intervals in terms the two endpoint curves, $\gamma_R(\lambda)$ and $\gamma_L(\lambda)$. We now turn to discussing the role of these observations in a more general framework.

%%%%%%%%%%%%%%%%%%%%%%%%%%%%%%%%%%%%%%%%%%%%%%%%%%%%%%%%%%%%%%%%%%%%%%%%%%%%%%%%
%%%%%%%%%%%%%%%%%%%%%%%%%%%%%%%%%%%%%%%%%%%%%%%%%%%%%%%%%%%%%%%%%%%%%%%%%%%%%%%%
%%%%%%%%%%%%%%%%%%%%%%%%%%%%%%%%%%%%%%%%%%%%%%%%%%%%%%%%%%%%%%%%%%%%%%%%%%%%%%%%
\section{General hole-ographic correspondence}
\labell{general}

In this section, we prove a theorem that establishes the
connection between differential entropy in the boundary theory and gravitational entropy
of bulk curves in a very general context. In particular, it allows for
higher dimensions, general holographic backgrounds depending on the
transverse coordinates (\ie both space and time dependence) as well as on the radial coordinate, 
certain higher-curvature bulk theories including Lovelock gravity,
and bulk surfaces that vary in time and radius. However, we should
note that, underlying our calculations, there is an assumption of a generalized planar symmetry, which we fully characterize in section \ref{planarS}.  Our construction makes use of two of the salient lessons coming from studying bulk surfaces that vary in time, as in the previous section:
\begin{itemize}
\item The boundary data that is input into the construction is a periodic family of boundary intervals $(\gamma_L(\lambda),\gamma_R(\lambda))$;
\item We can define the boundary intervals by finding an extremal surface
which is tangent to the bulk surface at each point.                               
\end{itemize}

The analysis of \cite{jun} also makes apparent that another essential ingredient in the hole-ographic construction
is that the holographic entanglement entropy is evaluated
by extremizing some geometric functional over various surfaces with fixed boundary conditions.  
In the examples in section \ref{time}, we evaluated the 
Bekenstein-Hawking formula on surfaces described by a specific ansatz depending on a single parameter $s$ and extremized the resulting expression subject to specific
boundary conditions at the endpoints. Hence the problem
of evaluating the holographic entanglement entropy is essentially reduced to
a standard problem in classical mechanics with a perhaps non-standard
Lagrangian. The proof therefore relies essentially on some standard tools from classical mechanics. Given its generality, as we will note in section \ref{discuss}, it has a broader applicability to probing the bulk geometry with other new `differential' observables besides the differential entropy.

\subsection{Proof}

We begin with a general action of the form 
 \be
S=\int_{s_i}^{s_f}\!ds
\ {\cL}(\gamma,\dot\gamma)
\,,
 \labell{action0}
 \ee
where, in applying classical-mechanics intuition, we will let the parameter $s$ play the role of `time'. The details of the Lagrangian $\cL$ will not be important in the
following except for two features. First, it is a function only of the position coordinates $\gamma^\mu$ and the corresponding `velocities' $\dot \gamma^\mu:=\partial_s\gamma^\mu$; no higher derivatives of $\gamma^\mu$ appear. While
this is, of course, conventional in classical mechanics, it is worth
emphasizing in the present context where one may wish to consider entropy
functionals for higher-curvature theories of gravity (\eg see appendix \ref{lovelock}).
Second, the action is reparameterization-invariant: if we reparameterize $s\to \tilde s=\tilde s(s)$, then
 \be
\int_{s_i}^{s_f}\!ds
\ {\cL}(\gamma,\partial_s\gamma)=\int_{\tilde s_i}^{\tilde s_f}\!d\tilde s
\ {\cL}(\gamma,\partial_{\tilde s}\gamma)
\,.
 \labell{action1}
 \ee
(For the moment, we require this to hold only for orientation-preserving reparametrizations.) A standard example is the length functional,
\begin{equation}
\cL(\gamma,\dot\gamma) = |\dot\gamma|:=\sqrt{g_{\mu\nu}(\gamma)\,\dot\gamma^\mu\,\dot\gamma^\nu}
\labell{length}
\end{equation}
(where, if the metric is Lorentzian, we require $\dot\gamma(s)$ to be everywhere spacelike).

Reparametrization invariance implies that $\cL$ is homogeneous of degree 1 in $\dot\gamma$, \ie
\begin{equation}
\cL(\gamma,\alpha \dot\gamma) = \alpha\cL(\gamma,\dot\gamma)\qquad(\alpha>0)\,.
\labell{homoL1}
\end{equation}
This in turn implies that
\begin{equation}
\cL(\gamma,\dot\gamma) = \dot\gamma^\mu\, p_\mu(\gamma,\dot\gamma)\,,
\labell{homoL}
\end{equation}
where $p_\mu$ is the canonical momentum,
\begin{equation}
p_\mu:=\pd\cL{\dot\gamma^\mu}\,.
\end{equation}
is the canonical momentum. An important consequence of this equality is that the canonical Hamiltonian vanishes identically. \eqref{homoL1} also implies that $p_\mu$ is homogeneous of degree 0 in $\dot\gamma$:
\begin{equation}
p_\mu(\gamma,\alpha\dot\gamma)=p_\mu(\gamma,\dot\gamma)\qquad(\alpha>0)\,.
\labell{homop}
\end{equation}
These facts will play an important role in what follows. It is easy to verify them in the case of the length functional \eqref{length}, for which
\begin{equation}
p_\mu = \frac{g_{\mu\nu}\dot \gamma^\nu}{|\dot\gamma|}\,.
\labell{lengthp}
\end{equation}

The second ingredient in our proof is the following classical-mechanics lemma (which does not rely on reparametrization invariance): Given a family of solutions $\Gamma^\mu(s;\lambda)$ of the equations of motion, that is continuous and periodic in the parameter $\lambda$, the quantity
\begin{equation}
R(s):=\oint d\lambda\left.\Gamma^{\prime\mu}\,p_\mu\right|_s\,,
\labell{Ecdef}
\end{equation}
where $\Gamma^{\prime\mu}:=\partial_\lambda\Gamma^\mu$, is independent of $s$ (\ie it is a conserved quantity). This can be proven as follows: Given any two
`times' $s_1<s_2$, we define\footnote{Implicitly below, we imagine that the solutions $\Gamma^\mu(s;\lambda)$ are defined on some longer interval $(s_1,s_3)$ but
we can restrict our attention to $(s_1,s_2)$ for any $s_1<s_2\le s_3$.}
\begin{equation}
S_{12}(\lambda) := \int_{s_1}^{s_2}ds\,\cL(\Gamma,\dot\Gamma)\,.
\end{equation}
A standard result in classical mechanics says that the derivative of the on-shell action with respect to the final position is the final momentum, and with respect to the initial position is the minus of the initial momentum. Hence
\begin{equation}
\frac{dS_{12}}{d\lambda} = \left.\Gamma^{\prime\mu}\,p_\mu\right|_{s_2} - \left.\Gamma^{\prime\mu}\,p_\mu\right|_{s_1}\,.
\end{equation}
Integrating over $\lambda$, and using the fact that $S_{12}(\lambda)$ is periodic, we find $R(s_1)=R(s_2)$. An alternative proof uses Noether's theorem: in the `field theory' for $\Gamma(s;\lambda)$ with ultralocal action $S_{\rm field}=\oint d\lambda\,S$, which reproduces the equations of motion derived from $S$, $R(s)$ is the conserved quantity associated with translations in $\lambda$.

With our classical-mechanics tools ready at hand, we now apply them to the hole-ography setup. The holographic entanglement entropy $S(\gamma_L,\gamma_R)$ for a given boundary interval is given by the action (conveniently also called $S$) of an extremal curve, \ie a solution of the `classical equations of motion,' with endpoints $\gamma^\mu_L,\gamma^\mu_R$. By reparametrization invariance, without loss of generality we can fix the initial and final times $s_L,s_R$. Then let $\Gamma(s;\lambda)$ be a (continuous, periodic) family of solutions with endpoints $\gamma_{L,R}(\lambda):=\Gamma(s_{L,R};\lambda)$, whose action equals $S(\gamma_L(\lambda),\gamma_R(\lambda))$. Then, from eq.~\reef{right},
the differential entropy is given by
\begin{equation}
E = \oint d\lambda\left.\pd{S(\gamma_L(\lambda),\gamma_R(\lambda'))}{\lambda'}\right|_{\lambda'=\lambda}
= \oint d\lambda\,\gamma_R^{\prime\mu}\,\pd{S(\gamma_L,\gamma_R)}{\gamma^\mu_R}
= \oint d\lambda\left.\Gamma^{\prime\mu}\,p_\mu\right|_{s_R} = R(s_R)\,,
\labell{Epf}
\end{equation}
where in the third equality we again used the fact that the derivative of the on-shell action with respect to the final position equals the final momentum. Since $R(s)$ is constant, we can thus calculate $E$ by evaluating $R(s)$ at any convenient value of $s$. In particular, evaluating this quantity at the initial endpoint $s_L$ and using $\partial S(\gamma_L,\gamma_R)/\partial\gamma^\mu_L=-p_\mu|_{s_L}$, we can demonstrate that the equivalence of the two expressions for differential entropy given in eqs.~\reef{left} 
and \reef{right}.

Now we assume that there exists a periodic bulk curve $\gamma_B(\lambda)$ that, for each $\lambda$, is tangent to the solution $\Gamma(s;\lambda)$ at some point $s=s_B(\lambda)$:
\begin{equation}
\Gamma^\mu(s_B(\lambda);\lambda)=\gamma^\mu_B(\lambda)\,,\qquad\dot\Gamma^\mu(s_B(\lambda);\lambda) = \alpha(\lambda)\,\gamma_B^{\prime\mu}(\lambda)\,,\qquad
\alpha(\lambda)>0\,.
\labell{tangent}
\end{equation}
We will refer to this condition as `tangent vector alignment'.
We can assume here without loss of generality that $s_B$ is constant.\footnote{That is, we can make use of reparameterization invariance to set $s_B$ to some prescribed value.
Alternatively, if we allow $s_B$ to vary with $\lambda$, we would note that the additional variation induced in the on-shell action would be proportional to the Hamiltonian.
However, as discussed below eq.~\reef{homoL}, reparametrization invariance implies that the Hamiltonian vanishes and so $R$ remains a constant. Further, let us note at this point that reparametrizations rescale $\dot\Gamma$, and therefore change the value but not the sign of $\alpha$ in eq.~\reef{tangent}.} We now evaluate $R$ at $s_B$:
\begin{equation}
R(s_B)=\oint d\lambda\left.\Gamma^{\prime\mu}\,p_\mu(\Gamma,\dot\Gamma)\right|_{s_B} = \oint d\lambda\,\gamma_B^{\prime\mu}\,p_\mu(\gamma_B,\gamma_B') = \oint d\lambda\,\cL(\gamma_B,\gamma_B')\,,
\labell{proof}
\end{equation}
where we made use of the homogeneity properties given by eqs.~\eqref{homop} and \eqref{homoL} in the second and third equalities, respectively. 
Note that we have also made the replacement $\Gamma^{\prime\mu}|_{s_B}=\gamma_B^{\prime\mu}|_{s_B}$ in the second equality.
Now, since $R$ is constant, we have found that $E$ equals the action evaluated on the curve $\gamma_B(\lambda)$ (which is not itself, in general, an extremal curve). That is, we have established that
the differential entropy for the family
of boundary intervals is equal to the gravitational entropy evaluated on the
corresponding bulk curve. Note this result applies for general surfaces in general
backgrounds, which may depend on the boundary coordinates $x$ and $t$ as well as the radial coordinate $z$,
and with general theories of gravity as long as the entropy functional
only produces first derivatives.

%%%%%%%%%%%%%%%%%%%%%%%%%%%%%%%%%%%%%%%
\subsection{Generalizations}
\labell{generalizations}
%%%%%%%%%%%%%%%%%%%%%%%%%%%%%%%%%%%%%%%

In this subsection, we will generalize the theorem proved in the previous subsection by relaxing the tangent vector alignment condition \eqref{tangent} in two ways. These generalizations will be useful when we develop the boundary-to-bulk construction in the next section. However, it is convenient to provide the proofs here since we have all the necessary machinery set up.

The first generalization is to allow the two tangent vectors $\dot\Gamma$ and $\gamma_B'$ to be oppositely oriented, in other words to remove the restriction $\alpha(\lambda)>0$ in \eqref{tangent}:
\begin{equation}
\Gamma(s_B(\lambda);\lambda)=\gamma_B(\lambda)\,,\qquad\dot\Gamma(s_B(\lambda);\lambda) = \alpha(\lambda)\,\gamma_B'(\lambda)\,.
\labell{tangent2}
\end{equation}
To do this, we need to assume that the action \eqref{action0} is invariant under orientation-reversing as well as orientation-preserving reparametrizations. (This clearly holds for the length functional \eqref{length}; 
on the other hand a gauge-field-type coupling $A_\mu(\gamma)\dot\gamma^\mu$, for example, is invariant only under orientation-preserving reparametrizations.) Then the homogeneity condition on the Lagrangian, \eqref{homoL1}, can be generalized to $\cL(\gamma,\alpha\dot\gamma)=|\alpha|\cL(\gamma,\dot\gamma)$, while eq.~\eqref{homop} becomes
\begin{equation}
p(\gamma,\alpha\dot\gamma)=\sgn(\alpha)\,p(\gamma,\dot\gamma)\,.
\end{equation}
Hence, combining eqs.~\reef{Epf} and \eqref{proof} now yields
\begin{equation}
E = \oint d\lambda\ \sgn(\alpha)\,\cL(\gamma_B,\gamma'_B)\,.
\labell{proof2}
\end{equation}
Thus the differential entropy is now equated to a generalized notion of gravitational entropy, where different parts of the bulk surface may contribute with
different signs. Further discussion of this point will be provided in section \ref{sign}.

The second generalization allows the tangent vectors not even to be collinear. For concreteness, we will specialize to the length functional \eqref{length}. For a three-dimensional bulk spacetime, the metric $g_{\mu\nu}$ appearing in the functional will be the spacetime metric (divided by $4G_N$), but in the higher-dimensional case the three-dimensional part of the spacetime metric gets multiplied by a Weyl factor depending on the metric in the extra dimensions (see the next section). The causal structure, which will play a key role in what follows, is of course unaffected by this Weyl factor.

Then substituting eq.~\eqref{lengthp} into eq.~\eqref{proof}, we have
\begin{equation}
E = R(s_B)=\oint d\lambda\,\frac{\gamma_B'\cdot\dot\Gamma}{|\dot\Gamma|}\,,
\end{equation}
where we have also used $\Gamma^{\prime\mu}|_{s_B}=\gamma_B^{\prime\mu}|_{s_B}$.
Previously, we had used the tangent vector alignment condition in eqs.~\reef{tangent} or \reef{tangent2} to replace $\dot\Gamma$ with $\gamma'_B$ in this
expression. However, here we note that eq.~\eqref{proof2} will still hold as long as
\begin{equation}
\gamma'_B\cdot\dot\Gamma = \pm|\gamma'_B|\,|\dot\Gamma|\,,
\labell{conull}
\end{equation}
if we now define
\begin{equation}
\alpha := \frac{\gamma_B'\cdot\dot\Gamma}{\gamma_B'^2} = \pm\frac{|\dot\Gamma|}{|\gamma'_B|}\,.
\labell{alphadef}
\end{equation}
This definition is chosen to agree with the previous one, $\dot\Gamma^\mu=\alpha\,\gamma_B^{\prime\mu}$, when the two vectors are collinear. As above, the condition \eqref{conull}, as well as the sign of $\alpha$, are invariant under reparametrizations, which simply rescale $\dot\Gamma$. Thus, although in the derivation we used a parametrization in which $s_B$ was constant, in fact eqs.~\eqref{conull} and \eqref{alphadef} imply eq.~\eqref{proof2} in any parametrization.

In the Euclidean context, eq.~\eqref{conull} implies that $\dot\Gamma^\mu\propto\gamma_B^{\prime\mu}$, bringing us back to the (generalized) condition of tangent vector alignment \reef{tangent2}. However, in a Lorentzian metric, eq.~\eqref{conull} is satisfied not only if the vectors are collinear but also if they span a null plane.  Hence the constraint \eqref{tangent2} can be significantly relaxed in this context.
We will refer to the generalized constraint as `null vector alignment'  ---  see the discussion in section \ref{generic}. The null vector alignment condition will prove useful when we develop a covariant boundary-to-bulk construction in section \ref{new}.

%%%%%%%%%%%%%%%%%%%%%%%%%%%%%%%%%%%%%%%
\subsection{Characterization of generalized planar symmetry}
\labell{planarS}
%%%%%%%%%%%%%%%%%%%%%%%%%%%%%%%%%%%%%%%

For the above analysis to apply in higher dimensions, we are implicitly making some assumptions about the
relevant surfaces and the background geometry. In particular, given a general holographic ($d+1$)-dimensional
spacetime with coordinates $q^i=\{t, x, z\}$ and $y^a=\{y^1, \cdots, y^{d-2}\}$, we would like to consider a co-dimension two surface in the bulk parameterized by $\{\la, \sigma^a\}$ with a simple embedding, which
factorizes as
\ban{
\gamma_B^\mu(\la,\sigma^b)=\lbrace q^i(\la, \sigma^b), y^a(\la, \sigma^b)\rbrace= \lbrace q^i(\la),\sigma^a\rbrace
 \labell{planar}
}
Implicitly to describe the gravitational entropy of this bulk surface, it must be that the extremal surfaces appearing
in the holographic evaluation of the differential entropy have a similar simple description, \ie
\ban{
\Gamma^\mu(s,\sigma^b;\la)=\lbrace q^i(s(\la), \sigma^b(\la)),\ y^a(s(\la), \sigma^b(\la))\rbrace= \lbrace q^i(s(\la)),\ \sigma^a\rbrace
 \labell{planar2}
}
However these extremal surfaces must be solutions to the equations of motion extremizing the given Lagrangian
and so this implicit property restricts the class of background spacetimes which we can consider.
If the surfaces admit the parameterization in eqs.~\reef{planar} and \reef{planar2}, we say that they have a 
`generalized planar symmetry' and we call $y^a$ the planar coordinates. Similarly, we say that the background geometry has generalized planar symmetry if the parameterization \reef{planar2} consistently applies for solutions of the equations of motion determining the extremal surfaces. 

Towards identifying the class of backgrounds which admit solutions with a generalized planar symmetry, we restrict our attention to the case of Einstein gravity in the bulk, for which appropriate entropy functional is simply the Bekenstein-Hawking entropy as in eq.~\reef{define}. In the language used above, the `Lagrangian' is simply $\sqrt{h}/(4G_N)$, where
$h$ is the determinant of the induced metric on the bulk surface. Now, we
show that spacetimes for which we can `factor out' the $y^a$ coordinates admit generalized planar symmetry. In particular, 
we consider spacetimes with a metric of the form
\ban{
ds^2 = g_{jk}(q^i)\, dq^j\, dq^k+ g_{bc}(q^i, y^a)\, dy^b\, dy^c \labell{pmetric}
}
and where the determinant of $g_{bc}$ can be written as 
\begin{equation}
\det\!\[g_{bc}(q^i, y^a)\] = F(q^i)\,\Sigma(y^a)\,.
\labell{pdet2}
\end{equation} 
We now show that the ansatz \reef{planar2} indeed provides a solution of the corresponding
equations of motion for metrics of this form. 

First, the determinant of the induced metric can be written as 
\ban{
h = \varepsilon^{\alpha_0\cdots \alpha_{d-2}}& \left(g_{ij} \partial_s q^i \partial_{\alpha_0}q^j + g_{ab}\partial_s y^a \partial_{\alpha_0} y^b\right)\notag\\
&\times\left(g_{ij} \partial_{\sigma^1} q^i \partial_{\alpha_1}q^j + g_{ab}\partial_{\sigma^1} y^a \partial_{\alpha_1} y^b\right)\times\cdots \notag\\
&\times\left(g_{ij} \partial_{\sigma^{d-2}} q^i \partial_{\alpha^{d-2}}q^j + g_{ab}\partial_{\sigma_{d-2}} y^a \partial_{\alpha_{d-2}} y^b\right)
}
where $\varepsilon^{\alpha_0\cdots \alpha_{d-2}}$ is the totally antisymmetric symbol on the surface.
Next given ${\cal L}\propto\sqrt{h}$, the equations of motion can be written as 
\ban{
\pd h{\xi^\mu} -\partial_\alpha \pd{h}{(\partial_\alpha \xi^\mu)}+\frac{\partial_\alpha{h}}{2h} \, \pd{h}{(\partial_\alpha \xi^\mu)}=0 \labell{eom99}
}
where $\xi^\mu=\{q^i, y^a\}$ and $\partial_\alpha = \lbrace\pd{}{s},\, \pd{}{\sigma^{\alpha_i}}\rbrace$. 
To simplify notation, we introduce $Q(s) = g_{ij}\,\partial_s q^i(s) \partial_s q^j(s)$. Next we evaluate each term in eq.~\reef{eom99} for the $q^i$ coordinates evaluated on the generalized planar symmetry ansatz \reef{planar2}:
\ban{
\left. \pd{h}{q^i}\right|_{\Gamma_{\mathcal P}}&= \partial_s q^j(\la) \partial_s q^k(\la)  \Sigma(\sigma^a)\pd{}{q^i}{\left[F(q^i(s))\,g_{jk}(q^i(s))\right]}\notag\\
\left. -\partial_\alpha\pd h {(\partial_\alpha q^i)}\right|_{\Gamma_{\mathcal P}}&=-2\Sigma(\sigma^a) \pd{}\la \left[F(\la)g_{ij}(\la) \partial_\la x^j(\la)\right]
 \labell{threep}\\
\left. \frac {\partial_\alpha h }{2h}\pd h {(\partial_\alpha q^i)}\right|_{\Gamma_{\mathcal P}}&=\frac {\Sigma(\sigma^a)}{Q(\la)} g_{ij}(\la)\partial_\la q^j (\la) \pd{}\la\left[Q(\la) F(\la)\right]  \notag
}
Summing the three above equations gives the equation of motion for $q^i$. Hence we see that all of the dependence on $\sigma^a$ is isolated in an overal factor of $\Sigma(\sigma^a)$. Hence, dividing out by this factor  (which we will assume
only vanishes at isolated points), all of the $\sigma^a$ dependence drops out of these three equations of motion for $q^i$. We can additionally assume that our original spacetime is well enough behaved so that these resulting equations have a solution.

Next, we examine the equations of motion \reef{eom99} for $y^a$. Similarly we can write
\ban{
\left. \pd{h}{y^a}\right|_{\Gamma_{\mathcal P}}&=F(\la)Q(\la)\pd{\Sigma}{\sigma^a}\notag\\
\left. -\partial_\alpha\pd h {(\partial_\alpha y^i)}\right|_{\Gamma_{\mathcal P}}&=- 2F(\la) Q(\la)\pd{\Sigma}{\sigma^a}
 \labell{gongshow}\\
\left. \frac {\partial_\alpha h }{2h}\pd h {(\partial_\alpha y^a)}\right|_{\Gamma_{\mathcal P}}&= F(\la) Q(\la)\pd{\Sigma}{\sigma^a} \notag}
and therefore we see that summing these three terms gives a vanishing result in eq.~\reef{eom99}. Hence we conclude that spacetimes with metrics of the form described by eq.~\reef{pmetric} and satisfying eq.~\reef{pdet2} have generalized planar symmetry and are accommodated by the construction in the previous section. 

Certainly, the simplest example of a background with generalized planar symmetry is  AdS space described by Poincar\'e coordinates, since this background is planar symmetric in a conventional sense. The hole-ographic construction also extends to a variety of other backgrounds with planar symmetry, such as those describing boundary field theories with Lifshitz or Schr\"odinger symmetries, the throat regions of general D$p$-brane solutions or planar black holes in any of the preceding backgrounds \cite{jun}. However, the generalized planar symmetry described here allows us to consider a much broader class of holographic backgrounds. For example, it encompasses spherical symmetry as a special case. As an amusing example, let us consider the AdS-Vaidya-Bonner geometry
\begin{equation}
ds^2=-\left(\frac{r^2}{L^2}+1-\frac{m(v)}{r^{d-2}}-\frac{q(v)^2}{r^{2d-4}} \right)dv^2+2\,dr\,dv+r^2\left(d\theta^2+
\sin^2\!\theta\,d\Omega^2_{d-2}\right)\,,
\labell{vaidya}
\end{equation}
which might describe the formation of a black hole by a collapsing shell of charged null dust. If we identify $q^i=\lbrace r,v,\theta\rbrace$ and $y^a$ as the coordinates on the $(d-2)$-sphere, this metric has the form \reef{pmetric} and also satisfies the constraint \reef{pdet2}. Hence this background has the desired generalized planar symmetry and we could use the hole-ographic construction to evaluate the gravitational entropy of spherical surfaces\footnote{We should mention that are various subtleties here. For example, if the radius is too close to the event horzion in this background, the extremal surfaces in the hole-ographic construction may not be minimal surfaces \cite{hole,extreme}. Further there are complications at the poles of the sphere, \ie $\theta=0$ and $\pi$ \cite{sully}. See section \ref{discuss}, for further discussion.} given by $\lbrace r(\lambda), v(\lambda), \theta(\lambda)\rbrace$, \eg these spheres in the bulk may have a radius that varies with $\theta$ and they may not lie in a constant time slice. On the other hand, the analysis presented here does come with limitations. For example, our ansatz \reef{pmetric} does not encompass the metric of a spinning AdS black hole --- although it may be possible to extend the discussion to include these backgrounds as well. 

%%%%%%%%%%%%%%%%%%%%%%%%%%%%%%%%%%%%%%%%%%%%%%%%%%%%%%%%%%%%%%%%%%%%%%%%%%%%%%%%
%%%%%%%%%%%%%%%%%%%%%%%%%%%%%%%%%%%%%%%%%%%%%%%%%%%%%%%%%%%%%%%%%%%%%%%%%%%%%%%%
%%%%%%%%%%%%%%%%%%%%%%%%%%%%%%%%%%%%%%%%%%%%%%%%%%%%%%%%%%%%%%%%%%%%%%%%%%%%%%%%
\section{Boundary-to-bulk construction} \labell{new}

In section \ref{time}, our analysis began with a bulk surface and we showed how to construct
a family of boundary intervals such that the differential entropy  evaluated on these intervals yields
the gravitational entropy of the bulk surface. It is natural to ask
if this construction can be reverse-engineered. That is,  given a family of boundary intervals, 
can we find a bulk surface for which the gravitational entropy matches the differential entropy? Of course,
there will be many bulk surfaces which yield the correct value of the gravitational entropy; however, implicitly
here we are demanding that a natural geometric construction produces the bulk surface from the extremal surfaces
determining the entanglement entropy of the boundary intervals. At first sight, it may
seem that the answer to this question is `no' since it is straightforward to find families of intervals for which the
corresponding extremal curves in the bulk simply do not intersect  ---  see section \ref{AdS}. However, we will show below
that in fact, a slight generalization of the hole-ographic construction, using the null vector alignment condition \reef{conull}, 
allows us to find a natural bulk surface for generic families of boundary intervals obeying natural geometric constraints.

To simplify the following discussion, we will limit our analysis to general holographic spacetimes in three dimensions (\ie
this discussion is not limited to AdS$_3$). However, for concreteness, we explicitly find the solution for the particular case of AdS$_3$ in section \ref{AdS}. We also restrict our attention
to the situation where the bulk is described by Einstein gravity, for which appropriate entropy functional is simply the 
Bekenstein-Hawking entropy, as in eq.~\reef{define}. But let us add that our generalized construction extends straightforwardly to higher-dimensional backgrounds with generalized planar symmetry. We also take some preliminary steps towards extending this construction to higher-curvature theories in appendix \ref{lovelock}.

Now, following the notation of the previous sections, we are given a family of boundary intervals defined by the endpoint curves, $\gamma_L(\la)$ and $\gamma_R(\la)$. For each $\lambda$, the corresponding geodesic is $\Gamma(s;\la)$, where $s$ is the parameter along the geodesic, which satisfies the boundary conditions $\Gamma(s_L;\la)=\gamma_L(\la)$ and $\Gamma(s_R;\la)=\gamma_R(\la)$. We wish to construct a bulk curve $\bulkc_B(\la)$ by taking a point $s_B(\la)$ from the extremal curve at each value of $\la$, in other words we would have $\bulkc_B(\la)=\Gamma(s_B(\la);\la)$ for some function $s_B(\la)$. Hence our goal is to show that for general families of boundary intervals (satisfying certain consistency conditions), we can find a function $s_B(\la)$ which yields a curve $\bulkc_B(\lambda)$ for which the gravitational entropy matches the differential entropy of the boundary intervals. For this purpose we will make use of the theorem proved in section \ref{general}, showing that either the tangent vector  \eqref{tangent2} or the null vector \eqref{conull} alignment condition is sufficient to produce this equality.

Let us make explicit an important assumption of our analysis. It is well known that the extremal surface whose area gives the entanglement entropy of a boundary region can change discontinuously under continuous changes in the region. We will assume that the extremal surface varies smoothly for the family of intervals $[\gamma_L(\lambda),\gamma_R(\lambda)]$. This implies in particular that any component of the extremal surface other than the one that reaches the boundary at $\gamma_{L,R}(\lambda)$ --- for example, one that wraps a horizon --- is the same for all $\lambda$. Such a component makes a $\lambda$-independent contribution to the entanglement entropy, and therefore does not contribute to the differential entropy. We will therefore neglect it; in particular, we define $\Gamma(s;\lambda)$, for each $\lambda$, as the curve beginning and ending at $\gamma_{L,R}(\lambda)$, regardless of the existence of any other components.

We will begin, in subsection \ref{flat intersection}, by working on a constant-time slice of a static spacetime. Here we will define $\gamma_B(\lambda)$ heuristically as the point, for each $\lambda$, where the neighbouring curves $\Gamma(s;\lambda)$ and $\Gamma(s,\lambda+d\lambda)$ cross, or more formally as the point where the deviation vector
\begin{equation}
v:=\Gamma'-\frac{\Gamma'\cdot\dot\Gamma}{\dot\Gamma^2}\dot\Gamma
\labell{deviation}
\end{equation}
vanishes. (The deviation vector is the projection of $\Gamma'$ orthogonal to $\dot\Gamma$, and is easily seen to be reparametrization-invariant.) We will give a necessary and sufficient condition, from the boundary point of view, for this crossing to exist, and we will show that, with this choice of $s_B$, the curve $\gamma_B$ satisfies the (generalized) tangent vector alignment condition \eqref{tangent2}. However, as we will discuss in subsection \ref{sign}, it turns out that the function $\alpha$ appearing in eq.~\eqref{tangent2} can take either sign, and can change sign as a function of $\lambda$. We will give examples of such behavior, which cannot be ruled out by any simple condition on the intervals $[\gamma_L(\lambda),\gamma_R(\lambda)]$. Therefore, in order to obtain agreement between the differential entropy and the gravitational entropy, it is necessary to generalize the definition of the latter to include the factor $\sgn(\alpha)$ appearing in eq.~\eqref{proof2}. This generalization allows for the fact that the differential entropy can take either sign, and we will show that it can be understood naturally from a geometrical point of view.

In subsection \ref{generic}, we explain how to covariantize the boundary-to-bulk construction of subsection \ref{flat intersection}. If we are not restricted to a constant-time slice, then generically the deviation vector does not vanish anywhere on the geodesic. However, as we show, under very simple conditions (again purely from a boundary point of view) it does become null, which is enough to guarantee that the null vector alignment condition \eqref{conull} is satisfied. Finally, in subsection \ref{AdS}, we will examine explicit examples in the context of planar AdS${}_3$.

\subsection{On a constant-time slice}\labell{flat intersection}

We begin by assuming that the entire family of geodesics $\Gamma(s;\lambda)$ lies on a constant-time slice of a static spacetime. We will work entirely within that slice. Since it carries a Euclidean metric, each geodesic is locally minimal on it. We will be interested in the displacements between the geodesic at $\lambda$ and the `neighbouring' one at $\lambda+d\lambda$. The vector $\Gamma'$ measures the displacement at a fixed value of $s$. However, this is not invariant under $\lambda$-dependent reparametrizations of $s$. An invariant vector is the deviation vector $v$ defined in eq.~\eqref{deviation}, which measure the displacement from a given point on one curve to the \emph{nearest} point on the neighbouring curve. In particular, $v=0$ precisely when the two curves cross.

We will address the issue of existence and uniqueness of such crossings below. For now, we assume that one exists for each $\lambda$, at a continuously-varying value of $s$. We define $s_B(\lambda)$ to be that value, $v(s_B(\lambda);\lambda)=0$, and define $\gamma_B(\lambda):=\Gamma(s_B(\lambda);\lambda)$. In fact, without loss of generality, we can parametrize the geodesics such that $s_B$ is a fixed constant. Then $\gamma_B'=\Gamma'$, and it is clear from the definition \eqref{deviation} that $v=0$ is equivalent to the tangent vector alignment condition \eqref{tangent2}. Heuristically, this can be seen as follows:\footnote{See appendix \ref{geom}, for a more rigorous analysis.} 
$\gamma_B(\lambda)$ is the crossing point of $\Gamma(s;\lambda)$ and $\Gamma(s;\lambda+d\lambda)$, while $\gamma_B(\lambda-d\lambda)$ is the crossing point of $\Gamma(s;\lambda-d\lambda)$ and $\Gamma(s;\lambda)$. Both of these crossings lie on the geodesic $\Gamma(s;\lambda)$, so their displacement $\gamma'_B(\lambda)d\lambda$, and hence $\gamma_B'(\lambda)$, must be proportional to its tangent vector $\dot\Gamma(s;\lambda)$.

We now address the issues of existence, uniqueness, and continuity of such crossings. First, if the deviation vector vanished at two different points on the same geodesic, then these would be conjugate points. However, locally minimally curves do not contain conjugate points. Therefore, $v$ can vanish at most at one point on each geodesic. Since $\Gamma$ is assumed smooth, $v$ is a smooth vector field and its vanishing locus is a continuous function of $s$.

To establish conditions for the existence of a crossing point, it is useful to invoke the so-called homology condition, which requires the existence, for each $\lambda$, of a bulk spatial region $r(\lambda)$ bounded on one side by the interval $[\gamma_L(\lambda),\gamma_R(\lambda)]$ in the asymptotic boundary and on the other by the geodesic $\Gamma(s;\lambda)$. For any point $(s,\lambda)$ along the geodesic, if the deviation vector $v$ does not vanish, then it can point either towards $r(\lambda)$ (`in') or away from it (`out'). Which way it points at the endpoints is determined by the sign of $x_{L,R}'(\lambda)$. In particular, if $x'_L(\lambda)>0$, then $v$ points in at $s_L$, and similarly, if $x'_R(\lambda)>0$, then it points out at $s_R$. If both are true, then $v$ must vanish at some intermediate value of $s$. Similarly if $x'_L(\lambda)<0$ and $x'_R(\lambda)<0$. Thus a crossing point must exist on the condition
\begin{equation}
x'_L(\lambda)\,x'_R(\lambda)>0\,.
\labell{constrain88}
\end{equation}
(This condition was previously derived in \cite{jun}.) We note that this is a condition that can be checked directly from knowing $\gamma_{L,R}(\lambda)$, without finding the geodesics.

In the opposite case, $x'_L(\lambda)x'_R(\lambda)<0$, $v$ points either in or out at both endpoints. Therefore it must vanish at an even number of intermediate points.\footnote{The possibility of a single intersection point where the curves are tangent to each other is also ruled out. Neighbouring curves cannot be tangent to each other, since the extremal curves satisfy a second-order ordinary differential equation and therefore have a unique solution for a given initial position and `velocity.' Therefore, wherever $v=0$, it must switch from pointing in to pointing out or vice versa.} However, we showed above that there cannot be more than one crossing, so there are none at all. (This is a special case of Theorem 4.3 of \cite{matt}, which says that if $[\gamma_L(\lambda_1),\gamma_R(\lambda_1)]\subset[\gamma_L(\lambda_2),\gamma_R(\lambda_2)]$ then $r(\lambda_1)\subset r(\lambda_2)$.) Thus our construction fails in this case to produce a curve $\gamma_B$. We leave it to future work to determine whether there exists an alternative construction that naturally produces a bulk curve whose gravitational entropy continues to match the differential entropy of a family of intervals for which eq.~\reef{constrain88} is not satisfied everywhere.

%%%%%%%%%%%%%%%%%%%%%%%%%%
\subsection{Signed areas}
\labell{sign}
%%%%%%%%%%%%%%%%%%%%%%%%%%

The construction described in the previous subsection guarantees that the curve $\gamma_B$ obeys the generalized tangent vector alignment constraint \eqref{tangent2}, \ie $\dot\Gamma=\alpha\,\gamma_B'$. However, it does not guarantee that the function $\alpha(\lambda)$ is positive. Indeed, as we will see in examples in this subsection, it can take either sign, and can even switch signs as a function of $\lambda$. Therefore, according to the theorem of subsection \ref{generalizations}, the differential entropy will not in general equal the area of $\gamma_B$, but rather a signed area in which certain segments contribute positively and others negatively.

For a simple example, consider the vacuum of a two-dimensional CFT on a circle of length $2\pi R$, where we denote the angular coordinate $\theta$ (having periodicity $2\pi$ as usual). The entanglement entropy of an interval $[\theta_L,\theta_R]$ is \cite{cardy1}
\begin{equation}
S(\theta_L,\theta_R) = \frac c3\log\[\frac{2R}{\delta}\,\sin\left(\frac{\theta_R-\theta_L}2\right)\]
\end{equation}
where $c$ is the central charge, 
$\delta$ is the short-distance cut-off and $(\theta_R-\theta_L))$ is taken to be between 0 and $2\pi$. Let us consider a family of intervals with a fixed angular size $2\Delta$ going once around the circle, \ie
\begin{equation}
\theta_L(\lambda) = \lambda-\Delta\,,\qquad
\theta_R(\lambda) = \lambda+\Delta\,,
\labell{goal}
\end{equation}
where $\lambda$ has periodicity $2\pi$. The differential entropy is easily computed as
\begin{equation}
E=\frac{\pi c}3\cot\Delta\,.
\labell{exampleE}
\end{equation}
Thus when $\pi/2<\Delta<\pi$, in other words when each interval covers more than half the circle, the differential entropy is negative.

\begin{figure}[h!]
\centering
\subfloat[]{\includegraphics[width=0.35\textwidth]{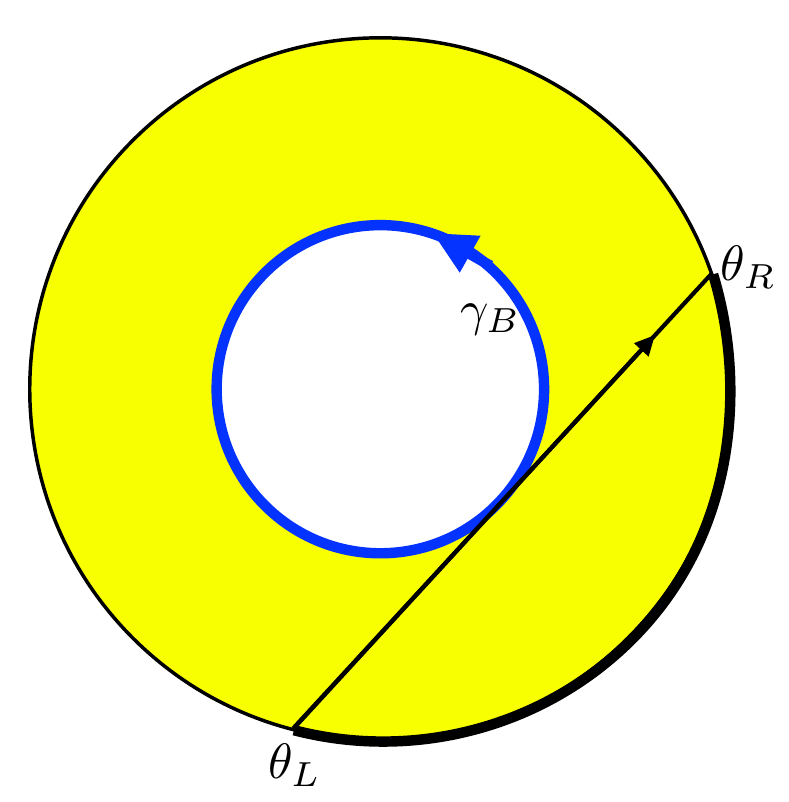}}\qquad
\subfloat[]{\includegraphics[width=0.35\textwidth]{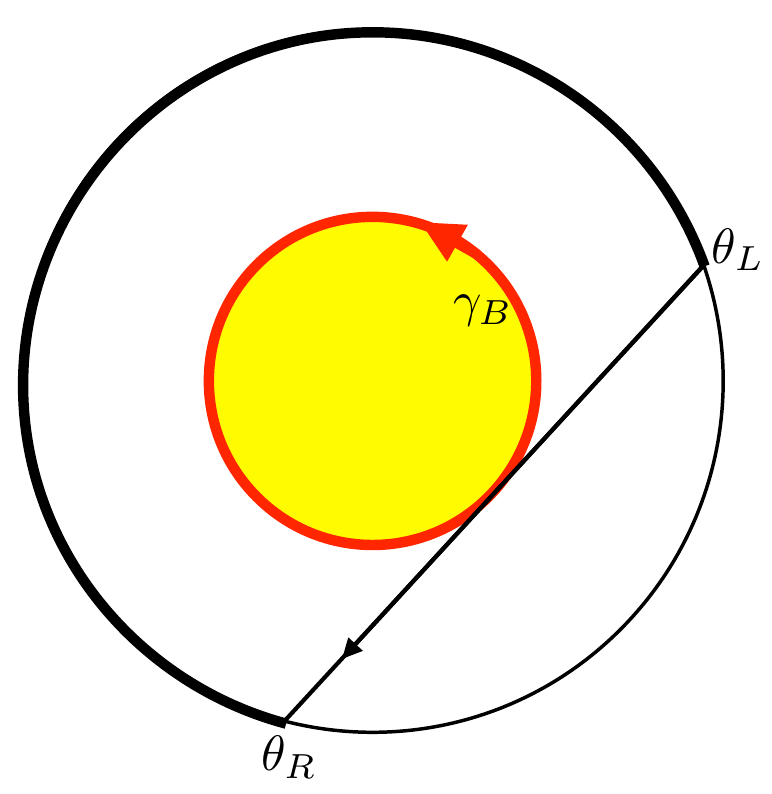}}
\caption{(Color online) A constant-time slice of AdS${}_3$, in the $(r,\theta)$ coordinate system described in the text. (a) A boundary interval $[\theta_L,\theta_R]$ with $\Delta<\pi/2$, and the corresponding bulk geodesic (black line). The yellow annulus is the union of the bulk regions corresponding to intervals with the same value of $\Delta$, and its boundary (the blue circle) is $\gamma_B$. (b) A boundary interval with $\Delta>\pi/2$, and the corresponding bulk geodesic. The yellow disc is the intersection of the bulk regions corresponding to intervals with that value of $\Delta$, and its boundary (the red circle) is $\gamma_B$.
}
\labell{signed}
\end{figure}
The holographic spacetime describing this state is global AdS${}_3$ (see fig.\ \ref{signed}). On a fixed time slice, the coordinates $(r,\theta)$ (with $0\le r<1$) can
be chosen so that the metric becomes
\begin{equation}
ds^2 = L^2\left(\frac1{(1-r^2)^2}\,dr^2+\frac{r^2}{1-r^2}\,d\theta^2\right)\,.
\end{equation}
The geodesics corresponding to the above intervals \reef{goal} are given implicitly by the equations
\begin{equation}
r\cos(\theta-\lambda) = \cos\Delta\,,\qquad r\sin(\theta-\lambda)=s\,,
\labell{geodesics}
\end{equation}
with the parameter $s$ taking the range $-\sin\Delta<s<\sin\Delta$. Since the endpoints satisfy the condition \eqref{constrain88}, we should expect to find a crossing point  ---  a solution to $v=0$  ---  on each geodesic. By symmetry, since there is a unique solution, it must lie at the point of symmetry, $s=0$. This is confirmed by an explicit computation of $v$, which shows that it carries an overall factor of $s$. 
Setting $s=0$ in \eqref{geodesics}, we find
\begin{equation}
\gamma_B(\lambda) = (r_B(\lambda),\theta_B(\lambda))=\begin{cases}(\cos\Delta,\lambda)\,,&\quad\Delta<\pi/2 \\(-\cos\Delta,\lambda+\pi)\,,&\quad\Delta>\pi/2
\end{cases}\,.
\end{equation}
This is a circle of proper length $2\pi L\,|\cot\Delta|$. A short computation shows that $\alpha(\lambda)=\sec\Delta$, which is positive for $\Delta<\pi/2$ and negative for $\Delta>\pi/2$. The sign of $\alpha$ is also intuitively clear from the fact that $\gamma_B^{\prime \theta}=\theta_B'>0$, while $\dot\Gamma^\theta$ is positive for $\Delta<\pi/2$ and negative for $\Delta>\pi/2$. (This can be seen in fig.\ \ref{signed}, where $\gamma_B'$ is parallel to $\dot\Gamma$ on the left but antiparallel on the right.) Finally, using the fact that $c=3L/2G_N$, we see that eq.~\eqref{proof2} is verified for both signs of $\alpha$.

The curve $\gamma_B(\lambda)$ bears an intriguing relation to the regions $r(\lambda)$ in this case. By definition, the boundary of $r(\lambda)$ is the geodesic $\Gamma(s;\lambda)$. For the above geodesics, $r(\lambda)$ consists of the set of points satisfying $r\cos(X-\lambda)\ge\cos\Delta$. For $\Delta<\pi/2$, the union of $r(\lambda)$ over all $\lambda$ is the annulus $r\ge\cos\Delta$, whose inner boundary is precisely $\gamma_B(\lambda)$. This is an example of the `outer envelope' construction of \cite{jun}. (The term `outer' there was used from the point of view of the boundary.) On the other hand, when $\Delta>\pi/2$, the union covers the entire slice. Instead, in this case $\gamma_B(\lambda)$ is the \emph{outer} boundary of the \emph{intersection} of the $r(\lambda)$, which is the disc $r\le-\cos\Delta$. Thus in this case the holographic hole is inside out: the `hole' is the annulus extending to the boundary (see fig.\ \ref{signed}). A similar picture in fact applies to general families of intervals, as we will discuss below.

In the vacuum, or any pure state, the entanglement entropy of an interval equals that of its complement. However, in the complement the roles of the left- and right-endpoints are switched, and it is easy to see from the definition in eqs.~\reef{left} and \reef{right} that the sign of the differential entropy is reversed under such a transformation. For example, taking the complement of all the intervals in the above example takes $\Delta\to\pi-\Delta$, and indeed we see from eq.~\eqref{exampleE} that $E$ switches its sign under this transformation. From the bulk point of view, switching the left- and right-endpoints flips the sign of $\dot\Gamma$ and therefore, of $\alpha$. Similarly, in any state (pure or mixed), an orientation-reversing reparametrization of $\lambda$, such as $\lambda\to-\lambda$, will reverse the sign of $E$. From the bulk perspective, in this case, it is $\gamma'_B$ that is reverses its sign and hence that of $\alpha$ also flips.

In the example discussed above, $\alpha(\lambda)$ was either positive or negative for all $\lambda$. However, $\alpha$ can also change sign as $\lambda$ varies. This happens when $\gamma'_B(\lambda)$ goes to zero and then reverses direction, leading to a cusp in the bulk curve $\gamma_B(\lambda)$ (while $\dot\Gamma$ remains finite; thus $\alpha$ passes through infinity rather than 0). 
One way this can happen is if $x'_L(\lambda)$ and $x'_R(\lambda)$ simultaneously switch sign at some value of $\lambda$. However, $\alpha$ can switch sign even when $x'_L$ and $x'_R$ maintaining constant signs. To see this, let us return for simplicity to planar AdS${}_3$ as in eq.~\reef{3metric}, with coordinates $(z,x)$ on the constant-time slice. Let us write the endpoints as
\begin{equation}
x_L = x_c-\Delta\,,\qquad x_R=x_c+\Delta\,,
\end{equation}
where both $x_c$ and $\Delta$ are functions of $\lambda$. Now,
the geodesic is a semicircle of radius $\Delta$ centered at $(0,x_c)$:
\begin{equation}
\Gamma = (z,x) = (\Delta\sqrt{1-s^2},x_c+\Delta s)\,,
\end{equation}
where $-1<s<1$. For a family of intervals, one easily computes
\beqa
&&s_B = -\frac{\Delta'}{x_c'}\,,\qquad
\gamma_B = \left(\Delta\sqrt{1-\frac{\Delta^{\prime2}}{x_c^{\prime2}}},x_c-\frac{\Delta\Delta'}{x_c'}\right),
\nonumber\\
&&\qquad\quad
\frac1\alpha = \frac{x_c^{\prime3}-\Delta^{\prime2}x_c'-\Delta\Delta''x_c'+\Delta\Delta'x_c''}{\Delta x_c^{\prime2}}
\labell{constanttime}
\eeqa
The condition\footnote{One can easily show that this constraint also is equivalent to demanding $|s_B|<1$.} \eqref{constrain88} requires $|\Delta'|<|x_c'|$. If both intervals are moving forward, $x_L'>0$, $x_R'>0$, so $x_c'>0$, then by a reparametrization of $\lambda$ we can set $x_c=\lambda$, and the above equations simplify to:
\beq
s_B = -\Delta'\,,\qquad
\gamma_B = \left(\Delta\sqrt{1-\Delta^{\prime2}},\lambda-\Delta\Delta'\right)\,,
\qquad
\frac1\alpha = \frac1\Delta-\frac{\Delta^{\prime2}}\Delta-\Delta''\,.
\eeq
Clearly, even subject to the constraint $|\Delta'|<1$, $\alpha$ can switch sign, due to the presence of the $\Delta''$ term. 
A short calculation confirms that the integrand of the differential entropy, $(\Delta'+1)/\Delta$, and the integrand of the signed area, $\sgn(\alpha)|\gamma_B'|=1/(\alpha(\Delta^{\prime2}-1))$, differ by a total derivative. Hence as expected, we recover precisely the result in eq.~\reef{proof2}.

\begin{figure}[h!]
\centering
\subfloat[]{\includegraphics[width=0.4\textwidth]{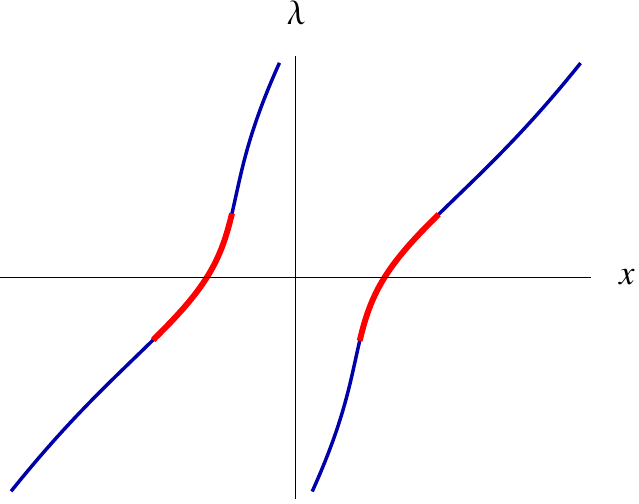}}\qquad
\subfloat[]{\includegraphics[width=0.4\textwidth]{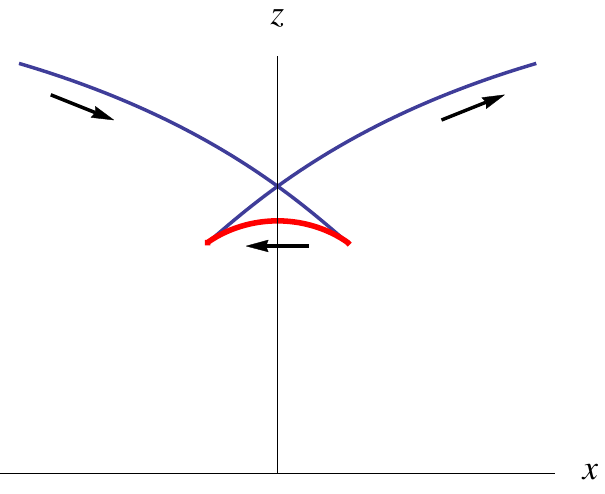}}
\caption{(Colour online) The (a) endpoints $x_L(\lambda)$, $x_R(\lambda)$ and (b) curve $\gamma_B(\lambda)$ for the family of intervals \eqref{reversalex}, with $\sigma=1$, for $-1.5<\lambda<1.5$. The parts of the curves for which $\alpha(\lambda)>0$ are shown in blue, and for which $\alpha(\lambda)<0$ in red. In (b), the parameter $\lambda$ increases in the direction shown by the arrows.
}
\labell{reversal}
\end{figure}
As a simple example, consider the family of intervals defined with\footnote{Since this is a local phenomenon, we are not concerned here with the periodicity in $x$ or $\lambda$; however, if desired, this function can be joined smoothly onto a periodic one.}
\begin{equation}
\Delta =2- \frac{\sigma}{\lambda^2+1}\,.
\labell{reversalex}
\end{equation}
where $\sigma$ is some constant, which must satisfy $|\sigma|\le \sigma_{max}=8\sqrt{3}/9\simeq 1.540$ in order that $|\Delta'|<1$. The intervals are plotted in figure \ref{reversal}a for $\sigma=1$. There we see that as $\lambda$ increases, the intervals are moving in the positive $x$ direction but their length decreases and then increases again in the vicinity of $\lambda=0$, as determined by eq.~\reef{reversalex}. However, this rather benign behaviour by the boundary intervals produces the bulk curve illustrated in figure \ref{reversal}b. In fact, a similar reversal in the bulk is produced for any $\sigma>\sigma_{min}=1-(1/\sqrt{2})\simeq0.293$.

\begin{figure}[h!]
\centering
\subfloat[]{\includegraphics[width=0.4\textwidth]{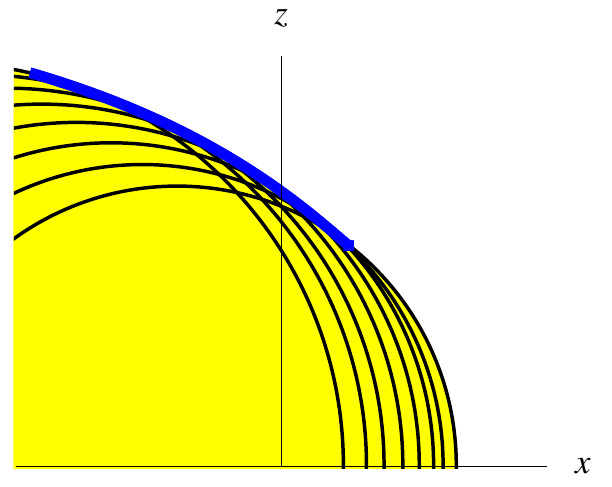}}\qquad
\subfloat[]{\includegraphics[width=0.4\textwidth]{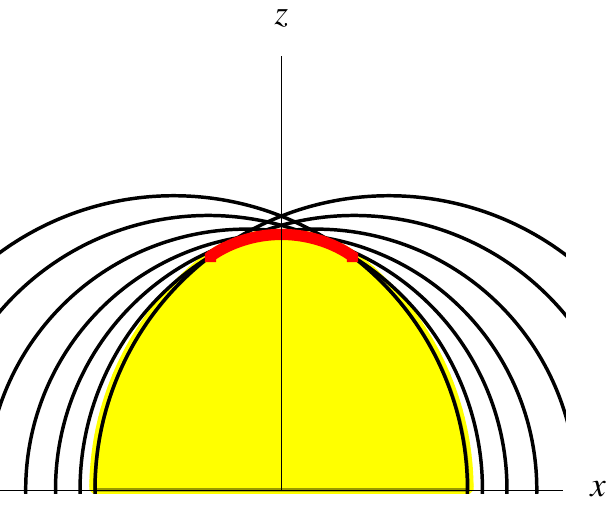}}
\caption{(Color online) (a) Segment of the curve $\gamma_B(\lambda)$ with $\alpha(\lambda)>0$ (blue), along with geodesics $\Gamma(s;\lambda)$ (black) and regions $r(\lambda)$ (yellow) for the corresponding values of $\lambda$. (b) Segment of the curve $\gamma_B(\lambda)$ with $\alpha(\lambda)<0$ (red), along with the corresponding geodesics (black) and intersection of the regions (yellow).}
\labell{regions}
\end{figure}
With this example in hand, we now return to the relation between the curve $\gamma_B$ and the regions $r(\lambda)$ that are bounded by the geodesics $\Gamma(s;\lambda)$. On the left side of figure \ref{regions}, we've plotted the first segment of the $\gamma_B$ curve shown in figure \ref{reversal}, for which $\alpha>0$, together with a sequence of the corresponding regions. It is clear that that segment is the boundary of the union of the regions. Similarly, on the right side of figure \ref{regions}, we've plotted the $\alpha<0$ segment of $\gamma_B$, together with the intersection of the corresponding regions; again, the former is the boundary of the latter. In fact, this is a general rule: When $\alpha>0$, $\gamma_B$ is locally the boundary of the union of the regions $r(\lambda)$, and when $\alpha<0$ it is locally the boundary of their intersection. In the former case, its extrinsic curvature points away from the regions, and in the latter case towards them. (More precisely, this rule applies when $x'_L$ and $x'_R$ are both positive; when they are negative the rule is reversed.)

%%%%%%%%%%%%%%%%%%%%%%%%%%
\subsection{Generic families of intervals}  \labell{generic}
%%%%%%%%%%%%%%%%%%%%%%%%%%

For a time-varying family of boundary intervals, we cannot restrict ourselves to a slice of the bulk. In the full bulk, the geodesics are codimension-two, and therefore neighbouring ones (\eg for $\lambda$ and $\lambda+d\lambda$) generically do not intersect. So we must generalize the previous construction. To do this, we will appeal to the second generalization described in subsection \ref{generalizations}, \ie  the null vector alignment condition. The latter states that the differential entropy will equal the gravitational entropy if \eqref{conull} is satisfied. Implicitly, this condition states that the vectors $\dot\Gamma$ and $\gamma_B'$ lie in a common null plane. That is, one
can easily verify that eq.~\reef{conull} is satisfied when
  \be
\frac{\bulkc_B'{}^{\!\!\mu}}{|\bulkc_B'|}=\pm\frac{\dot \Gamma^\mu}{|\dot \Gamma|}+k^\mu\qquad
{\rm with}\ \ k\cdot k=0 \,.
 \labell{sol3}
 \ee
Further, it is straightforward to show that the extra vector $k^\mu$ also satisfies:
  \be
k\cdot \bulkc_B' =0\quad
{\rm and}\quad k\cdot\dot \Gamma=0 \,.
 \labell{sol4}
 \ee
Since $\gamma_B'$ is a linear combination of $\Gamma'$ and $\dot\Gamma$, eqs.~\reef{conull} and \reef{sol3} can be expressed equivalently with $\gamma_B'$ replaced by $\Gamma'$.  That is, the null vector alignment condition can be seen as
demanding that  $\Gamma'$ and $\dot\Gamma$ lie in a common null plane, or equivalently, that the deviation vector $v$ in eq.~\reef{deviation} is null. To simplify the discussion, from this point we will assume that $\dot\Gamma\cdot\gamma_B'>0$ (\ie\ $\alpha>0$). The generalization to the opposite case will hopefully be clear.

Below, we will consider under what conditions there will exist a solution to the null vector alignment condition. But first we would like to ask, assuming a solution exists, what the analogue of the above `outer envelope' construction is, \ie the statement that $\gamma_B$ is the boundary of the union of the regions $r(\lambda)$. We will argue that, here, the bulk curve $\gamma_B$ emerges naturally in terms of the union of the `entanglement wedges' \cite{mattEW}. Therefore we should 
first comment on the definition and properties of entanglement wedges \cite{mattEW}: In general, given a boundary region and the corresponding extremal surface in the bulk, the entanglement wedge
is defined as the domain of dependence or causal development of the bulk spacelike codimension-one region extending between these two.  In our
case, we must consider the boundary $W(s,\tau;\la)$ of the entanglement wedge, which is formed by the (converging) light sheets sent out toward the boundary (in the direction of $r(\lambda)$) from each point on the extremal curve $\Gamma(s;\la)$. The light rays comprising these light sheets may reach the asymptotic boundary, however,  generically they will end with the formation of caustics, as illustrated in figure \ref{EW}. 
One remarkable feature of the entanglement wedge is that the intersection of $W(s,\tau;\la)$ with the asymptotic boundary is precisely the boundary of the causal development of the boundary region, as proved in \cite{mattEW}.
\begin{figure}[h!]
\begin{center}
\includegraphics[width=0.4\textwidth]{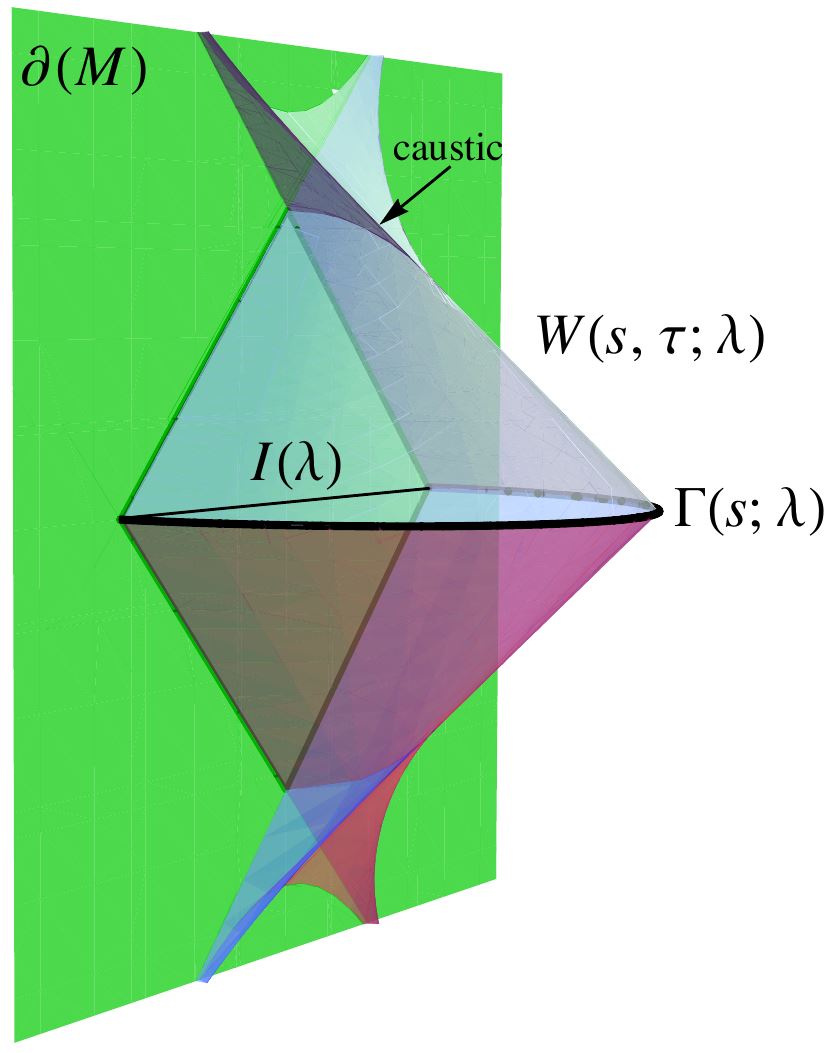}
\caption{(Colour online) The boundary of the entanglement wedge $W(s,\tau;\la)$ is shown above for the extremal curve $\Gamma(s;\la)$ corresponding to the interval $I(\la)$. The surface ends when the light rays emerging from $\Gamma(s;\la)$
either reach the asymptotic boundary or form caustics. }
\labell{EW}
\end{center}
\end{figure}
\begin{figure}[h!]
\begin{center}
\includegraphics[width=0.6\textwidth]{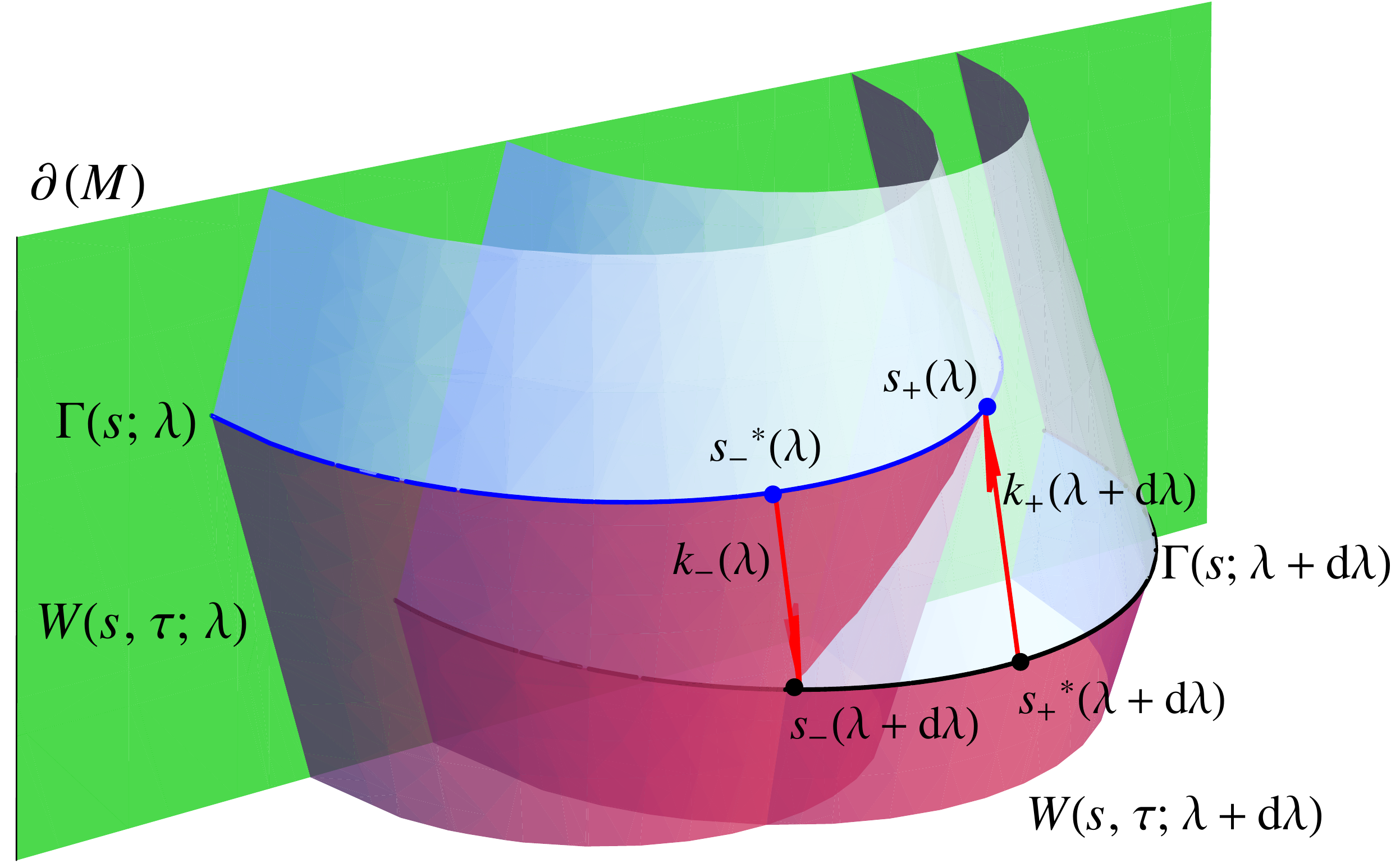}
\caption{(Colour online) The intersection of the surfaces $W(s,\tau;\la)$ and $W(s,\tau;\la+d\la)$ is shown above. The point $\Gamma(s^*_+(\la+d\la);\la+d\la)$ on the neighbouring extremal curve is identified as being separated from the intersection point $\Gamma(s_+(\la);\la)$ by the null vector $k^\mu_+(\la+d\la)$. Similarly, the point $\Gamma(s^*_-(\la);\la)$ on the neighbouring extremal curve is identified as being separated from the intersection point $\Gamma(s_-(\la+d\la);\la+d\la)$ by the null vector $k^\mu_-(\la)$. One can see intuitively that in the limit $d\la \to 0$, $s_-^*(\la)$ does not generically approach $s_+(\la)$.}
\labell{EWI}
\end{center}
\end{figure}
In analogy with the constant-time case, suppose that, for each $\lambda$, $\gamma_B(\lambda)$ is the intersection of the extremal curve $\Gamma(s;\la)$ with the boundary $W(s,\tau;\la+ d\la)$ emerging from $\Gamma(s;\la+d\la)$ (see figure \ref{EWI}). We will show that this definition reproduces the null vector alignment condition. Since the two extremal curves are only displaced by an infinitesimal amount, the relevant portion of $W(s,\tau;\la+ d\la)$ is null, \ie we do not expect any caustics to form in the vicinity of $\Gamma(s;\la)$. Meanwhile, $\gamma_B(\lambda-d\lambda)$ is the intersection point of $\Gamma(s;\lambda-d\lambda)$ with $W(s,\tau;\la)$. Thus both $\gamma_B(\lambda-d\lambda)$ and $\gamma_B(\lambda)$ lie on $W(s,\tau;\la)$, which we can approximate as a null plane, so the vector $\gamma_B'(\lambda)$ lies in that plane. The geodesic $\Gamma(s;\lambda)$ also lies in $W(s,\tau;\la)$, so its tangent vector $\dot\Gamma(s;\lambda)$ does as well. Thus, as promised, $\gamma_B'$ and $\dot\Gamma$ lie in a common null plane.

\begin{figure}[h!]
\centering
%\subfloat[]{\includegraphics[width=0.45\textwidth]{3DEnvelope}}\qquad
%\subfloat[]{\includegraphics[width=0.45\textwidth]{3Denvelope2}}
\subfloat[]{\includegraphics[width=0.45\textwidth]{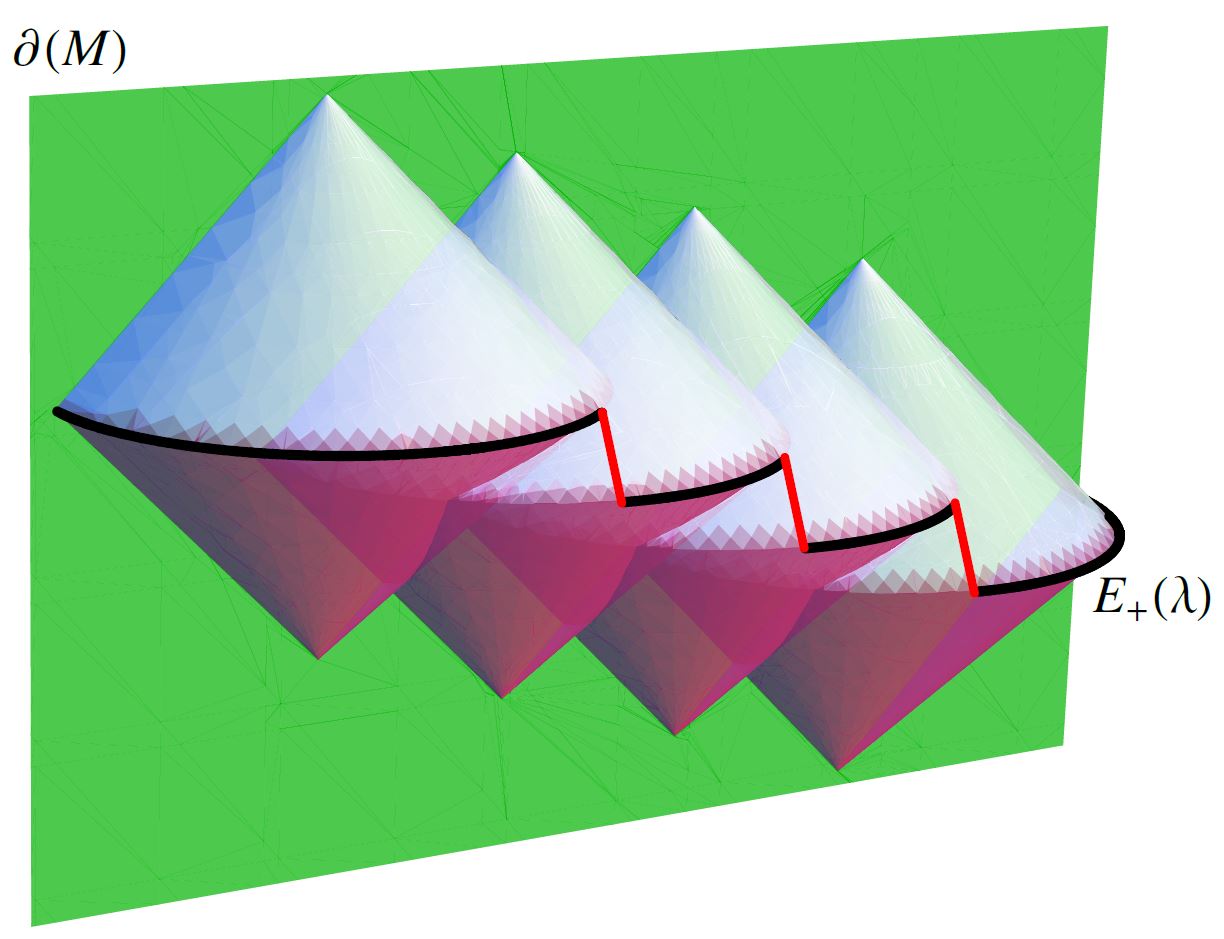}}\qquad
\subfloat[]{\includegraphics[width=0.45\textwidth]{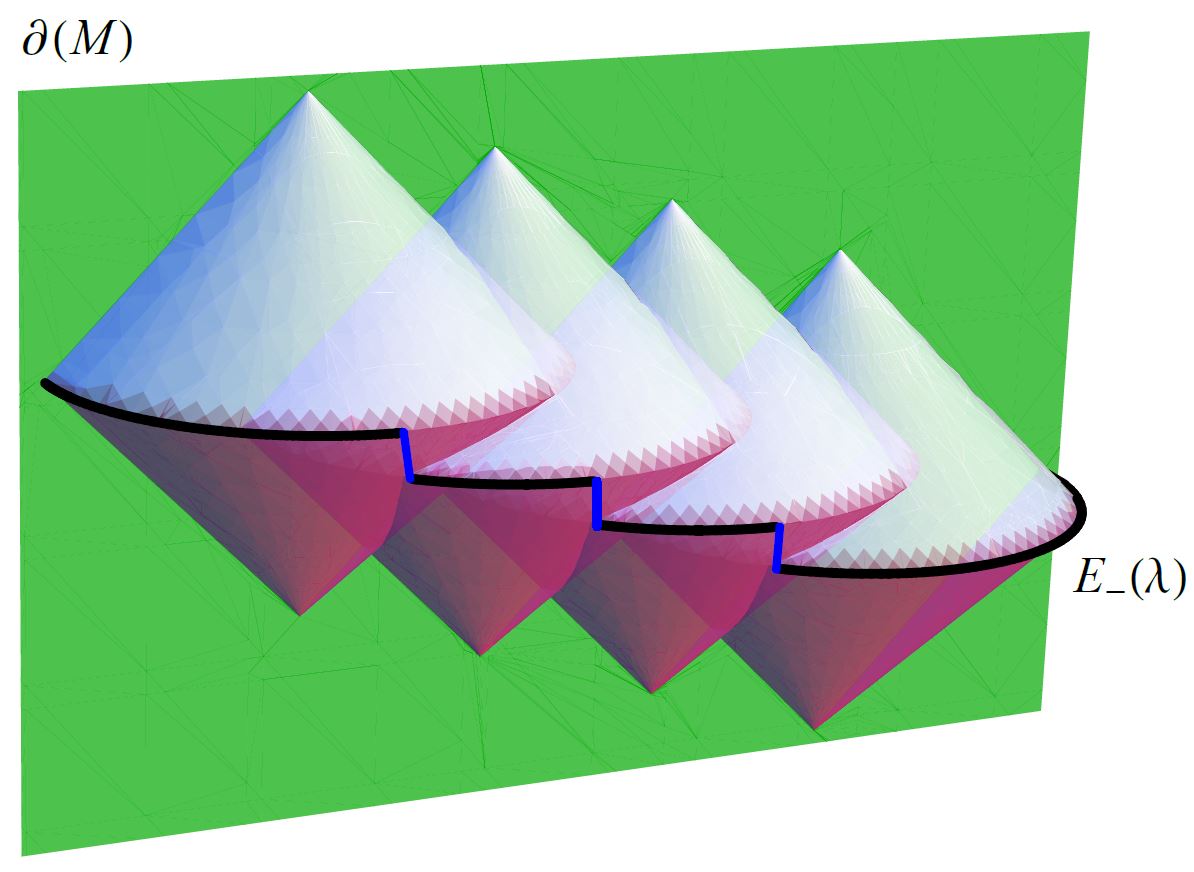}}
\caption{(Colour online) We picture the outer envelope $\gamma_B^+(\la)$ in (a) as being built from the pieces of the extremal curve between $s_+^*(\la)$ and $s_+(\la)$, connected by null segments on each entanglement wedge boundary. In the continuum limit this curve consists only of the intersection points $s_+(\la)$, and its gravitational entropy is equal to the differential entropy of the boundary intervals. As illustrated in (b), a similar curve $\gamma_B^-(\la)$ can be constructed using $s_-^*(\la)$ and $s_-(\la)$.  Generically the two curves, $\gamma_B^+(\la)$ and $\gamma_B^-(\la)$, remain distinct in the continuum limit.}
\label{3Denvelope}
\end{figure}

We can be slightly more explicit with this argument as follows: The intersection point $s_+(\la)$ on $\Gamma(s;\la)$ is connected to a point $s^*_+(\la+d\la)$ on $\Gamma(s;\la+d\la)$ by a null vector $k_+(\la+d\la)$. Further, this null vector is orthogonal to the  extremal curve $\Gamma(s;\la+d\la)$ at $s^*_+(\la+d\la)$, \ie  $k_+(\la+d\la)\cdot \dot\Gamma(s^*_+(\la+d\la);\la+d\la)=0$. Hence intuitively, this intersection point will produce the desired relation $\Gamma'(s;\la)|_{s_+(\la)} \propto \dot \Gamma (s;\la)|_{s_+(\la)}+ k_+(\la)$, in the continuum limit. We verify that this intuition is correct in appendix \ref{geom} and simply proceed here. Hence we are led to a generalized notion of the outer envelope in this case. The piece-wise construction of the bulk curve now includes segments of the extremal curves extending between intersections with the boundaries of the corresponding entanglement wedges. However, these segments alone do not form a contiguous curve but rather they are connected by infinitesimal null segments lying in the boundaries $W(s,\tau;\la)$.  A sketch illustrating this construction is given in figure \ref{3Denvelope}a.

\begin{figure}[h!]
\begin{center}
\includegraphics[width=0.55\textwidth]{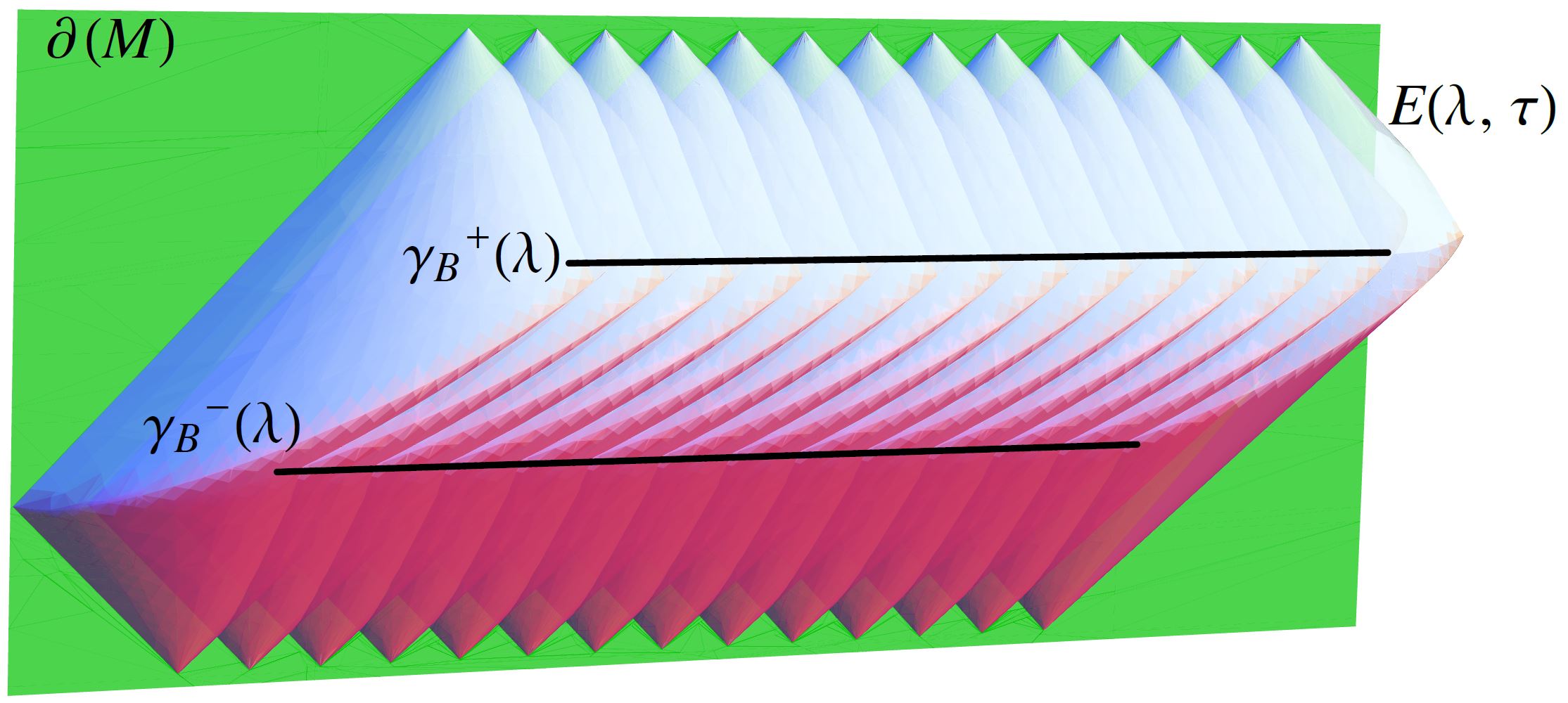}
\caption{(Colour online) The enveloping surface $E(\la, \tau)$ being built for a family of boundary intervals with a fixed width but slightly tilted in the ($t$,$x$)-plane  ---  see further discussion of this example in section \ref{AdS} and figure \ref{boosts}. The two bulk curves, $\gamma_B^+(\la)$ and $\gamma_B^-(\la)$, correspond to the lines across which the normal vector makes the transition between spacelike and null.}
\label{EWS}
\end{center}
\end{figure}

Figure \ref{EWI} also shows the intersection of the extremal curve $\Gamma(s;\la+d\la)$ with the boundary $W(s,\tau;\la)$ emerging from $\Gamma(s;\la)$. Our notation will be to label this intersection point $s_-(\la+d\la)$ on $\Gamma(s;\la+d\la)$
and it is connected to a point $s^*_-(\la)$ on $\Gamma(s;\la)$ by the null vector $k_-(\la)$. Here,  $k_-(\la)$ is orthogonal to $\Gamma(s;\la)$ at $s^*_-$, \ie  $k_-(\la)\cdot \dot\Gamma(s^*_-(\la);\la)=0$. Hence our intuition would again be that
this intersection point  produces the null vector alignment condition in the continuum limit, \ie $\Gamma'(s;\la)|_{s^*_-(\la)} \propto \dot \Gamma (s;\la)|_{s_-^*(\la)}+ k_-(\la)$, and again, we verify this result in appendix \ref{geom}.
Now an interesting feature of the present construction is that generally when both intersections exist, they do not coincide in
the continuum limit. That is, the difference $s_+(\la)-s_-^*(\la)$ is an order-one quantity. The reader may already find this feature 
evident from figure \ref{EWI}, but it will also become explicit in the examples studied in the following section. Therefore applying
the generalized notion of the outer envelope here, we are lead to a second distinct curve in the bulk, as
illustrated in figure \ref{3Denvelope}b. Hence for a broad
class of families of boundary intervals, the null vector alignment condition actually leads to the construction
of two bulk curves for which the gravitational entropy equals the differential entropy of the boundary intervals --- see also figure \ref{EWS}. Of course, as we
will discuss in a moment, both intersections may not exist or they may not both exist globally. That is, the boundary intervals must satisfy global constraints analogous to eq.~\reef{constrain88} in order to properly define a bulk surface.

Further insight comes from extending the outer envelope to the `enveloping surface' $E(\la,\tau)$ which can loosely be thought of as  the boundary
of the union of all of the entanglement wedges.\footnote{Similar to the discussion of the outer envelope in \cite{jun}, this picture is only precise for $\hat n_1(s_B(\la))\cdot a(\la)<0$, where $a^\mu(\la)$ is the proper acceleration along the bulk curve and the unit vector $\hat n_1$ is defined in subsection \ref{AdS}. 
This is a covariant generalization of the condition found for the constant-time case \cite{jun}. In higher dimensions, \ie bulk
dimensions greater than three, this condition becomes $\hat n_1(s_B(\la))\cdot K(\la)<0$, where $K^\mu(\la)$ is the trace of the extrinsic curvatures on the bulk curve.} More precisely, this enveloping surface should be thought of as being composed of all of the segments of $W(s,\tau;\la)$ between the lines of intersection with $W (s,\tau;\la\pm d\la)$, as illustrated in figure \ref{EWS}. The bulk curves constructed with null vector alignment are then the lines on the enveloping surface across which the normal vector makes a transition between being spacelike and null.\footnote{As the union of the entanglement wedges, the enveloping surface typically consists of five parts: First, `top' and `bottom' of the
entanglement wedges typically contains caustics --- see figure \ref{EW}. Hence the union of these cusps will produce regions at the top and bottom of the enveloping surface with a timelike normal. Second, the light sheets themselves make up sections of the enveloping surface with null normal vector. The regions with the future-pointing and past-pointing null normals correspond to the `upper' and `lower' parts of the enveloping surface respectively. Finally, the region between these null sections is comprised to the portions of the extremal surfaces running from $s_-(\la)$ to $s_+(\la)$. The union of all these geodesics will produce a surface with a spacelike normal vector. The bulk curves picked out by the null vector alignment condition form the boundary between this spacelike region and the two null regions. With tangent vector alignment, $s_+(\la)=s_-(\la)$ and thus the spacelike region shrinks to zero size. The bulk curve
is then the boundary between the upper and lower null regions. \labell{parcel0}} With tangent vector alignment, the `spacelike' region shrinks to zero size and the normal vector is not well defined on the resulting bulk curve, \ie the normal makes a  transition between being future-pointing null and past-pointing null. 

\begin{figure}[h!]
\begin{center}
\includegraphics[width=0.6\textwidth]{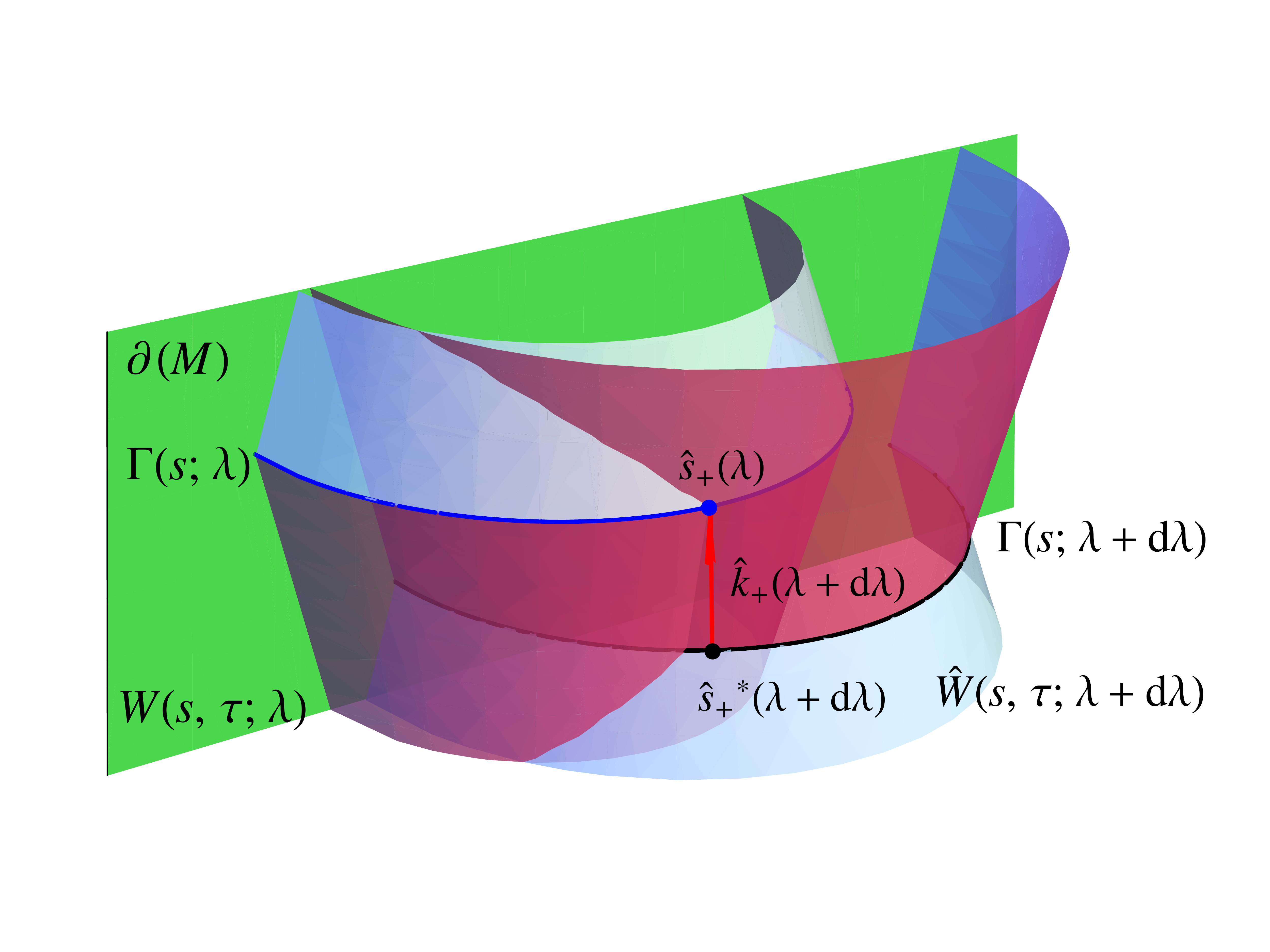}
\caption{(Colour online) The intersection of the surfaces $W(s,\tau;\la)$ and $\hat W(s,\tau;\la+d\la)$ is shown above. The point $\Gamma(\hat s^*_+(\la+d\la);\la+d\la)$ on the neighbouring extremal curve is identified as being separated from the intersection point $\Gamma(\hat s_+(\la);\la)$ by the null vector $\hat k^\mu_+(\la+d\la)$. One can see intuitively by comparison with figure \ref{EWI} that $\hat s_+(\la)\sim s^*_-(\la)$ as $\hat k_+(\la+d\la) \sim k_-(\la)$.}
\label{EWcomp}
\end{center}
\end{figure}

Before considering the global constraints, we point out a technical detail, illustrated in figure \ref{EWcomp}. Above, by focusing our attention on the intersections of extremal curves with the boundaries of the associated entangling wedges, we limited our attention to the converging light sheets shown in figure \ref{EWI}. However,
the `outward directed' light sheets traveling towards the interior of the bulk can also provide null vectors connecting two extremal
curves, $\Gamma(s;\la)$ and $\Gamma(s;\la+d\la)$. If the boundary theory is in a pure state, we can think that these light
sheets $\widehat W(s,\tau;\la)$ define the boundary of the entanglement wedge of the complement of the original intervals
considered in our previous discussion. To introduce some notation, figure \ref{EWcomp} 
illustrates the intersection of the extremal curve $\Gamma(s;\la)$ at $s=s_+(\la)$ with the boundary $\widehat W(s,\tau;\la+d\la)$ emerging from $\Gamma(s;\la+d\la)$. Here we have a null vector  $\hat k_+(\la+d\la)$ connecting the intersection point $\hat s_+(\la)$ on $\Gamma(s;\la)$ with the point $\hat s^*_+(\la+d\la)$ on $\Gamma(s;\la+d\la)$. In this case, the null vector is orthogonal to $\Gamma(s;\la+d\la)$ at $\hat s^*_+(\la+d\la)$, \ie  $\hat k_+(\la+d\la)\cdot \dot\Gamma(s^*_+(\la+d\la);\la+d\la)=0$ and as we verify in appendix \ref{geom}, this intersection also leads to null vector alignment in the continuum limit. 

However, as we discuss in appendix \ref{geom}, an important point is that this new intersection does not lead to the construction of a new bulk curve in the continuum limit. Intuitively, this occurs because the two light sheets,
$W(s,\tau;\la+d\la)$  and $\widehat W(s,\tau;\la)$, essentially coincide in the vicinity of the relevant intersections. The
result can also be made apparent quantitatively by noting that $s^*_-(\la)-\hat s_+(\la)\sim O(d\la)$ and therefore the bulk curve constructed from $\hat s_{+}(\la)$ is the same curve as the one constructed from $s_-(\la)$. 
That is, we previously were thinking of the two solutions arising from the `left' intersection point $s_-(\la)$ of $\Gamma(s;\la)$ with $W(s,\tau;\la-d\la)$ and the `right' intersection point $s_+(\la)$ of $\Gamma(s;\la)$ with $W(s;\tau;\la+d\la)$. With this new perspective, we can also interpret the same two solutions as arising from the `left' intersection point $\hat s_+(\la)$ of $\Gamma(s;\la)$ with $\widehat W(s,\tau;\la+d\la)$ and the `right' intersection point $\hat s_-(\la)$ of $\Gamma(s;\la)$ with $\widehat W(s;\tau;\la-d\la)$. In particular, it will be useful in the following discussion of `trajectories' to be aware that the intersections with both the inward and outward directed light sheets can be used to construct the same bulk surfaces. 

To better understand the possible intersections and the global constraints mentioned above, it is convenient to think of the `trajectory' of the deviation vector $v(s;\la)$, defined in eq.~\reef{deviation}, in the transverse plane along an extremal curve $\Gamma(s;\la)$ for a fixed $\la$. In figure \ref{transit2}, we illustrate a few different classes of possible trajectories.\footnote{Note that implicitly in the previous discussion of intersecting entanglement wedges, we were having in mind a situation like that of figure \ref{transit2}c. In appendix \ref{geom}, we discuss the more general case.}   In general, the trajectory starts at $v(s_{init},\la) = \gamma_L'(\la)$ and ends at $v(s_{fin},\la) = \gamma'_R(\la)$. In between, it wanders around in the transverse space in some way. Of course, we are particularly interested in the points where the trajectory crosses the light cone
since this corresponds to the condition for null vector alignment, $v^2=0$. In crossing the light cone, the trajectory
is passing between different quadrants in the transverse space and so one may expect that in fact the 
physically interesting trajectories will begin and end in different quadrants. However, we should then be able to rule out the possibility that a trajectory can begin and end in the same quadrant and simply cross the same null direction an even number of times, as illustrated in figure \ref{transit2x}. In fact, while trajectories which start and end in the same spacelike quadrant, as
shown in figure \ref{transit2x}a, can be ruled out, it seems that starting and ending in the same timelike quadrant, as
shown in figure \ref{transit2x}b, is allowed.
\begin{figure}[h!]
\centering
\subfloat[]{\includegraphics[width=0.4\textwidth]{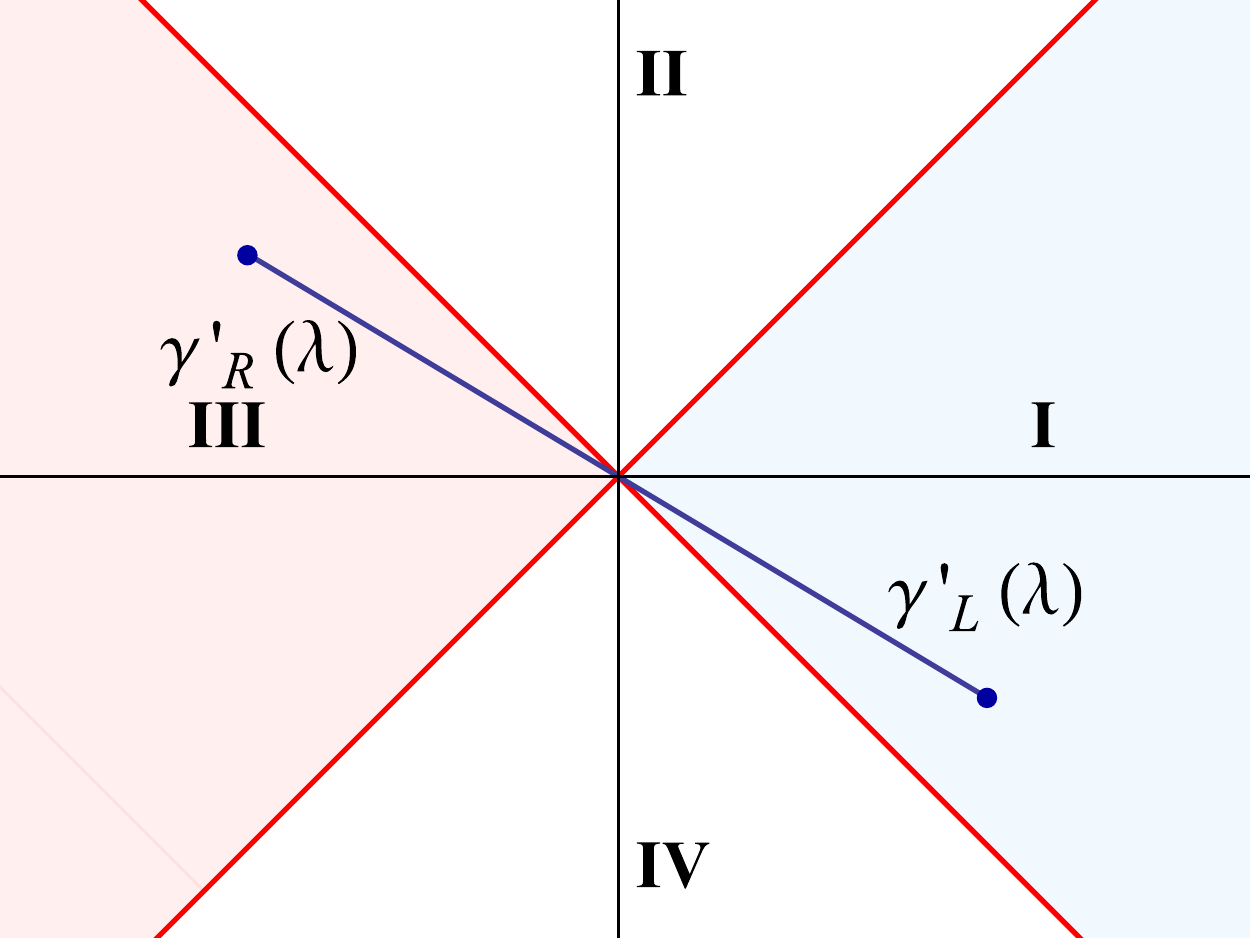}}\qquad
\subfloat[]{\includegraphics[width=0.4\textwidth]{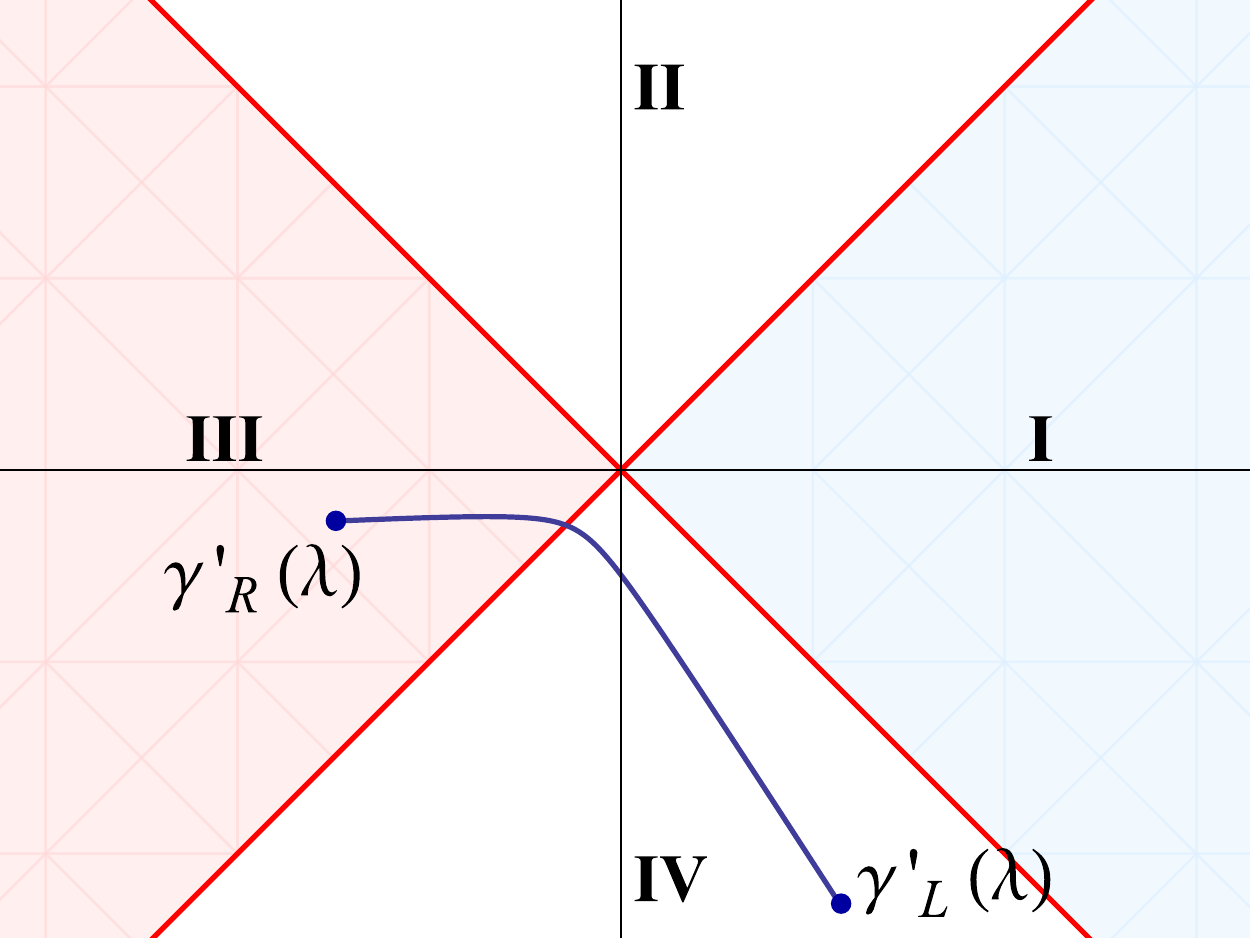}}\\
\subfloat[]{\includegraphics[width=0.4\textwidth]{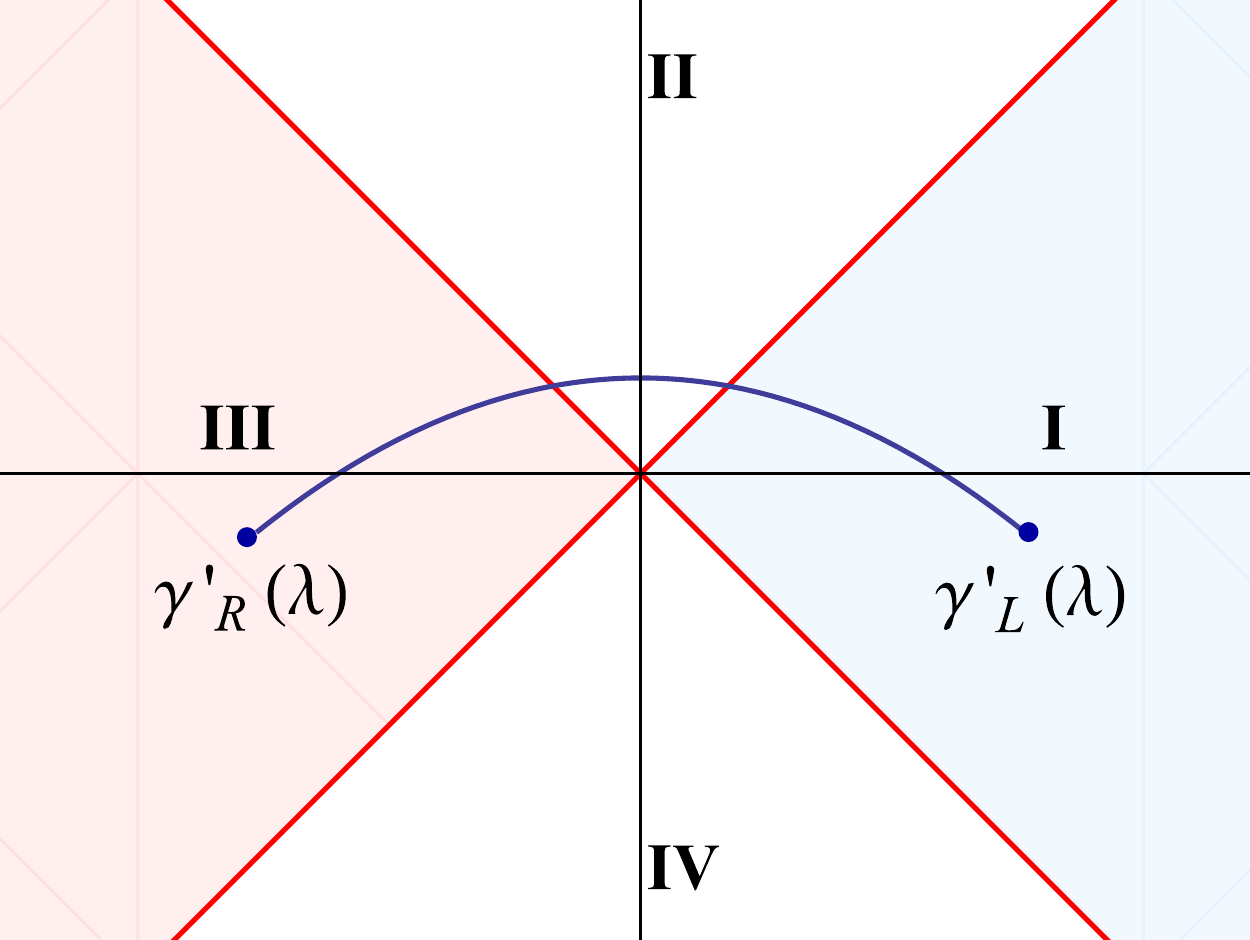}}\qquad
\subfloat[]{\includegraphics[width=0.4\textwidth]{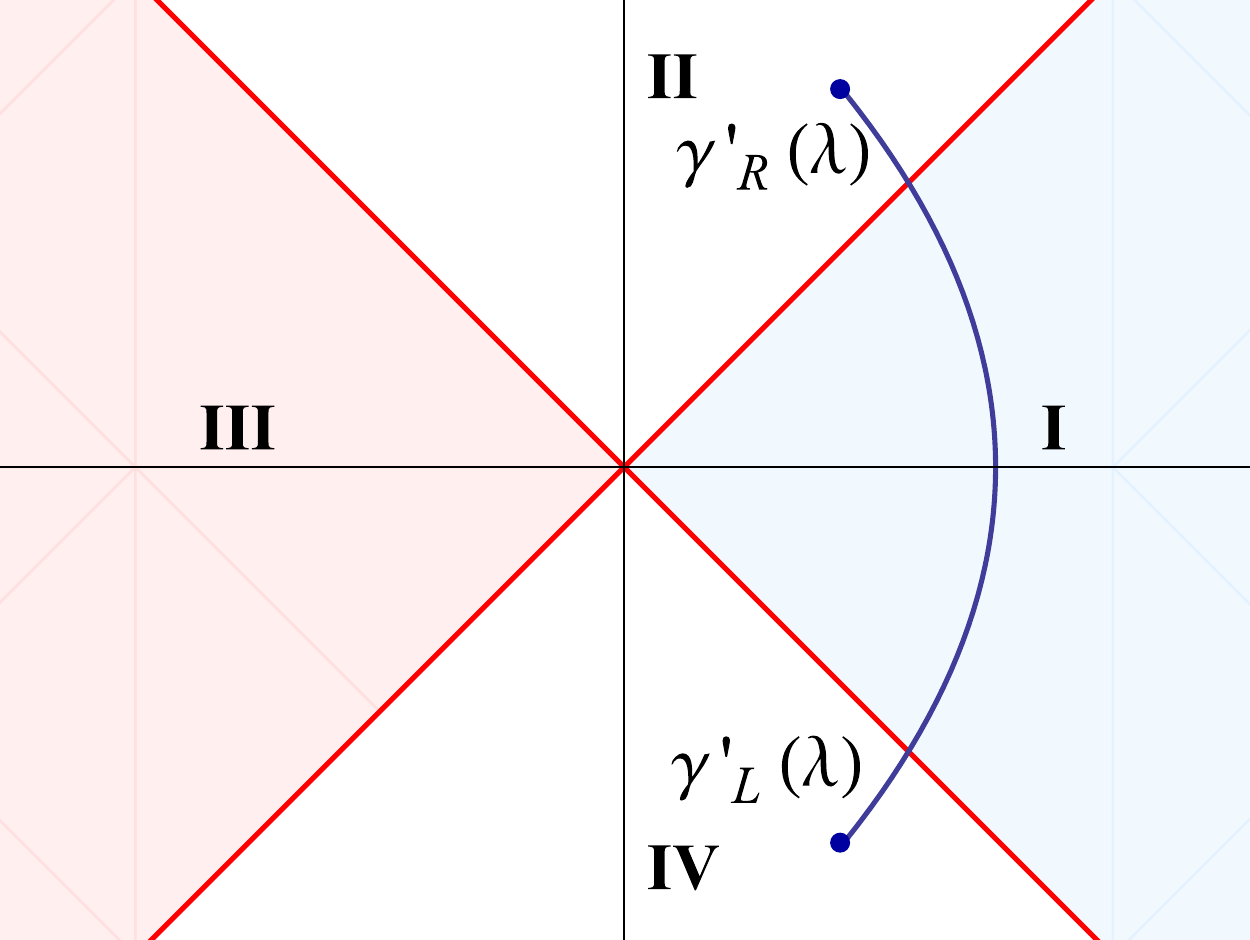}}
\caption{(Color online). We draw examples of possible trajectories of $v$ in the transverse plane. The blue shaded region I corresponds to the interior of the entanglement wedge, and the red shaded region III corresponds to the interior of the complement entanglement wedge. We can see there is a solution at the intersection of the trajectory with the boundary of either of these regions. Panel (a) illustrates an example of tangent vector alignment, where the solutions are degenerate. Panel (b) illustrates that it is possible for there to be only one solution, but without tangent vector alignment.}\labell{transit2}
\end{figure}
\begin{figure}[h!]
\centering
\subfloat[]{\includegraphics[width=0.4\textwidth]{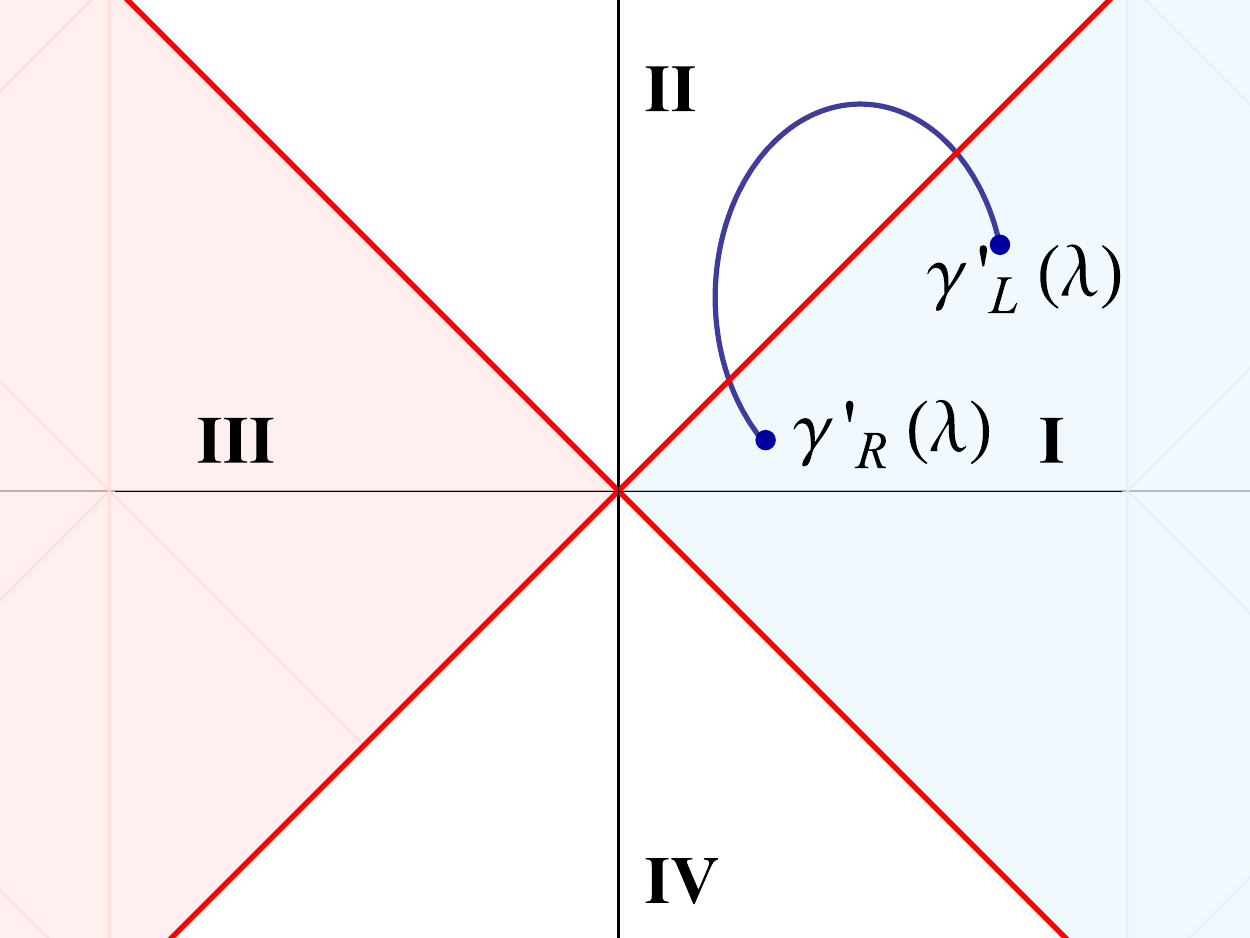}}\qquad
\subfloat[]{\includegraphics[width=0.4\textwidth]{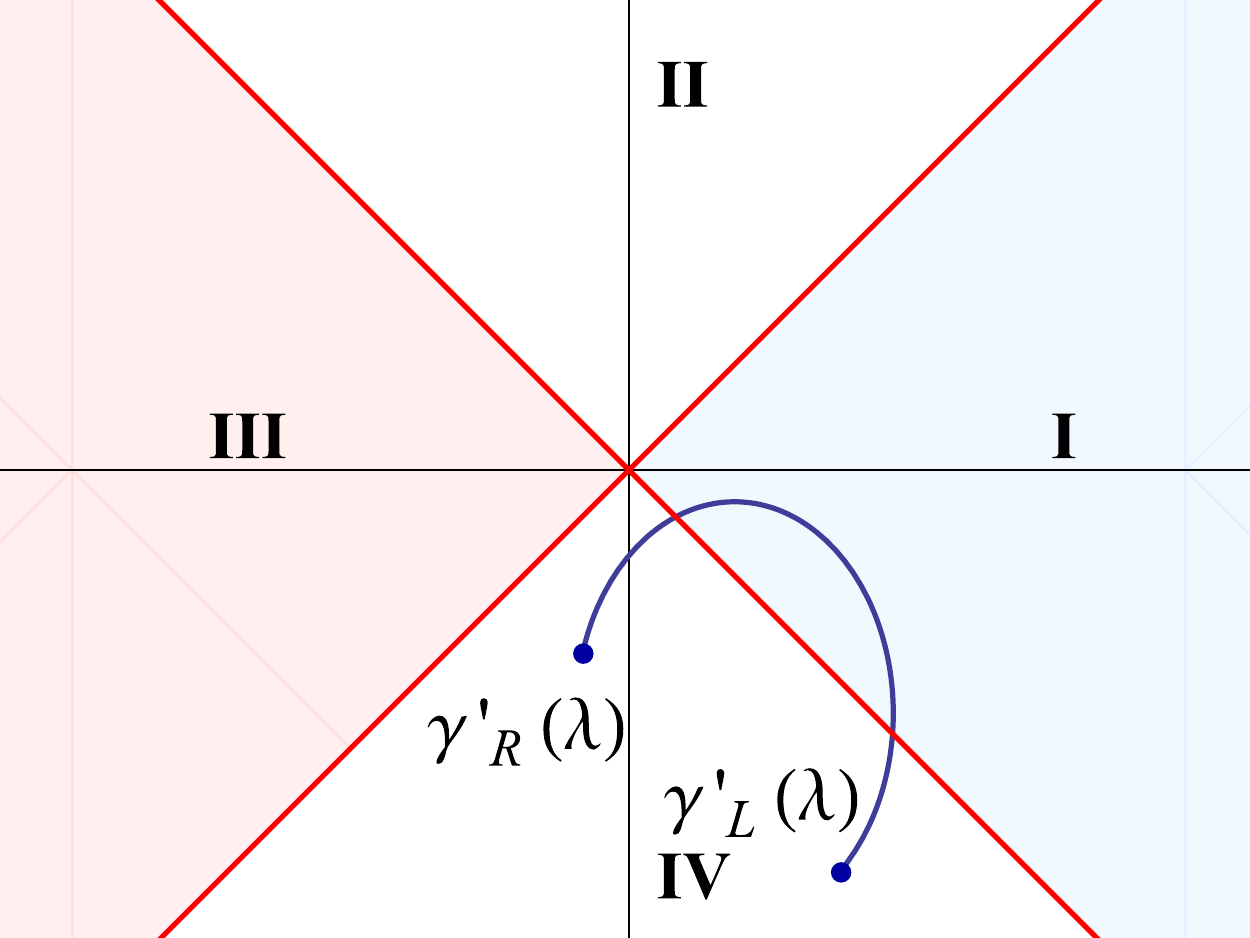}}
\caption{(Colour online) The trajectory drawn in (a) is ruled out by the covariant formulation \cite{aron2} of the previously mentioned argument from \cite{matt}. The trajectory drawn in (b), where both $\gamma_L'(\la)$ and $\gamma_R'(\la)$ are  timelike, is not ruled out.}\labell{transit2x}
\end{figure}
First we consider a trajectory like the one drawn in figure \ref{transit2x}a, where the endpoints $\gamma_L'(\la)$ and $\gamma_R'(\la)$ are both spacelike and in
the same quadrant. These trajectories can be ruled out using the results of \cite{aron2}, which provides a covariant formulation of the argument used in the discussion after eq.~\reef{constrain88}. In particular, in the situation
illustrated, the interval $I(\la+d\la)$ is entirely contained within $I(\la)$ on some time-slice in the boundary. Hence, the corresponding extremal curves, $\Gamma(\la+d\la)$ and $\Gamma(\la)$ are everywhere spacelike separated in the bulk.
Therefore $v^\mu$ must remain within the first quadrant along the entire trajectory and it cannot cross the
light cone, ruling out trajectories of the form illustrated in figure \ref{transit2x}a. 

This leaves us to consider trajectories where $\gamma_L'(\la)$ and $\gamma_R'(\la)$ are both timelike, as
shown in figure \ref{transit2x}b. However, we cannot generically rule out such trajectories rather it seems rather
simple to construct examples where this behaviour arises, as discussed in appendix \ref{geom}.

Hence we can see that if a trajectory begins and ends in different quadrants, then there is a solution of $v^2=0$, but 
the physically realizable trajectories seem to obey certain further constraints. For example, as discussed above, if either $\gamma_L'(\la)$ or $\gamma_R'(\la)$ is spacelike, then the trajectory must begin and end in different quadrants. The exception to the latter
rule seems to be when both $\gamma_L'(\la)$ and $\gamma_R'(\la)$ are timelike, \ie the boundary intervals are `moving in'
a timelike direction. However, the former rule demands that either of the following inequalities is satisfied:
\be
(x_R'-t_R')(x_L'+t_L')>0 \labell{constrA}
\ee
or
\be
(x_R'+t_R')(x_L'-t_L')>0 \labell{constrB}
\ee
These inequalities seem to provide a natural generalization of the global constraint given previously in eq.~\reef{constrain88} for families of boundary intervals in a fixed time slice. Certainly one sees that both eqs.~\reef{constrA} and \reef{constrB}  reduce to this previous constraint, \ie $x_L'(\la)x_R'(\la)>0$, when $t'_R=0=t'_L$. In general, given a family of boundary intervals, 
it is possible for both, one or neither of eqs.~\reef{constrA} or \reef{constrB} hold globally. If both are satisfied globally, then the new construction defines two (closed) bulk curves for which the gravitational entropy equals the differential entropy. If one holds everywhere, our generalized hole-ographic construction will certainly define a single (closed) bulk curve. In this case, a second curve may also exist but the corresponding family of boundary intervals must have both $\gamma_L'(\la)$ and $\gamma_R'(\la)$ timelike in the regime where the corresponding constraint does not hold.

We can gain further intuition by considering the evolution of the $v$ trajectory with the parameter $\la$. For a family of boundary intervals, it is possible that the number of solutions to the intersection equation $v^2=0$ changes for distinct values of $\la$. One scenario would be where the trajectories begin with two distinct crossings of the light cone as in figure \ref{transit2}c. As described above, in this situation, there are two distinct bulk curves corresponding to $s_+(\la)$
and $s_-(\la)$. Now as we vary $\la$, the trajectories could evolve smoothly such that the two null crossings shrink down
to the origin, at which point we have tangent vector alignment with $s_+(\la)=s_-(\la)$ as in figure \ref{transit2}a. From the bulk perspective, this case describes a situation where the two distinct bulk curves intersect at the special values of $\la$ where tangent
vector alignment is achieved. 

We can also consider trajectories which evolve from having two crossings to having a single crossing of a light cone, as in figure \ref{transit2}b. In this case, the family of boundary intervals is such that one of the endpoint
`velocities,' $\gamma_R'(\la)$ or $\gamma_L'(\la)$, crosses the light cone on the asymptotic boundary.
For example, if the trajectory evolves from that in figure \ref{transit2}b to that in figure \ref{transit2}c,   $\gamma_L'(\la)$ evolves from being timelike to being spacelike. In this situation, one of
the bulk intersection points approaches the asymptotic boundary, reaching infinity where $|\gamma_L'(\la)|=0$. 
Hence the corresponding bulk curve extends out to the boundary and terminates there. For spacetimes like AdS space, we may be concerned that as the curve hits the boundary, the `area' is infinite and hence the gravitational entropy should diverge.
However, the differential entropy remains finite!\footnote{The latter is clear since the bulk curve defined by the other crossing of the null cone shows no exceptional behaviour and integrating along both curves yields the same differential entropy.} This discrepancy arises because, as the explicit calculations in section
\ref{time} showed, the integrand in the differential entropy yields the area element on the bulk curve plus a total derivative. Of course, the latter is inconsequential if the bulk curve is closed. However, in the present situation the bulk curve terminates at the
boundary and the boundary contribution of the total derivative is responsible for canceling the divergence that appears in the gravitational entropy.

The above analysis holds true for a family of intervals on the boundary of an arbitrary holographic three-dimensional spacetime described by Einstein gravity, however, with minor modifications, it can be extended families of intervals on the boundary
of holographic backgrounds with generalized planar symmetry in any dimension.

%%%%%%%%%%%%%%%%%%%%%%%%%%%%%%%%%%
\subsection{AdS$_3$ as a case study}
\labell{AdS}
%%%%%%%%%%%%%%%%%%%%%%%%%%%%%%%%%%%

To build a better understanding of some of the generic properties of our
generalized hole-ographic construction, we explicitly solve for the bulk curves for AdS$_3$ in Poincar\'e coordinates \reef{3metric}. Given a set of spacelike boundary intervals with endpoints $\gamma_L(\la) = \coords{ x_L(\la), t_L(\la)}$ and $\gamma_R(\la) = \coords{ x_R(\la), t_R(\la)}$, first we change variables to a parameterization of the center $\{x_c,t_c\}$, the invariant length $\Delta$, and the boost angle $\beta$ (with respect to surfaces of constant $t$), for each of the intervals:
\ban{ \notag &x_c(\la) = \frac 12 (x_L(\la)+x_R(\la))\\ 
& t_c(\la)=\frac 12 (t_L(\la)+t_R(\la))\labell{param}\\ 
\notag & \Delta(\la) = \frac 12 \sqrt {(x_R(\la) - x_L(\la))^2-(t_R(\la)-t_L(\la))^2}\\ 
&\beta(\la)=\frac 12 \log \left[\frac{(x_R(\la)-x_L(\la))+(t_R(\la)-t_L(\la))}{(x_R(\la)-x_L(\la))-(t_R(\la)-t_L(\la))}\right]
\notag } 
and we choose $x_R(\la) \geq x_L(\la)$. Note that we are only considering spacelike intervals, \ie $|t_R(\la)-t_L(\la)|< x_R(\la)-x_L(\la)$, and hence the boost angle $\beta(\lambda)$ is everywhere finite and well-defined. For an interval at $\la$, with the parameterization $s\in[-1,1]$, the extremal curve has coordinates $\{z,x,t\}$ given by 
\ban{
\Gamma(s;\la)= \coords{\sqrt{1-s^2} \Delta(\lambda), \, x_c(\lambda) + s\,  \Delta(\lambda) \cosh \beta(\lambda), \, t_c(\lambda) + s \, \Delta(\lambda)\sinh \beta(\lambda)}\labell{geodB}
}
The following discussion is also facilitated by the introduction of an orthonormal basis at each point on the extremal curve consisting of the tangent vector $\hat u(s;\la)=\dot \Gamma(s;\la)/|\dot \Gamma(s;\la)|$ and two orthogonal unit vectors $\hat n_1(s;\la)$ and $\hat n_2(s;\la)$.\footnote{We choose $\hat n_1(s;\la)$ to be spacelike and to lie in the plane of the 
extremal curve, with $\hat n_1\cdot\hat n_1=1$ and $\hat n_1\cdot\hat u=0$. Further $\hat n_2(s;\la)$ is timelike
and orthogonal to the plane of the extremal curve,  with $\hat n_2\cdot\hat n_2=-1$ and $\hat n_2\cdot\hat u=0=
\hat n_2\cdot\hat n_1$.} 

For the general case in AdS$_3$, these basis vectors become
 \ban{
 \hat u^\mu &= \frac{\Delta(\la)}{2L}\,\sqrt{1-s^2}\,\coords{-s, \, \sqrt{1-s^2} \cosh\beta(\la), \, \sqrt{1-s^2}  \sinh \beta(\la)} \notag\\
 \hat n_1^\mu &= \frac{\Delta(\la)}{2L}\,\sqrt{1-s^2}\,\coords{-\sqrt{1-s^2}, \, -s  \cosh\beta(\la), \, -s\,  \sinh \beta(\la)} \labell{outside2} \\ 
 \hat n_2^\mu &=\frac{\Delta(\la)}{2L}\,\sqrt{1-s^2}\, \coords{0, \, \sinh \beta(\la), \,  \cosh\beta(\la)}\notag
 }
Next we evaluate eq.~\reef{deviation} to find
\ban{v(s;\la) \propto &-(s\, x_c'(\la) \cosh \beta(\la) - s\, t_c'(\la) \sinh \beta(\la) + \Delta'(\la)) \hat n_1 \notag\\
&\ \ +(t_c'(\la) \cosh \beta(\la) - x_c'(\la) \sinh \beta(\la) + s\,\Delta(\la) \beta'(\la)) \hat n_2
}
To determine when null vector alignment is achieved,  we solve for $v(s;\la)^2=0$, 
which yields the parameters $s_\pm(\la)$ as
\ban{s_\pm(\lambda)=-\frac{\Delta'(\la) \pm (t_c'(\la) \cosh \beta(\la) - x_c'(\la) \sinh \beta(\la)) }{x_c'(\la) \cosh\beta(\la)-t_c'(\la) \sinh\beta(\la)\pm \Delta(\la) \beta'(\la) } \,. \labell{nullcross}}
Note that this solution reduces to the constant time result \reef{constanttime} when $t'_c(\la)=0=\beta(\la)$. Further,
this explicit solution confirmd that $s_+(\la)$ and $s_-(\la)$ remain separate points in the continuum limit (as long as $t'_c(\la)\ne 0$ and/or $\beta(\la)\ne 0$). 

We can re-express these expressions for the intersection points \reef{nullcross} in terms of $\gamma_R(\la)$ and $\gamma_L(\la)$ using eq.~\reef{param}, but the
resulting formulae are rather lengthy and unilluminating. However, applying
the  constraints $|s_\pm(\la)|<1$, we see after some simplification that $|s_+(\la)|<1$ corresponds precisely to the inequality in eq.~\reef{constrA} and $|s_-(\la)|<1$ corresponds to that in eq.~\reef{constrB}. Therefore the global
constraint \reef{constrA} ensures that null vector alignment at $s_+(\la)$ 
produces a closed curve in the bulk. Similarly,
eq.~\reef{constrB} ensures the same at $s_-(\la)$. Furthermore,
we can interpret $s_+(\la)$ as the intersection with the null line $\hat n_1 + \hat n_2$ and $s_-(\la)$ as the intersection with the null line $\hat n_1 - \hat n_2$. These observations reveal that indeed for AdS$_3$, the previously mentioned trajectories in the transverse plane cross each light cone at most once, ruling out trajectories like the one in figure \ref{transit2x}b. We see explicitly why this happens in appendix \ref{geom}.

To further illustrate the situation of null vector alignment, we consider a simple family of boundary intervals with the same invariant width and boost angle and whose centers are all on a constant time slice. In particular, we choose $\beta(\la)=\beta_0$, $\Delta(\la)=\Delta_0$ and $t_c(\la)=0$. As shown in figure \ref{boosts}a, extremal curves corresponding
to neighbouring intervals do not intersect in this example and so one can not expect to
build the bulk curve with tangent vector alignment.  However, if we extend the geometry
to include the entanglement wedges, as illustrated in figure \ref{boosts}b, we see that two bulk curves can be constructed with null vector alignment by taking the `left'  or `right' intersection points. Considering eq.~\reef{nullcross} in this simple example,
the solution for the intersection points becomes
\ban{
s_\pm = \pm \tanh \beta_0\, . \labell{boostsol}
}
Hence, as is also clear in the figure, $s_\pm$ remain separate points in the
continuum limit and so the
left and right intersections yield two distinct bulk curves. In fact, substituting $s=s_\pm$ into 
eq.~\reef{geodB}, we find
\be
\gamma_B^\pm=\left\lbrace \frac{\Delta_0}{\cosh\beta_0},\, \ell\,\la,\,
\pm\Delta_0\,\frac{\sinh^2\beta_0}{\cosh\beta_0}\right\rbrace\,,
\labell{bcs9}
\ee 
where as in section \ref{time}, $\ell$ is the period in the $x$ direction and recall
that $\la\in[0,1]$. 
\begin{figure}[h!]
\centering
\subfloat[]{\includegraphics[width=0.45\textwidth]{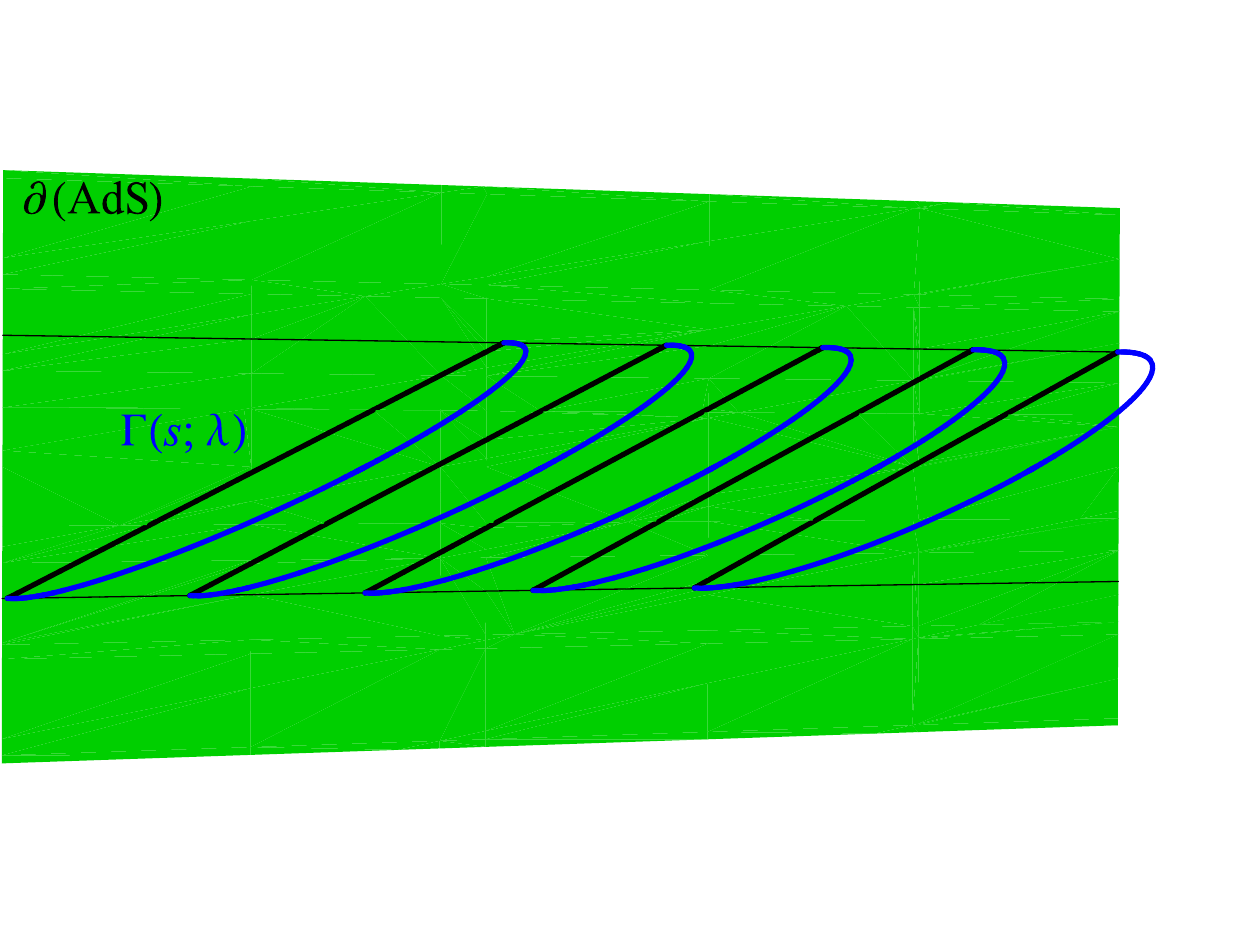}}\qquad
\subfloat[]{\includegraphics[width=0.45\textwidth]{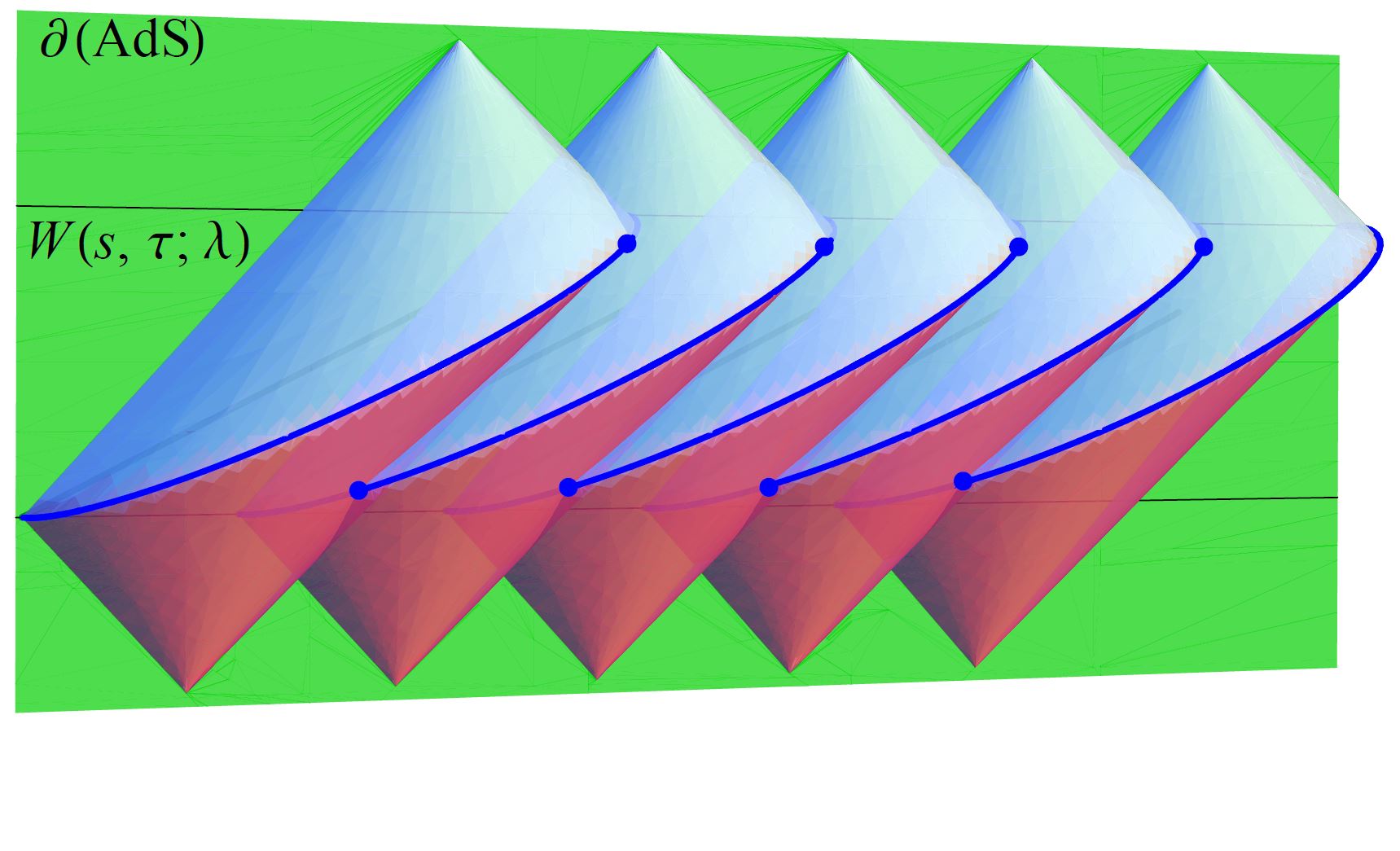}}\\
\subfloat[]{\includegraphics[width=0.5\textwidth]{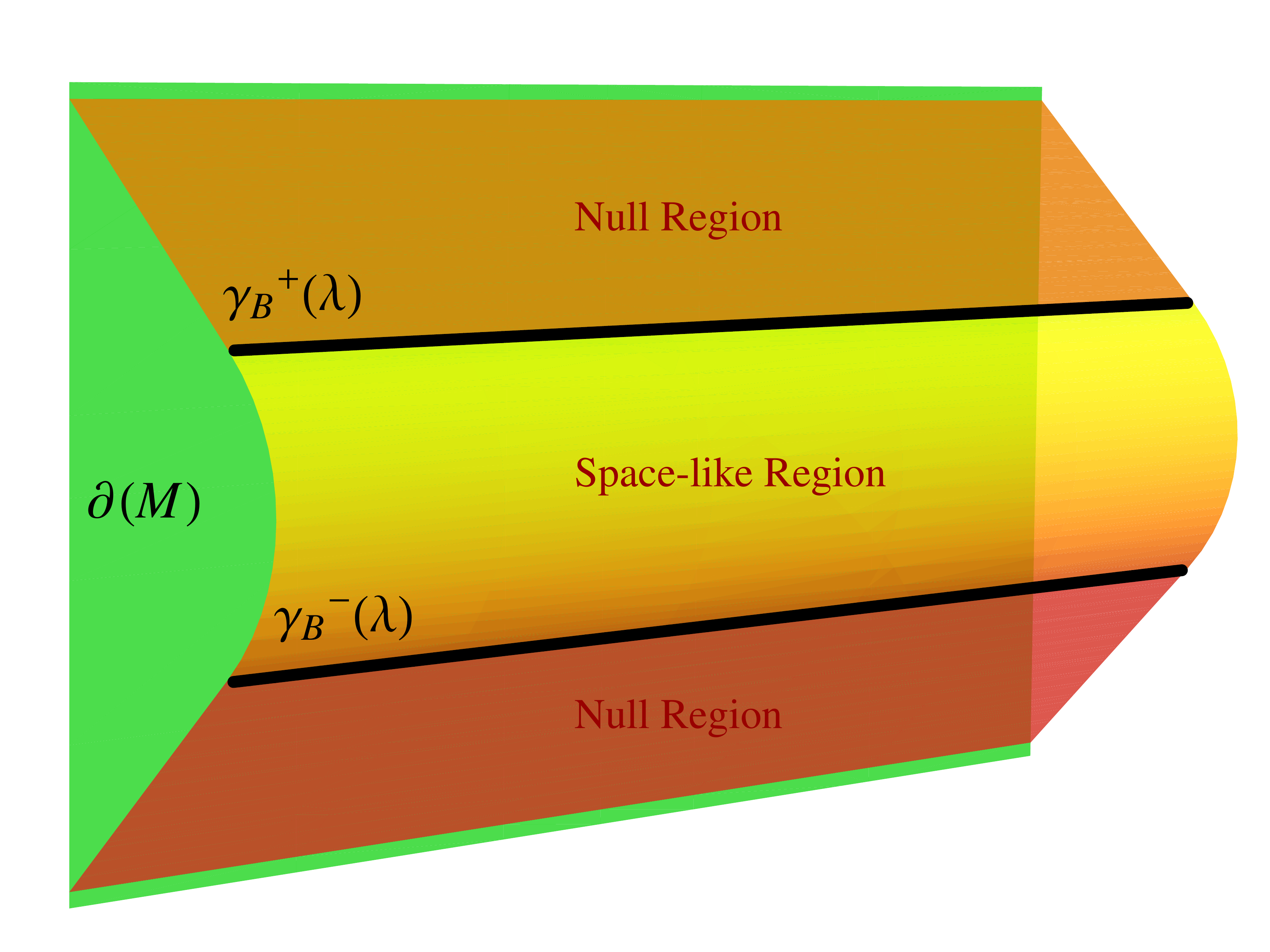}}\\
\caption{(Colour online) We consider a family of boundary intervals with constant invariant width and boost angle whose centers all lie on a constant time slice. In (a), we do not expect a bulk curve to arise from these intervals as the extremal curves do not intersect. However, we see in (b) that extending the geometry to the entanglement wedges yields two distinct sets of intersection points  ---  see also figure \ref{EWS}. In (c), we show the continuum enveloping surface
and we see that two distinct bulk curves emerge as the boundaries between the spacelike and null regions. }\labell{boosts}
\end{figure}

Since the two curves are only displaced from one another in the time direction,
it is clear that their gravitational entropy will be the same, as 
must be the case since both should match the same differential entropy in the boundary theory. Explicitly
evaluating eq.~\reef{prop0} using the AdS$_3$  metric \reef{3metric}, we find
\be
S_{BH}(\gamma_B^\pm)=\frac{L}{4G_N}\,\frac{\ell}{\Delta_0}\,\cosh\beta_0\,.
\labell{exit8}
\ee 

To calculate the differential entropy, first we note in general that using the formula \reef{left} for differential entropy and the formula \reef{Sboost} for holographic entanglement entropy in AdS$_3$, we can write the differential entropy in terms
of the new variables \reef{param} as 
\ban{
E&= \int_0^1 d\la \,\frac L{4G_N}\, \frac 1 {\Delta(\la)} \left(x_c'(\la)\cosh \beta(\la) -t_c'(\la)\sinh \beta (\la)+\Delta'(\la)\right) \notag \\
&= \int_0^1 d\la\, \frac L{4G_N}\, \frac 1 {\Delta(\la)} \left(x_c'(\la)\cosh \beta(\la) -t_c'(\la)\sinh \beta (\la)\right) \,. \labell{AdSdiffE}
}
where in the second line, we have set the total derivative to vanish by the periodic boundary conditions. Now for our simple example with the
bulk curves given in eq.~\reef{bcs9}, $x_c'=\ell$ and $t_c'=0$ (along with $\Delta(\la)=\Delta_0$ and $\beta(\la)=\beta_0$)
and therefore the above formula indeed yields
\ban{
E=\frac L{4G_N} \frac \ell{\Delta_0} \cosh \beta_0\,.
}
Hence we have $E=S_{BH}(\gamma^\pm_B)$ as desired.

%%%%%%%%%%%%%%%%%%%%%%%%%%%%%%%%%%%%%%%%%%%%%%%%%%%%%%%%%%%%%%%%%%%%%%%%%%%%%%%%
%%%%%%%%%%%%%%%%%%%%%%%%%%%%%%%%%%%%%%%%%%%%%%%%%%%%%%%%%%%%%%%%%%%%%%%%%%%%%%%%
%%%%%%%%%%%%%%%%%%%%%%%%%%%%%%%%%%%%%%%%%%%%%%%%%%%%%%%%%%%%%%%%%%%%%%%%%%%%%%%%
\section{Discussion}
\labell{discuss}

\subsection{Summary}

Hole-ography, or the interpretation of the gravitational entropy of bulk surfaces as an observable in the boundary 
theory, appears to be a robust
entry in the holographic dictionary. The original construction \cite{hole} was already extended
in \cite{jun} to higher dimensions, to other holographic backgrounds (\ie new backgrounds which may or may
not be asymptotically AdS) and to certain classes of higher curvature gravity theories (\ie Lovelock gravity). 
The present paper --- see also \cite{thesis} --- provides further extensions to the hole-ographic construction.
In particular, in section \ref{time}, these results are extended to bulk surfaces which varied both in space
and time. Further, the general proof presented
in section \ref{general} generalizes the construction to holographic backgrounds that are
themselves varying in the space and time directions of the boundary theory. Finally, the null vector alignment
approach of section \ref{new} indicates that a general family of boundary intervals, subject to some mild global constraints, naturally defines a bulk surface such that the differential entropy in the boundary and the gravitational
entropy in the bulk agree.

One of the lessons emerging from section \ref{time} is that in general, \ie for time varying surfaces in the bulk,
we should not be thinking of defining the corresponding intervals on some (possibly time varying) Cauchy surface
in the boundary geometry. Rather the appropriate boundary intervals may lie on completely different Cauchy surfaces,
and so it is best to define the boundary data in terms of two curves, $\gamma_R(\la)$ and $\gamma_L(\la)$, which
specify the endpoints of the intervals. Of course, this formalism was also essential to our formulation of eqs.~\reef{left}
and \reef{right}, which give natural continuum expressions for the differential entropy. As noted in footnote \ref{wacky}, with this endpoint data, we emphasize that these expressions do not require any modification to describe situations, \eg
where the bulk surface varies `too rapidly' in the radial direction and the boundary intervals progress in a `backwards
direction' --- see discussion in \cite{jun}. Further, we note that there is more freedom in the families of boundary intervals than one might have initially expected. In particular, in terms of the discussion surrounding figure \ref{transit2},  $\gamma_L'(\la)$ and $\gamma_R'(\la)$ can be in any quadrant in the space to the endpoints, and so the curves $\gamma_R(\la)$ and $\gamma_L(\la)$ may be either spacelike or timelike at different points. The only constraint
that we established in eqs.~\reef{constrA} and \reef{constrB} is that $\gamma_L'(\la)$ and $\gamma_R'(\la)$ should
be in different quadrants. However, even this constraint may be evaded when both $\gamma_L'(\la)$ and $\gamma_R'(\la)$ are timelike. Of course, an implicit assumption was also that for each $\lambda$ the interval $[\gamma_L(\lambda),\gamma_R(\lambda)]$ was itself spacelike (and furthermore lay on a Cauchy slice).

The calculations in section \ref{time} also illustrated another important lesson, which was that the gravitational entropy was properly reproduced if the extremal curves were chosen to be tangent to the bulk curve at each point. We denoted this configuration `tangent vector alignment,' which became an important ingredient in our general
proof in section \ref{general}. However, as we saw in section \ref{new}, constructing the bulk surface from a general family of boundary intervals required that we allow for a `looser' configuration, namely `null vector alignment,' which
was also allowed by the general proof.
This new approach has a geometric interpretation in terms of extremal curves intersecting the light sheets that defined the boundary of the entanglement wedges associated with neighbouring extremal curves (rather than the extremal curves
intersecting each other). With null vector alignment, we also have a new freedom in describing a given bulk surface --- a point that we return to below.

Another important lesson was that the function $\alpha$, appearing in eq.~\eqref{alphadef}, can take either sign, and in fact, it can change sign as a function of $\lambda$. The latter was demonstrated with an explicit example in section \ref{sign}, which illustrated that the sign changes were associated with cusps in the bulk curve. Further we note that these cusps in the bulk were not associated with any dramatic behaviour of the boundary intervals.\footnote{Similar geometries were discussed in \cite{verona}.} 
In any event, this behaviour requires that we associate the differential entropy with a generalized notion of the gravitational entropy, where an additional factor of $\sgn(\alpha)$ appears in the integration over the bulk surface, as in eq.~\eqref{proof2}. 

Of course, it is implicit in all of our analysis here (and in \cite{jun,thesis})  that the background geometries
exhibit a generalized planar symmetry, which was described in detail in section \ref{planarS}. 
The class of backgrounds where the metric takes the desired form \reef{pmetric} and satisfies the constraint \reef{pdet2} is quite broad, \eg including geometries dual to boundary field theories with Lifshitz or Schr\"odinger symmetries, or describing the formation of a black hole, as in eq.~\reef{vaidya}. An interesting exercise might be to  extend the analysis of section \ref{planarS} to even more general backgrounds, \eg to include stationary black hole backgrounds, which do not take the form given in eq.~\reef{pmetric}.
In any event, in the presence of generalized planar symmetry, the key feature is that the profile of the bulk surfaces has a nontrivial dependence on a single coordinate, \ie $\la$. An important direction for future research is extending these constructions to situations lacking the generalized planar symmetry and where the bulk surfaces depend on several coordinates independently. In fact, important progress in this
direction has already been made \cite{sully}.

\subsection{Open questions}

One cautionary note is that our analysis of the bulk surfaces dual to the boundary intervals is local. For example, the
general proof presented in section \ref{general} only relies on these surfaces being extremal. Hence the first caveat
is that these surfaces may not be minimal surfaces, \ie they may not be the surfaces that define the entanglement
entropy of the corresponding boundary interval according to the RT prescription. In many instances, there may be
multiple extremal surfaces for a given interval --- \eg see \cite{multiple}.  However, the hole-ographic construction
could be interpreted to suggest that these extremal but not minimal surfaces still have a role in holography. It is certainly another 
interesting direction for future research to better understand whether this is true and again, progress has been made in certain
cases \cite{extreme}. Related to this issue is the fact 
that there may be bulk regions which are not reached by extremal surfaces \cite{walls}. That is, it seems that no choice of intervals (or regions) in the boundary theory will yield a differential entropy that corresponds to the gravitational entropy of bulk surfaces entering such regions. Some progress in
overcoming this barrier can be made by relaxing the implicit assumption that the extremal surfaces are anchored
on a single asymptotic boundary and considering instead surfaces connecting two asymptotic regions \cite{progress}, 
\eg in the background of an eternal black hole. In this case, the bulk geometry is dual to multiple entangled copies of the boundary theory, and the entanglement entropies entering the hole-ographic construction involve regions in more than one boundary.  
However, one may still find that there
are `barrier' surfaces which no extremal surface can cross, \eg near a black hole singularity, irrespective of where the surfaces are anchored \cite{walls}.
The latter presents an important obstacle for the hole-ographic construction in reconstructing the entire bulk geometry for such a situation. 

As we noted above, with null vector alignment, there is a new freedom in associating the bulk surface with boundary intervals. In particular, if we choose a fixed spacelike curve $\gamma_B(\lambda)$ in the bulk, we can build boundary intervals by using extremal surfaces $\Gamma(s;\lambda)$ that satisfy null vector alignment \eqref{conull}
by choosing $\dot\Gamma(s_B;\lambda) = \gamma_B'(\lambda)+k(\lambda)$, where $k(\lambda)$ is an arbitrary null vector orthogonal to $\gamma_B'(\lambda)$. 
There is thus a large freedom in choosing boundary intervals for which the differential entropy equals the gravitational entropy of a given bulk curve. This defines some kind of symmetry in the space of families of boundary intervals, and in explicit examples, we can derive the transformation between the parameters characterizing each family. However, even for AdS$_3$, this transformation does not reveal itself as an obvious symmetry of the boundary theory. Of course, it is clear that amongst all of the possible families of intervals, the one which realizes tangent vector alignment is distinguished. We do not have a clear understanding of the significance of this observation at this point --- however, see below. We might also note that lifting the restriction to 
generalized planar symmetry will further expand the families of boundary regions which correspond to the same bulk curve.

\begin{figure}[h!]
\begin{center}
\includegraphics[width=0.5\textwidth]{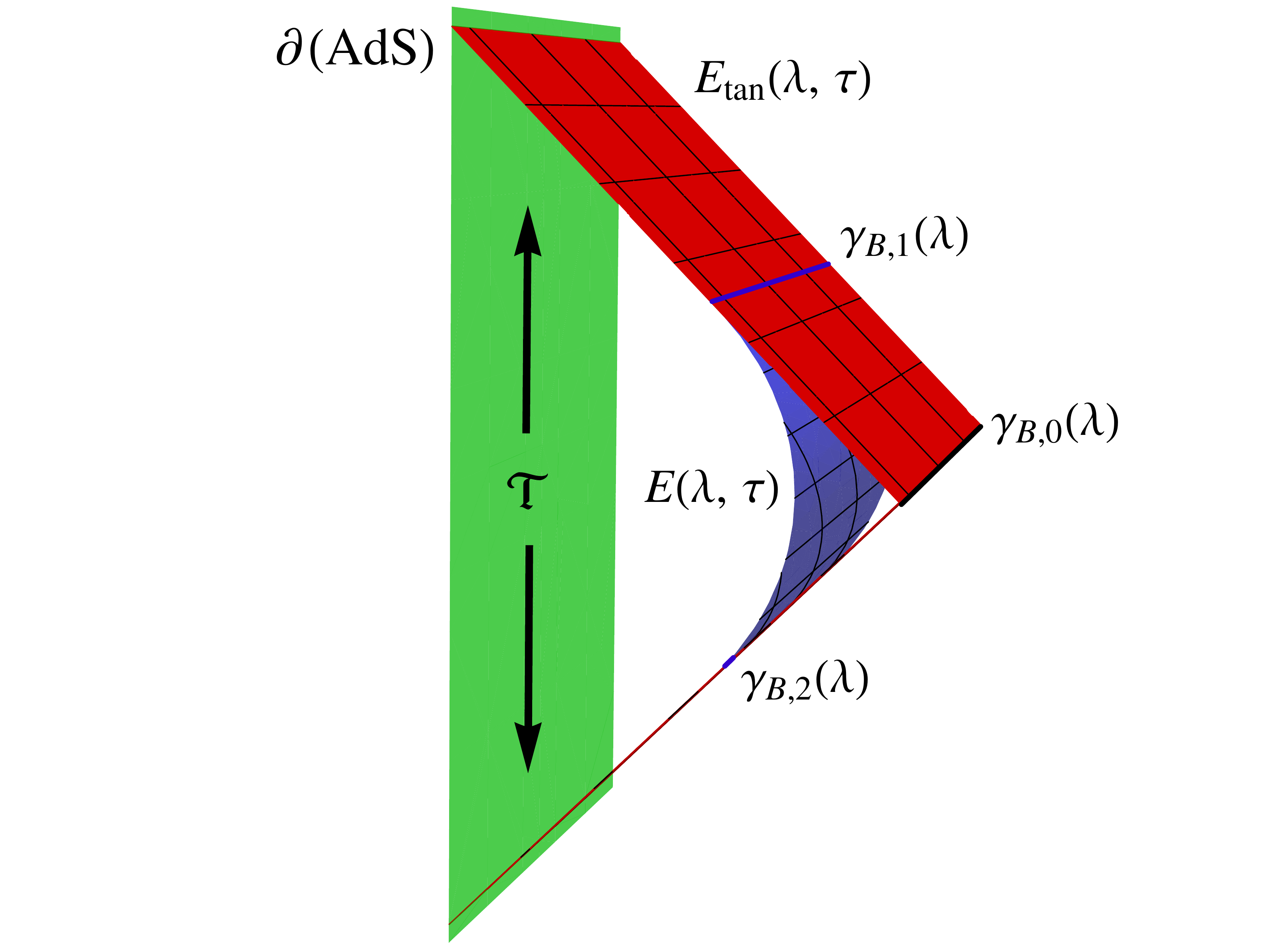}
\caption{(Colour online) We compare the enveloping surfaces for two families of boundary intervals defining the same time strip $\mathcal T$. The enveloping surface $E_{tan}(\la,\tau)$ corresponds to tangent vector alignment, which yields the
bulk surface $\gamma_{B,0}$ with the minimal gravitational entropy. We see that $E_{tan}(\la,\tau)$ bounds the enveloping surface constructed with null vector alignment and the corresponding bulk curves, $\gamma_{B,1}$ and $\gamma_{B,2}$, still lie on $E_{tan}(\la,\tau)$. }
\label{strip}
\end{center}
\end{figure}

The additional freedom allowed by null vector alignment is also manifest in other ways. Let us 
describe a certain bulk curve $\gamma_B$ with tangent vector alignment and then consider the corresponding enveloping
surface $E_{tan}(\la,\tau)$ constructed from the corresponding entanglement wedges, as described in section \ref{new}.\footnote{Note that $E_{tan}(\la,\tau)$ will not correspond to the lightsheets sent out from $\gamma_B$ to the boundary, because in general the top and bottom of the entanglement wedges are cut off by caustics --- see figure \ref{EW}. However, the two surfaces will coincide in the vicinity of the bulk curve. We also note
that AdS$_3$ is a special case where such caustics do not form. Further, we note that generally $E_{tan}(\la,\tau)$ will not coincide with either the `strip wedge' or the `rim wedge' defined in \cite{verona}.} 
Following the results of \cite{mattEW}, the intersection of this enveloping surface with the asymptotic boundary coincides with the envelope of the causal domains of the 
corresponding boundary intervals. Labeling this boundary region $\cal T$, we might denote the latter as a `time strip,' following the original discussion of hole-ography in \cite{hole}. However, when the intervals are not restricted to lie in a fixed time slice,
the same time strip can be defined using many different families of boundary intervals, as shown in figure \ref{tiling}. 
\begin{figure}[h!]
\centering
\subfloat[]{\includegraphics[width=0.45\textwidth]{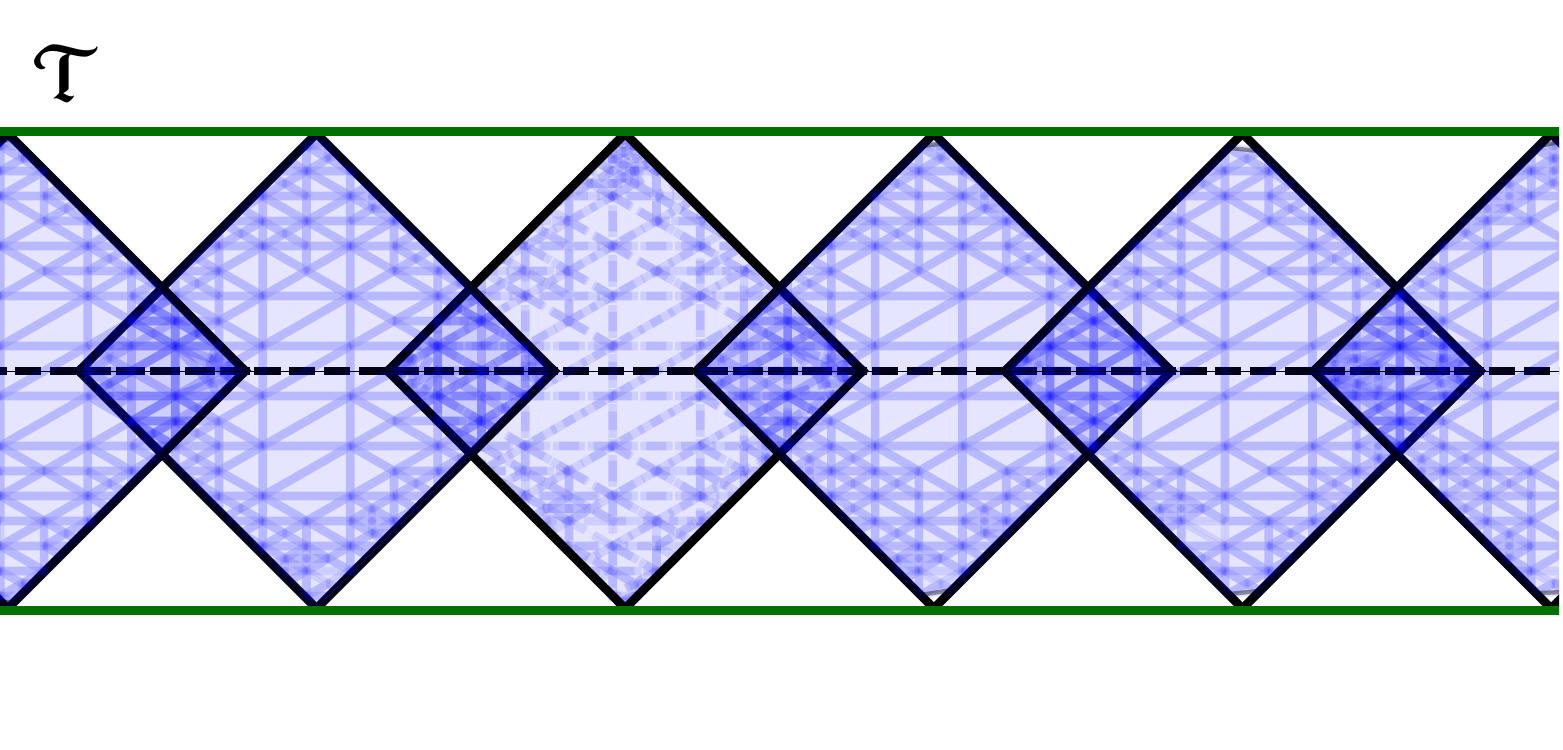}}\qquad
\subfloat[]{\includegraphics[width=0.45\textwidth]{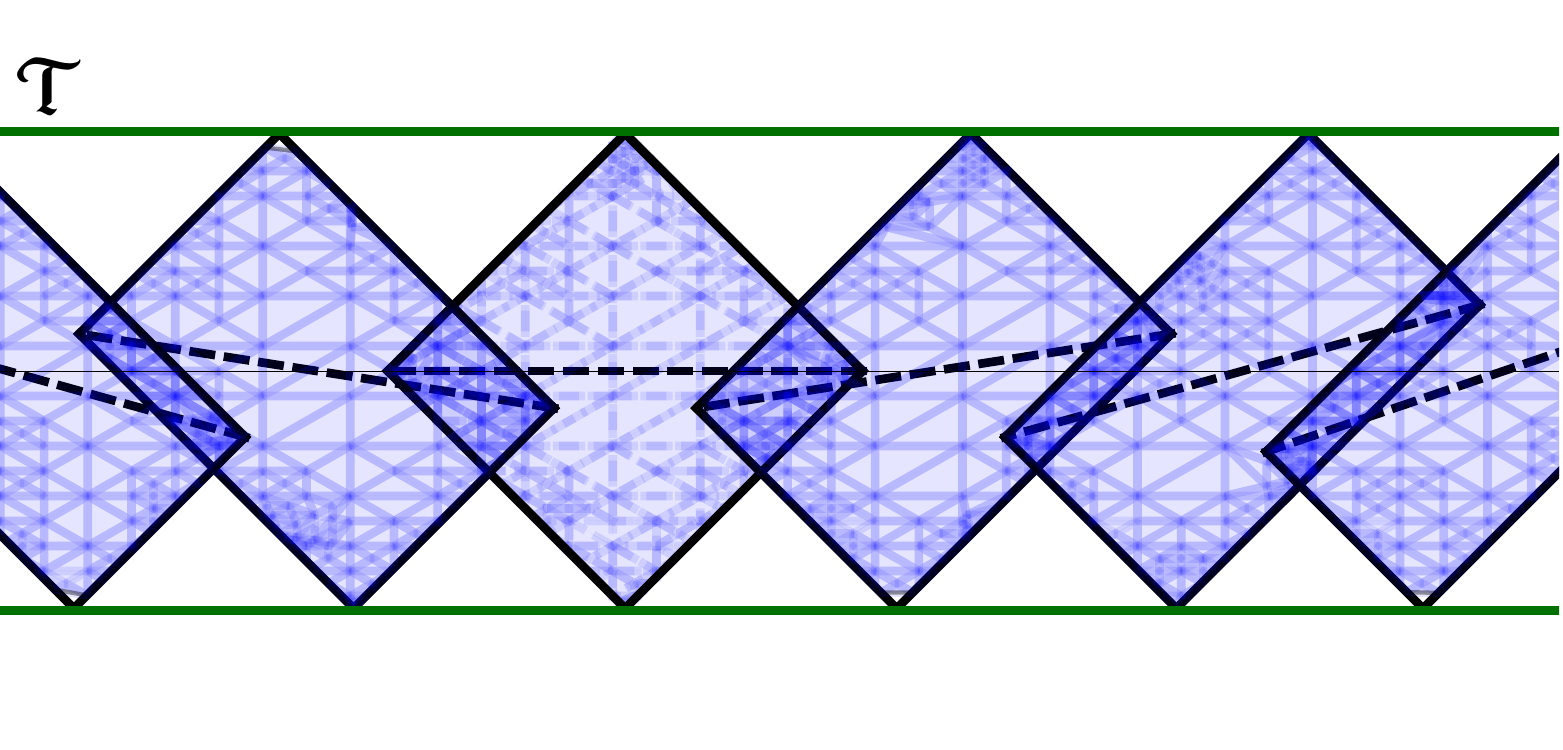}}\\
\caption{(Colour online) Given a time strip $\mathcal T$, we can `tile' it with a family intervals satisfying tangent vector alignment as in (a), or we can tile the same time strip with (many different) families of intervals for which the bulk curves
arise from null vector alignment as in (b).}\labell{tiling}
\end{figure}
Now for an alternate
family of time intervals which yield the same time strip $\cal T$, we can use null vector alignment to define
a new bulk curve. An interesting feature of any such bulk curve is that it will lie on $E_{tan}(\la,\tau)$,
the enveloping surface defined by tangent vector alignment, as shown in figure \ref{strip}. As illustrated in the
figure, the upper and lower portions of the enveloping surface (\ie the timelike and null regions) associated with the new choice of boundary intervals still match with $E_{tan}(\la,\tau)$, however, null vector alignment produces a spacelike region in between the upper and lower null regions --- see footnote \ref{parcel0} --- where the new enveloping surface departs from $E_{tan}(\la,\tau)$. Since the new bulk curves demarcate the boundary
between this spacelike region and the two null regions, they both lie on $E_{tan}(\la,\tau)$. Furthermore, as the bulk curve
defined by tangent vector alignment lies at the innermost limit of $E_{tan}(\la,\tau)$, its gravitational entropy will
be smaller than for any of the curves lying higher up on this enveloping surface. Hence from this perspective,
tangent vector alignment is distinguished since it selects out the boundary intervals with the minimal differential
entropy for a given time strip. We hope to return to the implications of these observations elsewhere.

Another noteworthy observation is that the `classical mechanics' theorem presented in section \ref{general} 
was not intrinsically linked to holographic entanglement entropy. Rather the essential ingredient was that
the calculation relied on extremizing an `area functional' in the bulk. However, the latter applies equally
well to many different holographic probes, at least to leading order in the large-$N$ limit \eg Wilson loops
\cite{wilson} and two-point correlators of some high dimension operators \cite{twop}. Hence our construction in section \ref{general} can easily be adapted to apply to these observables as well. For example, we could reconstruct the length of general curves in bulk
from a `differential version' of the two-point correlator of a high dimension operator,
 \be
\ell(\gamma) = \oint d\lambda
\ \frac{dq_L^a}{d\lambda}\,\frac{\partial_{q^a_L}
\langle {\cal O}(q_L^a)\, {\cal O}(q_R^a)\rangle}
{\Delta\,\langle {\cal O}(q_L^a)\, {\cal O}(q_R^a)\rangle}\,.
 \labell{long}
 \ee
It is interesting that in general nonlocal boundary observables, such as holographic entanglement entropy,
typically probe the bulk geometry at all scales from some minimum radius out to the boundary. However, 
the corresponding `differential observables' enable us to directly access information about the bulk at the
minimal radius. Hence these new observables are an exciting new tool towards the reconstruction of the bulk
geometry. This is certainly a topic to which we hope to return in future research.

\vskip 1cm

\section*{Acknowledgements} We would like to thank Aysha Abdel-Aziz,
Vijay Balasubramanian, Raphael Bousso, Bartlomiej Czech, Xi Dong, Netta Engelhardt, 
Damian Galante, Janet Hung, Juan Maldacena, 
Junjie Rao, Vladimir Rosenhaus, Misha Smolkin, and James Sully
for useful comments and discussions. MH would like to thank the Perimeter Institute for hospitality during the initial stages of this project. The work of MH was supported
in part by the National Science Foundation under CAREER Grant No. PHY10-53842. Research at Perimeter Institute is supported by the
Government of Canada through Industry Canada and by the Province of Ontario
through the Ministry of Research \& Innovation. RCM also acknowledges support
from an NSERC Discovery grant and funding from the Canadian Institute for
Advanced Research.

%%%%%%%%%%%%%%%%%%%%%%%%%%%%%%%%%%%%%%%%%%%%%%%%%%%%%%%%%%%%%%%%%%%%%%%%%%%%%%%%%%%%%%%%%%%%%%%%%
%%%%%%%%%%%%%%%%%%%%%%%%%%%%%%%%%%%%%%%%%%%%%%%%%%%%%%%%%%%%%%%%%%%%%%%%%%%%%%%%%%%%%%%%%%%%%%%%%
%%%%%%%%%%%%%%%%%%%%%%%%%%%%%%%%%%%%%%%%%%%%%%%%%%%%%%%%%%%%%%%%%%%%%%%%%%%%%%%%%%%%%%%%%%%%%%%%%

\appendix

\section{Derivation of the geometric interpretation}\labell{geom}

In this appendix we show that the geometric interpretation of section \ref{new} satisfies the condition for the holographic correspondence established in section \ref{general} between the differential entropy evaluated on a family of boundary intervals and the gravitational entropy of a bulk curve. Given a family of intervals with extremal curves $\Gamma(s;\la)$, the bulk curve constructed via $\bulkc_B(\la)=\Gamma(s_B(\la);\la)$ must satisfy the null vector alignment condition in eq.~\reef{conull} or equivalently,
\be
\dot\bulkc_B(\la) \cdot \bulkc'_B(\la) = |\dot \bulkc_B(\la)|\,|\bulkc'_B(\la)| \,.
\labell{aligned}
\ee
First we change variables to $\Gamma'(s;\la)$ and $\dot \Gamma(s;\la)$ with the relations:
\be
 \bulkc'_B{}^{\!\!\mu}(\la)=\left. \Gamma'{}^{\mu}(s;\la)\right|_{s_B(\la)}+ \dot \Gamma^\mu(s;\la)|_{s_B(\la)}\ s_B'(\la)
\quad {\rm and}\quad 
\dot \bulkc_B^\mu(\la)=\dot \Gamma^\mu(s;\la)|_{s_B(\la)}\,.
\labell{note99A}
\ee
Then it is straightforward to show that eq.~\reef{aligned} becomes
\ban{\left. \frac{\dot \Gamma(s;\la) \cdot \Gamma'(s;\la)} { |\Gamma'(s;\la)|{|\dot \Gamma(s;\la)|} }\right|_{s=s_B(\la)}
=1\,. \labell{alignedG}}

Now recall the basis of orthonormal vectors established for each extremal curve in section \ref{new}. This basis consists of the tangent vector $\hat u(s;\la)=\dot \Gamma(s;\la)/|\dot \Gamma(s;\la)|$ and two orthogonal unit vectors $\hat n_1(s;\la)$ and $\hat n_2(s;\la)$. With this formalism, we defined $v^\mu(s;\la)$, the projection of $\Gamma'{}^\mu$ into the
subspace transverse to $\hat u(s;\la)$ in eq.~\reef{deviation}. The condition \reef{alignedG} for null vector alignment then
became $|v(s_B(\la);\la)|=0$.

\vskip 0.2cm

Now we begin by showing that when the boundary intervals all lie on a constant time slice, at the intersection between the curve $\Gamma(s;\la)$ and $\Gamma(s;\la+d\la)$ the condition of null vector alignment \reef{aligned} is satisfied in the continuum limit \ie as $d\la \to 0$. In fact since all the intervals are on a constant time slice, in this case we have tangent vector alignment. Let $s_\pm(\la)$ denote the intersection of the extremal curve $\Gamma(s;\la)$ with $\Gamma(s;\la \pm d\la)$.\footnote{As discussed in section \ref{new}, we expect neighbouring curves to intersect at most once. However, in the situations where the curves are extremal but not minimal, it may be that they intersect more than once, as discussed in section \ref{discuss}. In this case, we can simply choose consecutive points such that eq. \reef{intersectp} holds.} By construction, the `right' intersection point for $\Gamma(s;\la)$ is equal to the `left' intersection point for $\Gamma(s;\la+d\la)$ so we have
\ban{
\Gamma(s_+(\la); \la) &= \Gamma(s_-(\la+d\la); \la+d\la)  \labell{intersectp}
}
We can expand this equation for $d\la \ll |\gamma_R(\la)-\gamma_L(\la)|$ to get
\ban{
\Gamma(s_+(\la);\la)=\Gamma(s_-(\la);\la)+\mathcal O(d\la)
}
And as we are assuming a bijective parameterization this equation implies that $s_+(\la)-s_-(\la)\sim \mathcal O(d\la)$. Therefore we can write
\ban{
s_+(\la) &= s_I(\la)+\delta s_+ (\la)d\la + \mathcal O(d\la^2)\\
s_-(\la) &= s_I(\la)+\delta s_- (\la)d\la + \mathcal O(d\la^2)
}
where we refer to $s_I(\la)$ as the `intersection point in the continuum limit.' Substituting these expressions into \reef{intersectp} we get
\ban{
&\Gamma(s_I(\la); \la)+\dot \Gamma(s_I(\la);\la) \delta s_+(\la) d\la + \mathcal O(d\la^2)\notag \\ &=\Gamma(s_I(\la);\la)+\dot \Gamma(s_I(\la);\la) \delta s_-(\la) d\la + \left( \Gamma'(s_I(\la); \la)+s'_I(\la)\dot \Gamma(s_I(\la);\la)\right)d\la+\mathcal O(d\la^2) \label{intersectdp}
}
\normalsize
And so we have that at the point $s_I(\la)$
\ban{
\alpha(\la) \dot \Gamma(s;\la)|_{s_I(\la)}= \Gamma'(s;\la)|_{s_I(\la)}
}
where $\alpha(\la)=\delta s_+(\la) - \delta s_-(\la)- s'_I(\la)$. In this way, at $s_I(\la)$ the curves satisfy tangent vector alignment, so the bulk curve can be thought of as being built from the intersection points between extremal curves and their neighbours, in the continuum limit. 

\vskip0.2cm

Next we repeat the above analysis for the general case. That is, we show that at the intersection of $\Gamma(s;\la)$ with the entanglement wedge boundary $W(s,\tau;\la\pm d\la)$, we have the desired relation $\Gamma'(s;\la)|_{s_\pm(\la)} \propto \dot \Gamma (s;\la)|_{s_\pm(\la)}+ k_\pm(\la)$. It is convenient to use the parameterization 
\ban{
W^\mu(s,\tau;\la)=\Gamma^\mu(s;\la)+  \tau\, k^\mu(s ;\la) \labell{Wparam}
}
where $k(s;\la)\cdot \dot \Gamma(s;\la)=0$, $|k(s;\la)|=0$, and $\tau \in [0,1]$. That is, the vector $k^\mu(s;\la)$ is the null separation between $\Gamma(s;\la)$ and the `cusp' of the entanglement wedge. Note that there are two such vectors which we denote $k_\uparrow$ and $k_\downarrow$, corresponding to the `upper' and `lower' parts of the entanglement wedge respectively. Hence they lie on two different light sheets and so they can not be smoothly deformed into one another while remaining null.

As above we denote the intersection of $\Gamma(s;\la)$ with $W(s,\tau;\la \pm d\la)$ by $s_\pm(\la)$. For concreteness we will assume that there exists one unique point for both $s_+(\la)$ and $s_-(\la)$, \ie the trajectory of $\Gamma'$ in the transverse plane can only cross each null line once, and we discuss the general case below. By construction we have
\ban{
\Gamma{}^\mu(s_+(\la);\la)&=\Gamma{}^\mu(s^*_+(\la+d\la); \la+d\la) + \tau^*_+(\la+d\la) k^\mu(s^*_+(\la+d\la); \la+d\la) \labell{EWintersect}
}
for some particular $\tau^*_+(\la+d\la)$ and $s^*_+(\la+d\la)$. Note that in general $s^*_+(\la)\neq s_-(\la)$ because the choice of $k_\uparrow$ or $k_\downarrow$ generically differs in the equation analogous to eq.~\reef{EWintersect} for $s_-(\la)$. We return to this point later in the discussion. The general setup and notation is illustrated in figure \ref{EWI}. Expanding eq. \reef{EWintersect} around $d\la$ we have
\ban{
\Gamma^\mu(s_+(\la);\la) = \Gamma^\mu(s^*_+(\la);\la) + \tau^*_+(\la)k^\mu(s^*_+(\la); \la) + \mathcal O(d\la)
}
This equation implies that the zeroth order separation between $\Gamma^\mu(s_+(\la);\la)$ and $\Gamma^\mu(s^*_+(\la);\la)$ is a null vector, but as the extremal curves are spacelike this must vanish. Therefore we can write 
\ban{
s^*_+(\la)&=s_+(\la)+\delta s_+(\la)\, d\la + \mathcal O(d\la^2)\\
\tau^*_+(\la)&= \delta\tau_+(\la)\, d\la + \mathcal O(d\la^2)
}
Plugging this expansion into eq. \reef{EWintersect} and keeping all terms to first order we have
\ban{
\Gamma{}^\mu(s_+(\la);\la)=&\Gamma^\mu(s_+(\la);\la)+\dot \Gamma^\mu(s_+(\la);\la) \delta s_+(\la) d\la+ \dot \Gamma^\mu(s_+(\la);\la)s'_+(\la) d\la \notag\\
&+ \Gamma'{}^\mu(s_+(\la);\la) d\la + \delta\tau_+(\la) k^\mu(s_+(\la);\la)d\la 
}
and we see explicitly 
\ban{
 \Gamma'{}^\mu |_{s_+(\la)}=\alpha \left( \dot \Gamma|_{s_+(\la)}+\tilde k_+\right) \labell{NVA}
}
where $\alpha(\la)= -(\delta s_+(\la)+s'_+(\la))$ and $\tilde k_+(\la)= -\frac{\delta\tau_+(\la)}{\delta s_+(\la)+s'_+(\la)} k(s_+(\la);\la)$. By construction $\tilde k_+$ is null and $\tilde k_+ \cdot \dot \Gamma =0$, and therefore at this point the condition of null vector alignment is satisfied. We can then identify $s_+(\la)$ with $s_B(\la)$ to form the bulk curve as the continuum limit of these intersection points. 

Additionally, we can write down equations analogous to eqs.~\reef{EWintersect} and \reef{NVA} for the intersection point $s_-(\la)$. By repeating the above arguments, we can show that at $s_-(\la)$ the extremal curves satisfy null vector alignment: 
\ban{
 \Gamma'{}^\mu |_{s_-(\la)}=\alpha \left( \dot \Gamma|_{s_-(\la)}+\tilde k_-\right) \labell{NVAm}
}
where $\alpha(\la)= -(\delta s_-(\la)+s'_-(\la))$ and $\tilde k_-(\la)= -\frac{\delta\tau_-(\la)}{\delta s_-(\la)+s'_-(\la)} k(s_-(\la);\la)$. 

Further, note from the above definitions $\tilde k_\pm$ is proportional either to $k_{\uparrow}$ or $k_{\downarrow}$, and we see from the null vector alignment equations \reef{NVA} and \reef{NVAm} that $\tilde k_\pm$ is additionally proportional to the projection of $\Gamma'$ into the transverse plane. We also note that by definition the two null directions in the transverse plane are given exactly by the vectors $k_\uparrow$ and $k_\downarrow$, and so crossings of each null direction in the transverse plane are characterized by the null vector alignment equations \reef{NVA} and \reef{NVAm}. As we assume the trajectories in the transverse plane can cross each light cone only once, then in the continuum limit there can be at most one point on the extremal curve satisfying null vector alignment for each $k_\uparrow$ and $k_\downarrow$. Therefore, we have that if $\tilde k_+(\la) \propto \tilde k_-(\la)$ then $s_+(\la)-s_-(\la) \sim \mathcal O(d\la)$. However, note that this situation can only arise when the extremal curves $\Gamma(s;\la\pm d\la)$ are either both `above' or both `below' $\Gamma(s;\la)$, and so at $\la$ the timelike separation between extremal curves is either a maximum or a minimum. Therefore in the continuum limit, $\delta\tau_\pm(\la)$ vanishes and in this case we additionally have tangent vector alignment, corresponding to a trajectory crossing through the origin as in figure \ref{transit2}a.

Next, we consider the intersection of $\Gamma(s;\la)$ with the complementary entanglement wedge boundary $\hat W(s,\tau;\la+d\la)$. We can similarly parameterize this surface as 
\ban{
\hat W^\mu(s,\tau;\la)=\Gamma^\mu(s;\la)+  \tau\, \hat k^\mu(s ;\la) \labell{Wparam2}
}
where $\hat k(s;\la)\cdot \dot \Gamma(s;\la)=0$, $|\hat k(s;\la)|=0$, and $\tau \in [0,1]$. These conditions do not uniquely fix $\hat k$, but rather pick out the two null rays `ingoing' to the bulk. One can use any non-vanishing vector along this ray, and for convenience we will take $\hat k_\uparrow =- k_\downarrow$ and $\hat k_\downarrow =- k_\uparrow$. Using notation shown in figure \ref{EWcomp}, at the intersection point $\hat s_+(\la)$ we have
\ban{
\Gamma^\mu(\hat s_+(\la);\la) = \Gamma^\mu(\hat s^*_+(\la+d\la);\la+d\la) + \hat \tau^*_+(\la+d\la)\hat k^\mu(\hat s^*_+(\la+d\la); \la+d\la) \labell{compint}
}
for some $\hat s^*_+(\la+d\la)$ and $\hat \tau ^*_+(\la+d\la)$. It is straightforward to repeat the usual analysis to show that at $\hat s_+(\la)$ the condition of null vector alignment is satisfied as $d\la \to 0$. 

Generically by construction if ${\tilde k_+}$ is proportional to $k_\uparrow$, then $\hat k$ in the above equation \reef{compint} is proportional to $k_\downarrow$ and vice versa. Following previously made arguments, the null vector alignment equation for the intersection of the complementary entanglement wedge boundary would involve a vector proportional to $k_\downarrow$, and thus at $\hat s_+(\la)$ the trajectory of $\Gamma'$ in the transverse plane would cross the null direction opposite to the one it crosses at $ s_-(\la)$. However, if $s_+(\la)$ and $s_-(\la)$ are distinct points as in figure \ref{transit2}c, by previous arguments, it is also true that at $s_-(\la)$ the trajectory of $\Gamma'$ in the transverse plane must cross the same null direction as $\hat s_+(\la)$. Finally, assuming the trajectory in the transverse plane can only cross each null direction at most once, we have $\hat s_+(\la) - s_-(\la) \sim \mathcal O(d\la)$. Therefore, as intuitively mentioned in section \ref{new}, considering complementary entanglement wedges does not yield new solutions to the null vector alignment equation. 

In the case of tangent vector alignment where $s_+(\la)-s_-(\la) \sim \mathcal O(d\la)$ as in figure \ref{transit2}a, we had that $\tilde k_+$ is proportional to $\tilde k_-$. Therefore from the argument in the previous paragraph, the analogous vectors for $\hat s_+(\la)$ and $\hat s_-(\la)$ are also proportional. We can make the same argument as we did previously to show that in this case $\hat s_+(\la) -\hat s_-(\la)\sim \mathcal O(d\la)$, and tangent vector alignment is satisfied. However as tangent vector alignment corresponds to the trajectory of $\Gamma'$ in the transverse plane crossing the origin, this point must be unique by assumption. So in this situation, we have that all four intersections are degenerate in the continuum limit, that is considering the intersection of $\Gamma(s;\la)$ with $W(s;\la\pm d\la)$ or $\hat W (s; \la \pm d\la)$ will give the same solution in the continuum limit. 

As mentioned in section \ref{new}, there are trajectories in the transverse plane for which there are multiple crossings of the light cones. In these cases, the proofs in this appendix show that $s_\pm(\la)$ and $\hat s_\pm(\la)$ are all solutions of null vector alignment. However, in these cases we do not expect it to be guaranteed that \eg $s_+-\hat s_-\sim \mathcal O(d\la)$. Instead, one only needs to piece together the solutions for each $\la$ such that the bulk curve formed from the intersection points is continuous. 

One example of multiple crossings that we considered in section \ref{new} was the case where both $\gamma'_L(\la)$ and $\gamma'_R(\la)$ are timelike, as illustrated in figure \ref{transit2x}b. 
In fact, we can generate a simple example of this type of trajectory. Consider extremal surfaces for a strip on the boundary
of higher dimensional AdS$_{d+1}$ space, which extend out to some maximal bulk depth, and return in a symmetric way. For an interval $I_1$ at $t=0$ and $x=0$ of half width $\Delta$, the maximal depth can be written as $z_\text{1,max}= c_d\, \Delta$ with \cite{rt1,rt2}
\be
c_d=\frac{\Gamma\(\frac{1}{2d-2}\)}{\sqrt{\pi}\,\Gamma\(\frac{d}{2d-2}\)}\,.
\labell{const77}
\ee 
Considering an interval $I_2$ shifted up in time by $dt$ and with half width $\Delta -dx$, this shift corresponds to a trajectory starting and ending in the same timelike quadrant when $dt>dx$. We have $z_\text{2,max}=c_d\,(\Delta-dx)$. The general setup is shown in figure \ref{counterexample}
\begin{figure}[h!]
\centering
\subfloat[]{\includegraphics[width=0.45\textwidth]{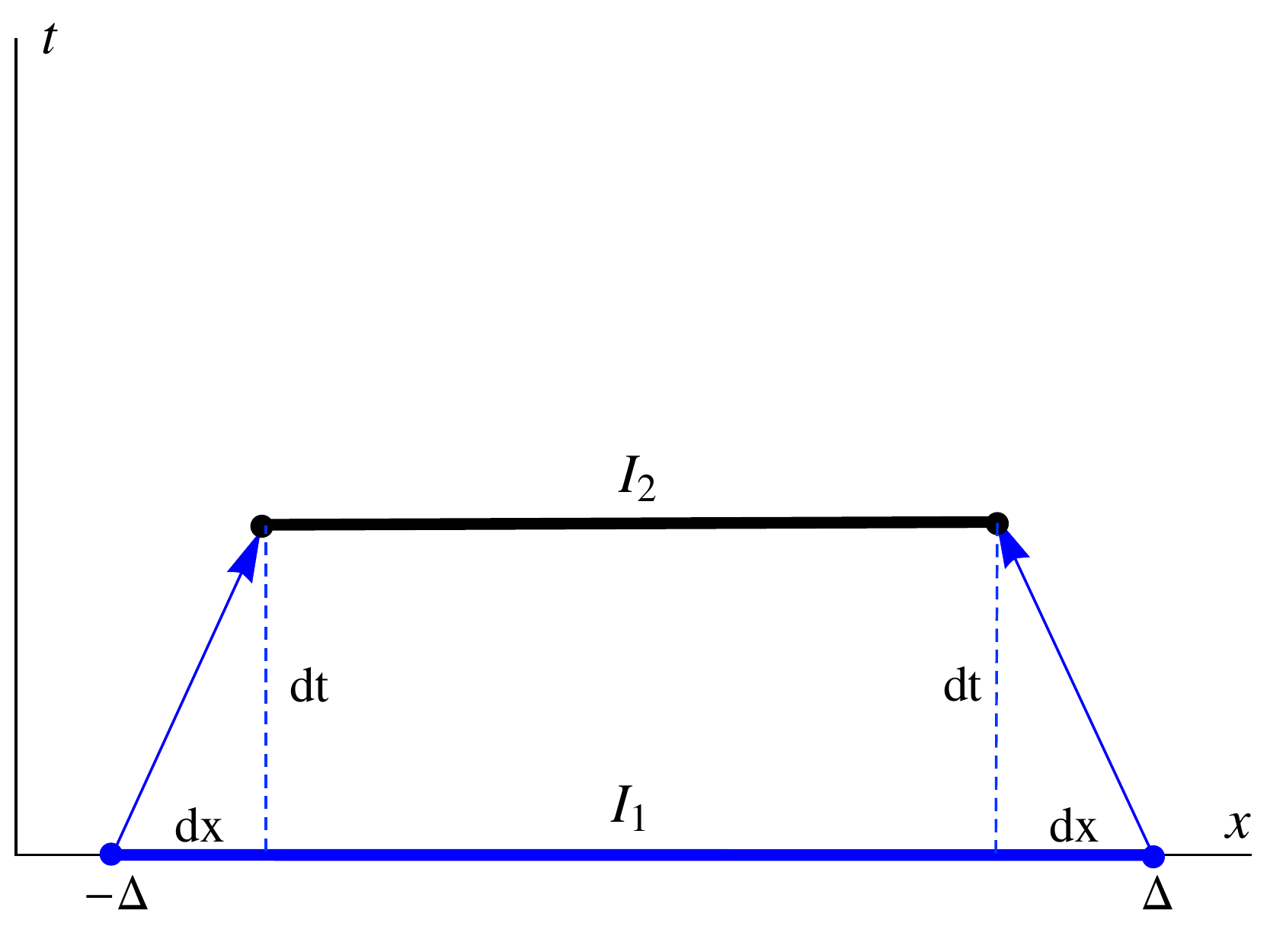}}\qquad
\subfloat[]{\includegraphics[width=0.45\textwidth]{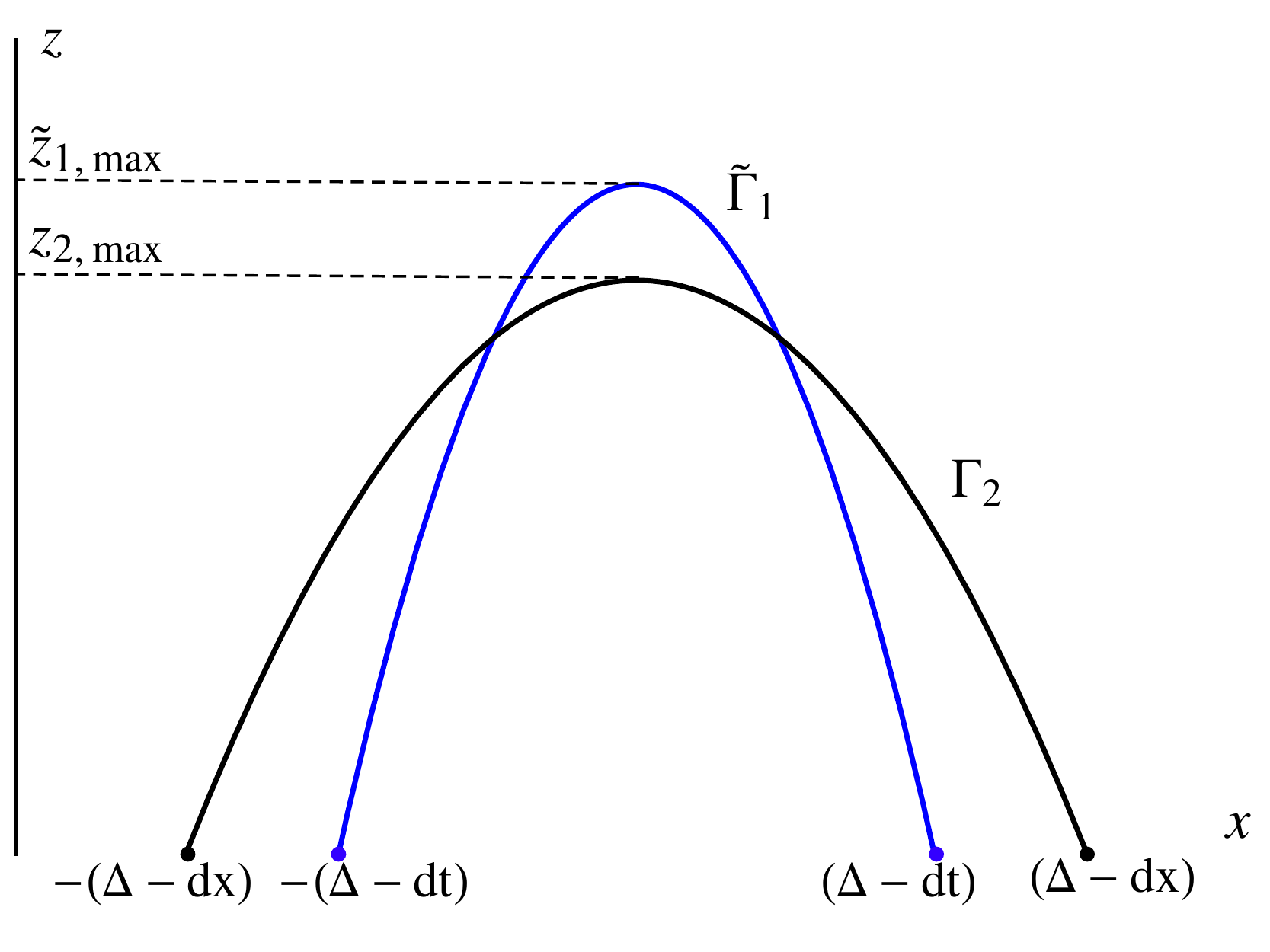}}
\caption{(Color online) In (a) we show the interval $I_2$ whose endpoints are shifted up from the interval $I_1$ by $dt$ in time and inward by $dx$ on each end. As $dt>dx$, this corresponds to a trajectory with endpoints in quadrant II as in figure \ref{transit2x}b. In (b), we show the time slice $dt$ with the extremal surface for $\Gamma_2$ for $I_2$, as well as the surface $\tilde \Gamma_1$, which is the intersection of the entanglement wedge for $I_1$ with the time slice $dt$. In both figures, all planar directions are suppressed.}
\labell{counterexample}
\end{figure}

To consider the intersection of the extremal surface for $ I_2$ with the entanglement wedge boundary of $I_1$, we first characterize the intersection of entanglement wedge boundary of $I_1$ with the time slice $dt$. This surface is formed by following null rays from the extremal surface for $I_2$, and so it has maximum bulk depth $\tilde z_\text{max}= c_d\, \Delta - dt$. Additionally, the endpoints of this surface are at $x=\pm(\Delta -dt)$,  which lie inside the interval $I_2$ with endpoints $x=\pm (\Delta -dx)$ as $dt>dx$. Therefore, the extremal surface for $I_2$ must intersect the entanglement wedge boundary of $I_1$ if $\tilde z_\text{1,max}>z_\text{2,max}$ or if $c_d>dt/dx$. Evaluating eq.~\reef{const77} for $d=2$, \ie AdS$_3$,
we find $c_d=1$, and therefore this trajectory is ruled out as $dt/dx>1$. However in general we find $c_d>1$ for $d\ge3$ and so trajectories that cross the same null line twice are indeed allowed in higher dimensions.

\section{Extension to Lovelock gravity} \labell{lovelock}

The hole-ographic construction was extended in \cite{jun} to the situation where the bulk is described by Lovelock gravity \cite{lovel}.\footnote{For holographic studies of Lovelock gravity, see for example \cite{highc,gb2,highc2}.} This extension accommodated a broad class of holographic backgrounds but was limited to the case where the bulk curve lies in a fixed time slice. The key point was to show that the entropy functional did not depend on higher derivatives of the coordinates. Here, we will further generalize the hole-ographic construction for Lovelock gravity using the general discussion in section \ref{general}. Again we must show that the entropy functional or the Lagrangian in eq.~\reef{action0} is a function only of the canonical coordinates and their first derivatives. The proof of section \ref{general} then automatically applies and we have that
the gravitational entropy is given by the differential entropy of a family of boundary intervals. However, this new proof also
accommodates time varying bulk curves and as we will see, it also allows for more general holographic backgrounds that depend on time and the spatial boundary coordinate, as well as the radial coordinate in the bulk. We note though that while the following discussion applies for a broad class of holographic backgrounds (see eq.~\reef{conon} below), we have not extended it to the most general metrics \reef{pmetric} considered in section \ref{planarS}. We expect that our analysis could be extended to incorporate this more general situation but we leave this for some future work.

Holographic entanglement entropy was first studied in the context of higher curvature gravity theories in the bulk
in \cite{highc,highc2} and there has been a great deal of recent progress \cite{xifriend} by applying the techniques of \cite{aitor,squash}. One simple result is that  the RT prescription is extended 
for Lovelock gravity by replacing the Bekenstein-Hawking entropy in eq.~\reef{define} with
the following entropy functional \cite{tedJ} for a ($d+1$)-dimensional spacetime:
\ban{
S_{JM}= \frac{1}{4\,G_N} \int_a d^{d-1} x\ \sqrt{h} \left[ 1+ \sum _{p=2}^{\left\lfloor \frac{d+1}2\right\rfloor}p\, c_p L^{2p-2} \cL_{2p-2}(\mathcal R)\right]
\labell{sjm}
}
where $h$ is the determinant of the induced metric on the horizon and $c_p$ are the dimensionless couplings associated with the higher curvature interactions  ---  for example, see \cite{highc}. The curvature dependence is given by
\ban{
\cL_{2p}(\mathcal R)= \frac 1{2^p} \delta ^{\nu_1 \cdots \nu_{2p}}_{\mu_1 \cdots \mu_{2p}}\mathcal R^{\mu_1\mu_2}{}_{\nu_1\nu_2}\cdots \mathcal R^{\mu_{2p-1}\mu_{2p}}{}_{\nu_{2p-1}\nu_{2p}}
\labell{euler}
}
where $\mathcal R_{\mu_1 \mu_2}{}_{\nu_1\nu_2}$ denotes the intrinsic curvature tensor on the extremal surface
in the bulk, and $\delta^{\nu_1 \cdots \nu_{n}}_{\mu_1\cdots \mu_n}$ is the totally antisymmetric product of $n$ Kronecker delta symbols. 

First, we consider a general holographic background with coordinates $\{t,x,z,y_i\}$, where $\{y^i\}$ denote the $d-2$ spatial coordinates with generalized planar symmetry. Let $x^\mu$ denote the coordinates $\{t,x,z\}$. Then we consider background metrics of the form  
\ban{
ds^2=\tilde g_{\mu\nu}(t,x,z)\, dx^\mu dx^\nu +\sum_{i=1}^{d-2} g_i(t,x,z)\,(dy^i)^2
\labell{conon}
}
where $\tilde g_{\mu\nu}$ denotes an arbitrary metric on the coordinates $\{x^\mu\}$ with one timelike direction. We consider a  ($d-1$)-dimensional surface with generalized planar symmetry 
given by the embedding: $\{t(\la), x(\la), z(\la), \sigma_i=y^i\}$. The induced metric is then
\ban{
ds^2_\text{ind}= Q(\la)\, d\la^2 + \sum_{i=1}^{d-2} g_i(\la)\,d\sigma_i^2
\labell{induct9}
}
where $Q(\la) = \tilde g_{\mu\nu} \partial_\la x^\mu \partial_\la x^\nu$. 

Now we would like to show that evaluating the the entropy functional \reef{sjm} yields an effective Lagrangian
that depends only on the three coordinates $\{t,x,z\}$ and their first derivatives. In fact, we will find second
derivatives of these coordinate functions but we will explicitly show that all second derivatives 
are simply removed by integrating by parts. 

The Riemann tensor of the induced metric \reef{induct9} has non-vanishing components
\ban{
\mathcal R ^{\la\sigma_i}{}_{\la \sigma_i}&= \frac 1{2 g_i(\la) \sqrt{Q(\la)}}\left[ \frac{g_i'(\la)^2}{2 g_i(\la) \sqrt{Q(\la)}}- \left(\frac{g_i'(\la)}{\sqrt{Q(\la)}}\right)'\,\right] \notag \\
\mathcal R^{\sigma_k\sigma_l}{}_{\sigma_k \sigma_l}&= -\frac 14 \frac {g_k'(\la) g_l'(\la)}{g_k(\la) g_l(\la) Q(\la)}
\labell{getty}
}
where there are no sums over the repeated indices on the left-hand side. Now the only dangerous contribution
is the second term of the first line, as it contains second derivatives of $x^\mu$ through the first derivative $Q'(\la)$ 
and the second derivatives $g_i''(\la)$. Now this contribution will appear in many terms in the entropy functional because of the sums in eqs.~\reef{sjm} and \reef{euler}.  However, we will show that each of these terms can be written in the form
\be
F(g_i(\la))\ \partial_\la {G(g_i(\la),Q(\la), g_i'(\la))}
\labell{formm}
\ee
which we can then integrate by parts to remove all second derivatives, as differentiating the function $F(g_i(\la))$ above only generates first derivatives.

Now the dangerous terms in eq.~\reef{euler} will contain a factor of $R^{\la\sigma_i}{}_{\la\sigma_i}$ but because of the anti-symmetric delta symbol, they can only contain one such factor. Hence in the Lagrangian given by \reef{sjm}, for a given $p$ the terms containing second derivatives in the sum \reef{euler} will all be of the form
\ban{
&\sqrt{h} \mathcal R ^{\la\sigma_{i_1}}{}_{\la \sigma_{i_1}}\mathcal R^{\sigma_{i_2}\sigma_{i_3}}{}_{\sigma_{i_2}\sigma_{i_3}}\cdots \mathcal R^{\sigma_{i_{2p-2}}\sigma_{i_{2p-1}}}{}_{\sigma_{i_{2p-2}}\sigma_{i_{2p-1}}}\notag\\
&={ F(g_i)} \left(\frac{g_{i_1}'(\la)}{\sqrt{Q(\la)}}\right)'  \frac 1 {Q(\la)^{p-1}} \prod _{k={i_2}}^{i_{2p-1}} g_k'(\la)+\cdots \labell{badterms}
}
where $F(g_i)$ is a function of only the coordinate functions as desired\footnote{Explicitly, $F(g_i)=\frac 1{2^{2p-1}}\frac{\sqrt{g_1(\la) \cdots g_{d-2}(\la)}}{g_{i_1}(\la)\cdots g_{i_{2p-1}}(\la)}$} and the omitted terms, denoted by the ellipsis in the second line, depend only on first derivatives.

For a given $p$, we must sum \reef{badterms} over all possible ways to choose $2p-1$ planar coordinates as well as all possible permutations of these coordinates. First, we will simply fix our choice of $2p-1$ coordinates and sum over all possible permutations of the $\sigma_i$ as in eq.~\reef{euler} for these coordinates. Noting that the distinct permutations are identified by which $\sigma_i$ is grouped with $\la$ in the first term, we can write the contribution to the entropy functional from the terms like those in eq.~\reef{badterms} for all possible permutations of a $2p-1$ subset of the planar coordinates as 
\ban{
{F(g_i)}& \sum_{i=i_1}^{i_{2p-1}}\left[ \left( \frac {g_i''(\la)}{\sqrt{Q(\la)}}- \frac 12 \frac {Q'(\la) g_i'(\la)}{Q(\la)^{3/2}}\right) \frac 1{Q(\la)^{p-1}} \prod _{k \neq i} g_k'(\la)\right]\notag \\
=&{F(g_i)} \left( \frac 1 {Q(\la)^{p-1/2}} \sum_{i=i_1}^{i_{2p-1}} \left[g_i''(\la) \left(\prod_{k \neq i} g_k'(\la)\right) \right]- \frac 12 (2p-1) \frac{Q'(\la) \prod_k g_k '(\la)}{Q(\la)^{p+1/2}}\right) \notag \\
=&{F(g_i)} \left(\frac{\prod_k g_k'(\la)}{Q(\la)^{p-1/2}}\right)'
}
Now we can integrate this final term by parts to eliminate all of the second derivatives. For every subset of $2p-1$ coordinates, we can apply the same trick, and so in this way we can write the entropy functional for Lovelock gravity completely in terms of the coordinate functions and their first derivatives only. Hence the general proof in section \ref{general} for the equality of the gravitational entropy of the bulk curve and the differential entropy of the corresponding family of boundary intervals can be applied here as well, and
thus we have extended the hole-ographic construction to Lovelock gravity for the class of holographic backgrounds described by eq.~\reef{conon}.

Implicitly, the above discussion assumes tangent vector alignment \reef{tangent2}. We would also like to consider the analogy of null vector alignment \reef{conull} for the present higher curvature theories of gravity.
Recall that eq.~\reef{conull} arose from requiring the second equality in eq.~\reef{proof}, \ie 
\ban{
\pd{\cL}{\dot \gamma^\mu_B} \gamma'{}^\mu_B =\pd{\cL}{\gamma'{}^\mu_B} \gamma'{}^\mu_B\,. \labell{crucial}
}
The simplest solution to this constraint comes from requiring 
\be
\pd{\cL}{\dot \gamma^\mu_B} =\pd{\cL}{\gamma'{}^\mu_B}\,.
\labell{simpleG}
\ee
Note that these two expressions are evaluated at the same point in the bulk and so eq.~\reef{simpleG} becomes a
constraint relating the two velocities $\dot \gamma^\mu_B$ and $\gamma'{}^\mu_B$ and certainly the simplest
solution is simply $\dot\gamma_B\propto\gamma_B'$, \ie tangent vector alignment.\footnote{Recall the reparametrization invariance of the action ensures the same for the momenta and hence $\dot\gamma_B\propto\gamma_B'$ (rather than $\dot\gamma_B=\gamma_B'$) is sufficient to ensure the
equality \reef{simpleG}.} However, following the discussion in section \ref{new}, eq.~\reef{crucial} is also
satisfied if we impose the conditions
\ban{
\pd{\cL}{\dot \gamma^\mu_B} =\pd{\cL}{\gamma'{}^\mu_B}+k_\mu
\quad{\rm with}\ \ k_\mu\,\gamma'{}^\mu_B  = 0\,. \labell{ortho}
}
This generalized condition also ensures that eq.~\reef{proof} holds and
therefore the result of section \ref{general} is still valid. Recall that for Einstein gravity $\pd{\cL}{\dot \gamma^\mu_B}=\frac{ g_{\mu\nu}\,\dot \gamma^\nu_B}{|\dot \gamma_B|}$ and so eq.~\reef{sol3} is equivalent to
the general solution \reef{ortho} for this specific case. However, recall that the nomenclature `null vector alignment'
in this case arose because it also followed that $|k|=0$ and $k_\mu\,\dot\gamma_B^\mu=0$. However, the latter
constraints, in particular $|k|=0$ and hence the connection to entanglement wedges, do not obviously arise in the more general case of Lovelock gravity. The latter is perhaps not surprising because the `null cone' for linearized gravitons
is modified in these higher curvature theories, \eg \cite{cones}. 
It would certainly be interesting to investigate further the implications of the generalized alignment condition \reef{ortho}
for these theories.

%%%%%%%%%%%%%%%%%%%%%%%%%%%%%%%%%%%%%%%%%%%%%%%%%%%%%%%%%%%%%%%%%%%%%%%%%%%%%%%%%%%%%%%%%%%%%%%%%
%%%%%%%%%%%%%%%%%%%%%%%%%%%%%%%%%%%%%%%%%%%%%%%%%%%%%%%%%%%%%%%%%%%%%%%%%%%%%%%%%%%%%%%%%%%%%%%%%


\begin{thebibliography}{9}

\bibitem %[hole]
 {hole} V.~Balasubramanian, B.~D.~Chowdhury, B.~Czech, J.~de Boer and M.~P.~Heller,
  ``A hole-ographic spacetime,''
  Phys.\ Rev.\ D {\bf 89}, 086004 (2014)
  [arXiv:1310.4204 [hep-th]].
  %%CITATION = ARXIV:1310.4204;%%

\bibitem %[jun]
 {jun} R.~C.~Myers, J.~Rao and S.~Sugishita,
  ``Holographic Holes in Higher Dimensions,''
  arXiv:1403.3416 [hep-th].
  %%CITATION = ARXIV:1403.3416;%%

\bibitem %[beks]
 {beks} J.~D.~Bekenstein,
``Black holes and the second law,''
  Lett.\ Nuovo Cim.\  {\bf 4}, 737 (1972);\\
  %%CITATION = NCLTA,4,737;%%
 J.~D. Bekenstein, ``{Black holes and entropy},''
    Phys. Rev. D {\bf 7}, 2333 (1973);\\
%%CITATION = PHRVA,D7,2333;%%
 J.~D.~Bekenstein,
  ``Generalized second law of thermodynamics in black hole physics,''
  Phys.\ Rev.\ D {\bf 9}, 3292 (1974).
  %%CITATION = PHRVA,D9,3292;%%

\bibitem %[hawk0]
 {hawk0} S.~W. Hawking, ``{Black hole explosions},'' {Nature} {\bf 248}, 30 (1974);\\
%%CITATION = NATUA,248,30;%%.
S.~W. Hawking, ``{Particle Creation by Black Holes},'' {Commun. Math.
  Phys.} {\bf 43}, 199 (1975).
%%CITATION = CMPHA,43,199;%%.

\bibitem %[areaent]
 {areaent} R. D. Sorkin, ``On the Entropy of the Vacuum Outside a Horizon,'' in
{\it General Relativity and Gravitation}, Volume 1, B. Bertotti, F. de Felice
and A. Pascolini, ed., p. 734 (1983);\\
 L.~Bombelli, R.~K.~Koul, J.~Lee and R.~D.~Sorkin,
  ``A Quantum Source of Entropy for Black Holes,''
  Phys.\ Rev.\ D {\bf 34}, 373 (1986);\\
  %%CITATION = PHRVA,D34,373;%%
 M.~Srednicki,
  ``Entropy and area,''
  Phys.\ Rev.\ Lett.\  {\bf 71}, 666 (1993)
  [hep-th/9303048];\\
  %%CITATION = HEP-TH/9303048;%%
L.~Susskind and J.~Uglum,
 ``Black hole entropy in canonical quantum gravity and superstring theory,''
  Phys.\ Rev.\ D {\bf 50}, 2700 (1994)
  [hep-th/9401070];\\
  %%CITATION = HEP-TH/9401070;%%
C.~G.~Callan, Jr. and F.~Wilczek,
  ``On geometric entropy,''
  Phys.\ Lett.\ B {\bf 333}, 55 (1994)
  [hep-th/9401072];\\
  %%CITATION = HEP-TH/9401072;%%
  E.~Bianchi,
  ``Horizon entanglement entropy and universality of the graviton coupling,''
  arXiv:1211.0522 [gr-qc].
  %%CITATION = ARXIV:1211.0522;%%

\bibitem %[new1]
 {new1} E.~Bianchi and R.~C.~Myers,
  ``On the Architecture of Spacetime Geometry,'' to appear in {\sl Classical and Quantum gravity} [arXiv:1212.5183 [hep-th]].
  %%CITATION = ARXIV:1212.5183;%%

\bibitem %[misha9]
 {misha9} R.~C.~Myers, R.~Pourhasan and M.~Smolkin,
  ``On Spacetime Entanglement,''
  JHEP {\bf 1306}, 013 (2013)
  [arXiv:1304.2030 [hep-th]].
  %%CITATION = ARXIV:1304.2030;%%

\bibitem %[WaldEnt]
 {WaldEnt} R.~M.~Wald, ``Black hole entropy is the
    Noether charge,''
  Phys.\ Rev.\  D {\bf 48}, 3427 (1993)
  [arXiv:gr-qc/9307038];\\
  %%CITATION = PHRVA,D48,3427;%%
T.~Jacobson, G.~Kang and R.~C.~Myers,
  ``On Black Hole Entropy,''
  Phys.\ Rev.\  D {\bf 49}, 6587 (1994)
  [arXiv:gr-qc/9312023];\\
  %%CITATION = PHRVA,D49,6587;%%
V.~Iyer and R.~M.~Wald, ``Some properties of Noether charge and a proposal for
dynamical black hole entropy,''
  Phys.\ Rev.\  D {\bf 50}, 846 (1994)
  [arXiv:gr-qc/9403028].
  %%CITATION = PHRVA,D50,846;%%

\bibitem %[revue]
 {revue} O.~Aharony, S.~S.~Gubser, J.~M.~Maldacena, H.~Ooguri and Y.~Oz,
``Large N field theories, string theory and gravity,'' Phys.\ Rept.\  {\bf 323}, 183 (2000) [arXiv:hep-th/9905111].
  %%CITATION = PRPLC,323,183;%%

\bibitem %[rt1]
 {rt1} S.~Ryu and T.~Takayanagi,
  ``Holographic derivation of entanglement entropy from AdS/CFT,''
  Phys.\ Rev.\ Lett.\  {\bf 96}, 181602 (2006)
  [arXiv:hep-th/0603001].
  %%CITATION = PRLTA,96,181602;%%

\bibitem %[rt2]
 {rt2}  S.~Ryu and T.~Takayanagi,
  ``Aspects of holographic entanglement entropy,''
  JHEP {\bf 0608}, 045 (2006)
  [arXiv:hep-th/0605073].
  %%CITATION = JHEPA,0608,045;%%

\bibitem %[rt3]
 {rt3}  T.~Nishioka, S.~Ryu and T.~Takayanagi,
  ``Holographic Entanglement Entropy: An Overview,''
  J.\ Phys.\ A  {\bf 42}, 504008 (2009)
  [arXiv:0905.0932 [hep-th]].
  %%CITATION = JPAGB,A42,504008;%%

\bibitem %[head]
 {head} M.~Headrick,
  ``Entanglement Renyi entropies in holographic theories,''
  arXiv:1006.0047 [hep-th].
  %%CITATION = ARXIV:1006.0047;%%

\bibitem %[fur06a]
 {fur06a} D.~V.~Fursaev,
  ``Proof of the holographic formula for entanglement entropy,''
  JHEP {\bf 0609}, 018 (2006)
  [hep-th/0606184].
  %%CITATION = HEP-TH/0606184;%%

\bibitem %[highc]
 {highc} L.-Y.~Hung, R.~C.~Myers and M.~Smolkin,
  ``On Holographic Entanglement Entropy and Higher Curvature Gravity,''
  JHEP {\bf 1104}, 025 (2011)
  [arXiv:1101.5813 [hep-th]].
  %%CITATION = ARXIV:1101.5813;%%

\bibitem %[aitor]
{aitor} A.~Lewkowycz and J.~Maldacena,
  ``Generalized gravitational entropy,''
  JHEP {\bf 1308}, 090 (2013)
  [arXiv:1304.4926 [hep-th]].
  %%CITATION = ARXIV:1304.4926;%%

\bibitem %[hrt]
 {hrt} V.~E.~Hubeny, M.~Rangamani and T.~Takayanagi,
  ``A Covariant holographic entanglement entropy proposal,''
  JHEP {\bf 0707}, 062 (2007)
  [arXiv:0705.0016 [hep-th]].
  %%CITATION = ARXIV:0705.0016;%%

\bibitem %[matcov]
 {matcov} R.~Callan, J.-Y.~He and M.~Headrick,
  ``Strong subadditivity and the covariant holographic entanglement entropy formula,''
  JHEP {\bf 1206}, 081 (2012)
  [arXiv:1204.2309 [hep-th]].
  %%CITATION = ARXIV:1204.2309;%%

\bibitem %[aron2]
 {aron2} A.~C.~Wall,
 ``Maximin Surfaces and the Strong Subadditivity of the Covariant Holographic Entanglement Entropy,''
  arXiv:1211.3494 [hep-th].
  %%CITATION = ARXIV:1211.3494;%%

\bibitem %[chm]
 {chm} H.~Casini, M.~Huerta and R.~C.~Myers,
  ``Towards a derivation of holographic entanglement entropy,''
  JHEP {\bf 1105}, 036 (2011)
  [arXiv:1102.0440 [hep-th]].
  %%CITATION = ARXIV:1102.0440;%%

\bibitem %[eom1]
 {eom1} T.~Faulkner, M.~Guica, T.~Hartman, R.~C.~Myers and M.~Van Raamsdonk,
  ``Gravitation from Entanglement in Holographic CFTs,''
  arXiv:1312.7856 [hep-th].
  %%CITATION = ARXIV:1312.7856;%%

\bibitem %[also]
 {also} T.~Faulkner, A.~Lewkowycz and J.~Maldacena,
  ``Quantum corrections to holographic entanglement entropy,''
  JHEP {\bf 1311}, 074 (2013)
  [arXiv:1307.2892].
  %%CITATION = ARXIV:1307.2892;%%

\bibitem %[mark1]
 {mark1} V.~Balasubramanian, M.~B.~McDermott and M.~Van Raamsdonk,
  ``Momentum-space entanglement and renormalization in quantum field theory,''
  Phys.\ Rev.\ D {\bf 86}, 045014 (2012)
  [arXiv:1108.3568 [hep-th]].
  %%CITATION = ARXIV:1108.3568;%%

\bibitem %[causal0]
 {causal0} V.~E.~Hubeny and M.~Rangamani,
  ``Causal Holographic Information,''
  JHEP {\bf 1206} (2012) 114
  [arXiv:1204.1698 [hep-th]];\\
  %%CITATION = ARXIV:1204.1698;%%
B.~Freivogel and B.~Mosk,
  ``Properties of Causal Holographic Information,''
  JHEP {\bf 1309}, 100 (2013)
  [arXiv:1304.7229 [hep-th]];\\
  %%CITATION = ARXIV:1304.7229;%%
V.~E.~Hubeny, M.~Rangamani and E.~Tonni,
  ``Global properties of causal wedges in asymptotically AdS spacetimes,''
  JHEP {\bf 1310} (2013) 059
  [arXiv:1306.4324 [hep-th]];\\
  %%CITATION = ARXIV:1306.4324;%%
W.~R.~Kelly and A.~C.~Wall,
  ``Coarse-grained entropy and causal holographic information in AdS/CFT,''
  arXiv:1309.3610 [hep-th].
  %%CITATION = ARXIV:1309.3610;%%

\bibitem %[thesis]
 {thesis} J.~Wien, ``A Holographic Approach to Spacetime Entanglement," Master's Thesis, Perimeter
Scholars International, University of Waterloo (2014),
arXiv:1408.6005 [hep-th].
  %%CITATION = ARXIV:1408.6005;%%

\bibitem %[matt]
 {matt} M.~Headrick,
  ``General properties of holographic entanglement entropy,''
  JHEP {\bf 1403}, 085 (2014)
  [arXiv:1312.6717 [hep-th]].
  %%CITATION = ARXIV:1312.6717;%%

\bibitem %[cardy1]
 {cardy1} P.~Calabrese and J.~L.~Cardy,
  ``Entanglement entropy and quantum field theory,''
  J.\ Stat.\ Mech.\  {\bf 0406}, P06002 (2004)
  [hep-th/0405152].
  %%CITATION = HEP-TH/0405152;%%

\bibitem %[mattEW]
 {mattEW} M.~Headrick, V.~E.~Hubeny, A.~Lawrence and M.~Rangamani,
  ``Causality \& holographic entanglement entropy,''
  arXiv:1408.6300 [hep-th].
  %%CITATION = ARXIV:1408.6300;%%

\bibitem %[verona]
 {verona} V.~E.~Hubeny,
  ``Covariant Residual Entropy,''
  arXiv:1406.4611 [hep-th].
  %%CITATION = ARXIV:1406.4611;%%

\bibitem %[sully]
 {sully}  B.~Czech, X.~Dong and J.~Sully,
  ``Holographic Reconstruction of General Bulk Surfaces,''
  arXiv:1406.4889 [hep-th].
  %%CITATION = ARXIV:1406.4889;%%
  
\bibitem %[multiple]
 {multiple} 
   I.~R.~Klebanov, D.~Kutasov and A.~Murugan,
  ``Entanglement as a probe of confinement,''
  Nucl.\ Phys.\ B {\bf 796}, 274 (2008)
  [arXiv:0709.2140 [hep-th]]; \\
  %%CITATION = ARXIV:0709.2140;%%
  T.~Albash and C.~V.~Johnson,
  ``Evolution of Holographic Entanglement Entropy after Thermal and Electromagnetic Quenches,''
  New J.\ Phys.\  {\bf 13}, 045017 (2011)
  [arXiv:1008.3027 [hep-th]];\\
  %%CITATION = ARXIV:1008.3027;%%
V.~E.~Hubeny, H.~Maxfield, M.~Rangamani and E.~Tonni,
  ``Holographic entanglement plateaux,''
  JHEP {\bf 1308}, 092 (2013)
  [arXiv:1306.4004 [hep-th]];\\
  %%CITATION = ARXIV:1306.4004,;%%
 R.~C.~Myers and A.~Singh,
  ``Comments on Holographic Entanglement Entropy and RG Flows,''
  JHEP {\bf 1204}, 122 (2012)
  [arXiv:1202.2068 [hep-th]].
  %%CITATION = ARXIV:1202.2068;%%

\bibitem %[extreme]
 {extreme} V.~Balasubramanian, B.~D.~Chowdhury, B.~Czech and J.~de Boer,
  ``Entwinement and the emergence of spacetime,''
  arXiv:1406.5859 [hep-th].
  %%CITATION = ARXIV:1406.5859;%%

\bibitem %[walls]
 {walls} N.~Engelhardt and A.~C.~Wall,
  ``Extremal Surface Barriers,''
  JHEP {\bf 1403}, 068 (2014)
  [arXiv:1312.3699 [hep-th]];\\
  %%CITATION = ARXIV:1312.3699;%%
V.~E.~Hubeny,
  ``Extremal surfaces as bulk probes in AdS/CFT,''
  JHEP {\bf 1207}, 093 (2012)
  [arXiv:1203.1044 [hep-th]];\\
  %%CITATION = ARXIV:1203.1044;%% 
S.~S.~Pal,
  ``Extremal Surfaces And Entanglement Entropy,''
  Nucl.\ Phys.\ B {\bf 882}, 352 (2014)
  [arXiv:1312.0088 [hep-th]].
  %%CITATION = ARXIV:1312.0088;%%

\bibitem %[progress]
 {progress} R.~C.~Myers and J.~Rao, in preparation.

\bibitem %[wilson]
 {wilson} J.~M.~Maldacena,
  ``Wilson loops in large N field theories,''
  Phys.\ Rev.\ Lett.\  {\bf 80}, 4859 (1998)
  [hep-th/9803002];\\
  %%CITATION = HEP-TH/9803002;%%
S.-J.~Rey, S.~Theisen and J.~-T.~Yee,
  ``Wilson-Polyakov loop at finite temperature in large N gauge theory and anti-de Sitter supergravity,''
  Nucl.\ Phys.\ B {\bf 527}, 171 (1998)
  [hep-th/9803135].
  %%CITATION = HEP-TH/9803135;%%

\bibitem %[twop]
 {twop}  V.~Balasubramanian and S.~F.~Ross,
  ``Holographic particle detection,''
  Phys.\ Rev.\ D {\bf 61}, 044007 (2000)
  [hep-th/9906226];\\
  %%CITATION = HEP-TH/9906226;%%
J.~Louko, D.~Marolf and S.~F.~Ross,
  ``On geodesic propagators and black hole holography,''
  Phys.\ Rev.\ D {\bf 62}, 044041 (2000)
  [hep-th/0002111].
  %%CITATION = HEP-TH/0002111;%%

\bibitem %[lovel]
 {lovel} D.~Lovelock, ``The Einstein tensor and its generalizations,''
  J.\ Math.\ Phys.\  {\bf 12}, 498 (1971);\\
  %%CITATION = JMAPA,12,498;%%
D.~Lovelock, ``Divergence-free tensorial concomitants,'' Aequationes
Math. {\bf 4}, 127 (1970).

\bibitem %[gb2]
 {gb2} A.~Buchel, J.~Escobedo, R.~C.~Myers, M.~F.~Paulos, A.~Sinha and M.~Smolkin,
  ``Holographic GB gravity in arbitrary dimensions,''
  JHEP {\bf 1003}, 111 (2010)
  [arXiv:0911.4257 [hep-th]];\\
  %%CITATION = ARXIV:0911.4257;%%
J.~de Boer, M.~Kulaxizi and A.~Parnachev,
  ``Holographic Lovelock Gravities and Black Holes,''
  JHEP {\bf 1006}, 008 (2010)
  [arXiv:0912.1877 [hep-th]];\\
  %%CITATION = ARXIV:0912.1877;%%
X.~O.~Camanho, J.~D.~Edelstein and M.~F.~Paulos,
  ``Lovelock theories, holography and the fate of the viscosity bound,''
  JHEP {\bf 1105}, 127 (2011)
  [arXiv:1010.1682 [hep-th]].
  %%CITATION = ARXIV:1010.1682;%%

\bibitem %[highc2]
 {highc2} J.~de Boer, M.~Kulaxizi and A.~Parnachev,
  ``Holographic Entanglement Entropy in Lovelock Gravities,''
  JHEP {\bf 1107}, 109 (2011)
  [arXiv:1101.5781 [hep-th]].
  %%CITATION = ARXIV:1101.5781;%%

\bibitem %[xifriend]
 {xifriend} X.~Dong,
  ``Holographic Entanglement Entropy for General Higher Derivative Gravity,''
  JHEP {\bf 1401}, 044 (2014)
  [arXiv:1310.5713 [hep-th]];\\
  %%CITATION = ARXIV:1310.5713;%%
 J.~Camps, ``Generalized entropy and higher derivative Gravity,''
  arXiv:1310.6659 [hep-th];\\
  %%CITATION = ARXIV:1310.6659;%%
A.~Bhattacharyya, A.~Kaviraj and A.~Sinha,
  ``Entanglement entropy in higher derivative holography,''
  JHEP {\bf 1308}, 012 (2013)
  [arXiv:1305.6694 [hep-th]];\\
  %%CITATION = ARXIV:1305.6694;%%
A.~Bhattacharyya, M.~Sharma and A.~Sinha,
  ``On generalized gravitational entropy, squashed cones and holography,''
  JHEP {\bf 1401}, 021 (2014)
  [arXiv:1308.5748 [hep-th]];\\
  %%CITATION = ARXIV:1308.5748;%%
A.~Bhattacharyya and M.~Sharma,
  ``On entanglement entropy functionals in higher derivative gravity theories,''
  arXiv:1405.3511 [hep-th].
  %%CITATION = ARXIV:1405.3511;%%

\bibitem %[squash]
 {squash} D.~V.~Fursaev, A.~Patrushev and S.~N.~Solodukhin,
  ``Distributional Geometry of Squashed Cones,''
  Phys.\ Rev.\ D {\bf 88}, no. 4, 044054 (2013)
  [arXiv:1306.4000 [hep-th]].
  %%CITATION = ARXIV:1306.4000;%%

\bibitem %[tedJ]
 {tedJ} T.~Jacobson and R.~C.~Myers,
  ``Black hole entropy and higher curvature interactions,''
  Phys.\ Rev.\ Lett.\  {\bf 70}, 3684 (1993)
  [hep-th/9305016].
  %%CITATION = HEP-TH/9305016;%%

\bibitem %[cones]
 {cones} M.~Brigante, H.~Liu, R.~C.~Myers, S.~Shenker and S.~Yaida,
  ``Viscosity Bound Violation in Higher Derivative Gravity,''
  Phys.\ Rev.\ D {\bf 77}, 126006 (2008)
  [arXiv:0712.0805 [hep-th]];\\
  %%CITATION = ARXIV:0712.0805;%%
C.~Aragone,
  ``Stringy Characteristics of Effective Gravity,''
  in {\it SILARG VI : proceedings}, edited by M.~Novello, {\it World Scientific, Singapore, (1988)};\\
Y.~Choquet-Bruhat, J.\ Math.\ Phys.\ {\bf 29}, 1891 (1988).




\end{thebibliography}
\end{document}